\pdfoutput=1
\documentclass{JINST}
\let\ifpdf\relax
\usepackage{graphicx}

\newcommand{\uboone}{$\mu$BooNE}

\title{Summary of the Second Workshop on Liquid Argon Time Projection Chamber Research and Development in the United States}         

\author{R.~Acciarri$^a$,
M.~Adamowski$^a$,
D.~Artrip$^b$,
B.~Baller$^a$,
C.~Bromberg$^c$,
F.~Cavanna$^{a,d}$,
B.~Carls$^a$,
H.~Chen$^e$,
G.~Deptuch$^a$,
L.~Epprecht$^f$,
R.~Dharmapalan$^g$
W.~Foreman$^h$
A.~Hahn$^a$,
M.~Johnson$^a$,
B.~J.~P.~Jones $^i$,
T.~Junk$^a$,
K.~Lang$^j$,
S.~Lockwitz$^a$,
A.~Marchionni$^a$,
C.~Mauger$^k$,
C.~Montanari$^l$,
S.~Mufson$^m$,
M.~Nessi$^n$,
H.~Olling~Back$^o$,
G.~Petrillo$^p$,
S.~Pordes$^a$,
J.~Raaf$^a$,
B.~Rebel$^a$,
G.~Sinins$^k$,
M.~Soderberg$^{a,q}$, 
N.~Spooner$^r$,
M.~Stancari$^a$,
T.~Strauss$^s$,
K.~Terao$^t$,
C.~Thorn$^e$,
T.~Tope$^a$,
M.~Toups$^i$,
J.~Urheim$^m$,
R.~Van de Water$^k$,
H.~Wang$^u$,
R.~Wasserman$^v$,
M.~Weber$^s$,
D.~Whittington$^m$,
T.~Yang$^a$\\
\llap{$^a$}Fermi National Accelerator Laboratory, Batavia, IL 60510, USA\\
\llap{$^b$}Research Catalytics, USA\\
\llap{$^c$}Michigan State University, East Lansing, MI 48824, USA\\
\llap{$^d$}Yale University, New Haven, CT 06520, USA\\
\llap{$^e$}Brookhaven National Laboratory, Upton, NY 11973, USA\\
\llap{$^f$}ETH Z\"urich, 8092 Z\"urich, Switzerland\\
\llap{$^g$}Argonne National Laboratory, Argonne, IL, 60439, USA\\
\llap{$^h$}University of Chicago, Chicago, IL 60637, USA\\
\llap{$^i$}Massachusetts Institute of Technology, Cambridge, MA 02139, USA\\
\llap{$^j$}University of Texas at Austin, TX 78712, USA\\
\llap{$^k$}Los Alamos National Laboratory, Los Alamos, NM 87545, USA\\
\llap{$^l$}Istituto Nazionale di Fisica Nucleare, Pavia 6-27100, Italy\\
\llap{$^m$}Indiana University, Bloomington, IN 47405, USA\\
\llap{$^n$}CERN, 1217 Meyrin, Switzerland\\
\llap{$^o$}Princeton University, Princeton, NJ 08544, USA\\
\llap{$^p$}University of Rochester, Rochester, NY 14627, USA\\
\llap{$^q$}Syracuse University, NY 13210, USA\\
\llap{$^r$}University of Sheffield, Sheffield, South Yorkshire S10 2TN, UK\\
\llap{$^s$}University of Bern, 3012 Bern, Switzerland\\
\llap{$^t$}Columbia University, New York, NY 10027, USA\\
\llap{$^u$}University of California Los Angeles, Los Angeles, CA 90095, USA\\
\llap{$^v$}Colorado State University, Fort Collins, CO 80523, USA}

\abstract{The second workshop to discuss the development of liquid argon time projection chambers (LArTPCs) in the United States was held at Fermilab on July 8-9, 2014.  The workshop was organized under the auspices of the Coordinating Panel for Advanced Detectors, a body that was initiated by the American Physical Society Division of Particles and Fields.  All presentations at the workshop were made in six topical plenary sessions:  $i)$ Argon Purity and Cryogenics, $ii)$ TPC and High Voltage, $iii)$ Electronics, Data Acquisition and Triggering, $iv)$ Scintillation Light Detection, $v)$ Calibration and Test Beams, and $vi)$ Software.  This document summarizes the current efforts in each of these areas.  It primarily focuses on the work in the US, but also highlights work done elsewhere in the world.
}


\begin{document}

\section{Introduction}
\label{sec:Introduction}

Liquid argon time projection chambers (LArTPCs) have found wide acceptance as detectors for neutrino experiments to study neutrino oscillations and cross sections.  These detectors are expected to be the primary detectors for future short- and long-baseline oscillation experiments.  The basic principal behind the LArTPC is that a minimum ionizing charged particle passing through the liquid argon produces 55,000 electrons and 80,000 scintillation photons for every centimeter traversed.  The total number of ionization electrons and scintillation photons are proportional to the energy deposited in the medium.  The electrons are drifted in a uniform electric field toward a readout; the uniform field provides a constant drift velocity, allowing one to determine with excellent resolution the location in the drift direction at which the ionization occurred.  This resolution provides neutrino experiments the ability to digitally record bubble-chamber-like images of the neutrino interactions, allowing researchers to distinguish between different interaction processes with certainty.  The scintillation light is also proportional to the energy deposited by the particle and a combination of the charge and light information provides the best measurement of the calorimetric energy of the interaction.



This document is the summary of the second workshop held at Fermilab to discuss the current efforts to develop LArTPC detectors in the US~\cite{workshop}.  The workshop, held July 8-9, 2014, was organized under the auspices of the Coordinating Panel for Advanced Detectors (CPAD), a standing panel that was empowered by the American Physical Society Division of Particles and Fields (DPF).  CPAD is intended to promote national detector R\&D programs and stimulate new ideas in instrumentation development.  CPAD also facilitates coordination among the national HEP laboratories and university groups engaged in detector R\&D and the use of targeted resources at the national laboratories.  

This second workshop was attended by 85 individuals from 32 institutions.  All presentations at the workshop were made in plenary sessions organized into these topical categories:  $i)$ Argon Purity and Cryogenics, $ii)$ TPC and High Voltage, $iii)$ Electronics, Data Acquisition and Triggering, $iv)$ Scintillation Light Detection, $v)$ Calibration and Test Beams, and $vi)$ Software.  This document summaries the information presented in those sessions in the sections below.  The work for several efforts can be placed in more than one of the topical categories.  As such, the reader will find aspects of those projects described throughout the document.  Important lessons learned in any of those areas that are key to building LArTPC experiments are preceded by a "{\bf Learned Lesson}" tag in the text.

\section{Physics Requirements for Neutrino Experiments}
\label{sec:PhysicsRequirements}

The major physics measurements proposed for large LArTPCs in long baseline experiments are the determination of the mass ordering and the precision measurement of the parameters defining the PMNS mixing matrix, particularly the CP violating phase $\delta$.  The detection of a supernova core-collapse event, relic supernova neutrinos, and the determination of the least upper bound for the proton partial lifetime in one or more important decay modes are also major goals that can be accomplished with very large LArTPCs located deep underground. Smaller LArTPCs can be used to resolve the conflicting observations of short baseline neutrino oscillations and the observation of coherent neutrino scattering and measurements of cross sections for this process.  Successfully achieving these goals with LArTPCs will require additional studies to develop a precise understanding of cross sections for neutrino and anti-neutrino interactions in argon, and a detailed knowledge of the final state interactions and other nuclear effects.

The successful achievement of these physics goals places many demands on the performance of LArTPCs, and on our understanding of the processes involved in producing the observed charge and light signals and on the backgrounds that accompany the desired signals.  For the precise determination of cross sections, clean identification of the final reaction products is required.  This goal also demands precise energy loss and range determination, precise total energy measurement for isolated tracks as well as showers, and high spatial resolution, especially near the vertex.  In general, the high spatial resolution and ability to separate multiple closely spaced tracks gives LArTPC the ability to use topological classification to help identify events in the presence of final state interactions and to reconstruct total energy. To properly separate neutral and charged current interactions, good separation of electron- and photon-induced showers obtained by observing the charge density at the beginning of the showers is required.  The identification of final state interactions will require low energy thresholds in order to observe, for example, Compton scattered photons from nuclear de-excitation and low energy argon recoils from neutrons.  To separate neutrino and anti-neutrino charge current interactions, charge determination of the outgoing muon is required.  This separation can be done by creating a magnetic field in the detector, or alternatively through statistical means by observing $\mu^{-}$ capture.  In the latter case, the cross section for $\mu^{-}$ capture needs to be well determined.  For supernova neutrino detection and coherent neutrino-nucleus scattering low energy detection thresholds will be required.  For coherent neutrino scattering, in particular, the thresholds are so low (keV) that scintillation light is the only useful signal available.

\section{Argon Purity and Cryogenic Systems}
\label{sec:Cryogenics}

Fermilab has the most experience in designing and operating liquid argon systems in the US. Cryogenic solutions and argon purification systems have been developed, built and successfully operated at Fermilab in a suite of dedicated test stand devices and physics experiments. At present, some of these devices have completed their program, like ArgoNeut~\cite{Anderson:2011ce} and the LAPD (Liquid Argon Purity Demonstrator)~\cite{ref:LAPD}, some other are in the course of operation, such as the LBNE 35 ton prototype~\cite{lbnedune}, or about to start, like MicroBooNE~\cite{MicroBooNETDR}, and new devices are currently in the designing phase, such as LAr1-ND~\cite{lar1ndsbnd}. In the following subsections some aspects of the cryogenics and purification features of these devices and of the results, where available from their operation, are briefly outlined. 

Removal of oxygen for the ultra-purification of liquid argon is of primary importance and oxygen getters are an integral part of the purification system for all these devices. New, higher-capacity materials are being developed mainly for industrial applications. Possible application of these new products for science is of great interest and Fermilab is currently involved in a test campaign in collaboration with specialized industries. A short report of the current status along this line of development is also provided at the end of this section.

All systems described below are reusing or improving designs initially developed with the LAPD. The primary set of information from the LAPD were how to efficiently construct the systems and the operation of the cryogenic pumps and filtration system.  The LAPD also provided the first measurement of the loading capacity for an oxygen filtering media used in liquid argon operation~\cite{ref:LAPD}.

\subsection{LBNE 35 ton Prototype Cryogenics} 
\label{sec:lbne35t}
\begin{figure}[htbp]
\begin{center}
	\includegraphics[width=4.5in]{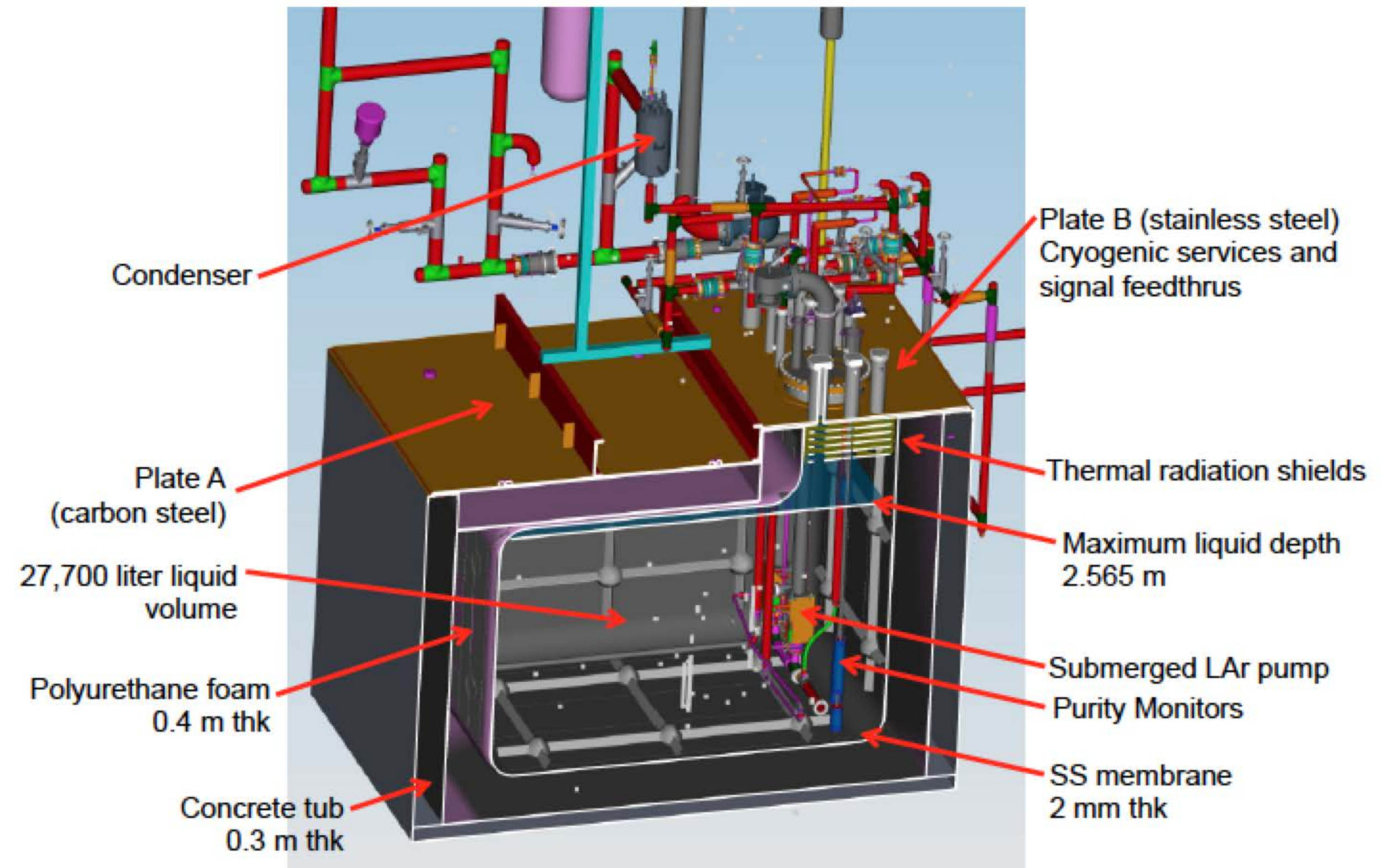}
\caption{Overview of the LBNE 35 ton prototype cryostat and the portion of the cryogenic system that was not part of LAPD.}
\label{fig:35T}
\end{center}
\end{figure}

The LBNE 35 ton prototype, shown in Fig.~\ref{fig:35T}, was built in 2013 at Fermilab.  First operation of the prototype was a success and met its goals.  The prototype demonstrated the contractual business model with a membrane cryostat supplier for the design and construction of a membrane cryostat.  The initial operation also demonstrated the membrane cryostat technology for liquid argon service with respect to thermal performance, feasibility for liquid argon, and leak tightness.  This operation also achieved and sustained the purity required for the LBNE Far Detector which is an equivalent oxygen contamination of less than 229 ppt, corresponding to an electron lifetime greater than 1.4 milliseconds.  

The LBNE 35 ton prototype utilizes the LAPD purification system which is located in the Proton Center area at Fermilab.  The cryostat liquid volume is 27,700 liters with a maximum depth of 2.57 meters.  The cryostat is insulated with 0.4 meters of polyurethane foam.  A 0.3 meter thick concrete shell provides the mechanical strength to resist the pressure load which is transferred from the liquid containing stainless steel membrane through the foam insulation.  The tank contains two submerged liquid argon pumps which direct the liquid argon up and out of the tank and through transfer lines to the LAPD purification system.  The required liquid purity was achieved one month after system startup.  Following the procedure developed for the LAPD~\cite{ref:LAPD}, the first purification step was an argon piston purge where room temperature argon gas was injected at a rate of 2.43 volume changes per hour into the cryostat which initially contained breathable air.  The argon piston purge reduced the oxygen concentration to from 21\% to 6.3 ppm, water from 550 ppm to 1.2 ppm, and nitrogen from 78\% to 10.6 ppm.  The concentration of contaminants in a room temperature tank filled with gas is reduced by a factor of 840 due to the liquid to vapor mass ratio such that that argon piston purge need not achieve purity lower than a few ppm.  

A room temperature gas recirculation then pulled argon gas out of the tank at a rate of 7.5 volume changes per day and sent it through the water and oxygen filters.  This phase revealed a leaking vacuum relief valve which would have been far more difficult to identify during cryogenic operation.  The gas recirculation phase reduced oxygen concentration to less than 100 parts per billion (ppb), water was reduced to less then 400 ppb, and the nitrogen concentration was 48 ppm.  Nitrogen is not actively filtered and can only be reduced by the dilution effect of argon gas added to the system to replace the argon gas sent to the gas analyzers.  Nitrogen was higher at the end of the gas recirculation phase than at the end of the argon piston purge due to the leaking vacuum relief valve discovered during the gas recirculation phase.  

The cryostat was cooled down using a system of liquid argon atomizing sprayers over a period of 28 hours and then filled with liquid argon. Refrigeration is provided by a liquid nitrogen powered condenser.  The condensed argon boil off, which contains the majority of the contamination due to outgassing from warm surfaces located in the ullage, is sent to the pump suction such that it is filtered prior to returning to the tank bulk liquid.  Pumped liquid filtration achieved electron lifetimes in excess of 2.5 milliseconds.  The electron lifetime trend was upward such that a longer period of filtration would likely have achieved larger electron lifetimes. 

One submersible pump suffered premature bearing wear at 750 hours.  After the run was over the second pump was also found to have excessive bearing wear.  The pump manufacturer has not determined the reason for excessive bearing wear. The electron drift lifetime was measured in the liquid argon using a purity monitor.  The purity monitor was based on the ICARUS design~\cite{ref:purityMonitor}. A thorough description of the purity monitor and the data acquisition hardware and software can be found in Reference~\cite{ref:LAPD}. As described in Reference~\cite{ref:LAPD}, a measure of electronegative impurities can be determined from looking at the fraction of electrons generated at the cathode that arrived at the anode. The purity monitors exhibited microphonic noise due to being supported from a large flat plate that also supports the cryogenic system. Installation of a time projection chamber (TPC) is planned for the second operation of the system in 2015.   

{\bf Lessons Learned}
\begin{itemize}
\item The gas recirculation step when commissioning a cryostat is very important as it is a final system integrity check before committing to cryogenic operation.  
\item The purity monitor devices should be isolated from even small amounts of vibration to ensure their stable operation. 
\end{itemize}

\subsection{MicroBooNE Cryogenics}

The first cryogenic system for LArTPC detectors at the scale of 100 tons built in the US is for the MicroBooNE experiment. Fabrication of the system was successfully completed in 2014 as was the installation in its final  experimental enclosure. The major cryogenic subsystems of MicroBooNE are: i) the initial cool down system, ii) the purification system, and iii) the nitrogen refrigeration system. A partially outfitted test was carried out and the results are summarized here below.  The test incorporated a few hundred liter capacity cryostat, which is not part of the final system, to test dielectric breakdown as a function of electronegative contamination~\cite{ref:HVC}. 

The MicroBooNE cryogenic system consists of two filter skids, each comprised of two pressure vessels, two pumps to circulate the argon, although only one will be in use at any given time, and a liquid nitrogen-based refrigeration system.  In each filter skid, one pressure vessel contains a 4A molecular sieve supplied by Sigma-Aldrich to eliminate water~\cite{ref:sigmaaldrich} and one contains BASF CU-0226~S, a dispersed copper oxide impregnated on a high surface area alumina to eliminate oxygen~\cite{ref:basf} from the argon.  The filtration system consistently produced better than 100~ppt oxygen equivalent contamination to the cryostat used for breakdown studies as well as the rest of the system.  
\begin{figure}[]
  \centering
  \includegraphics[width=0.495\textwidth]{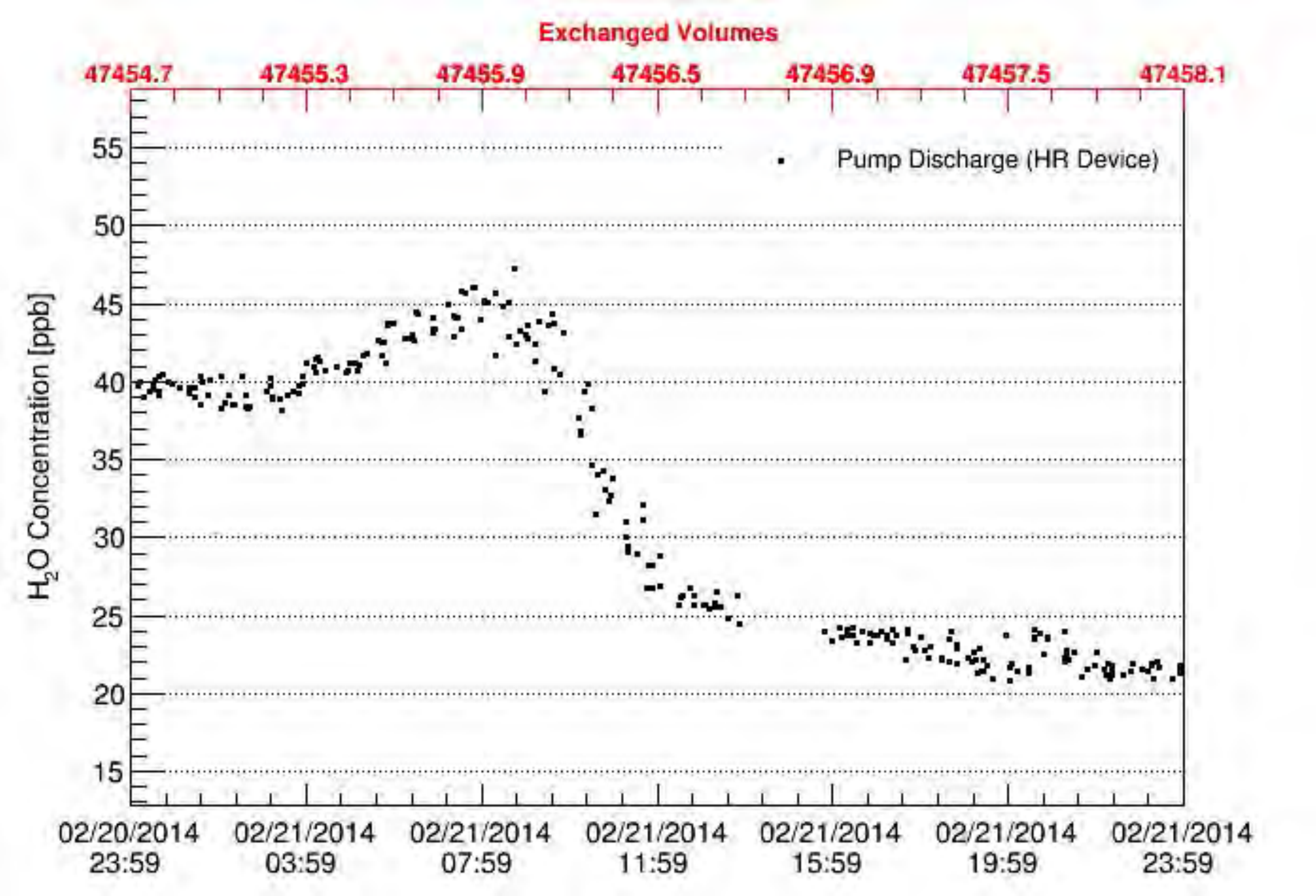}\label{fig:water}
  \includegraphics[width=0.495\textwidth]{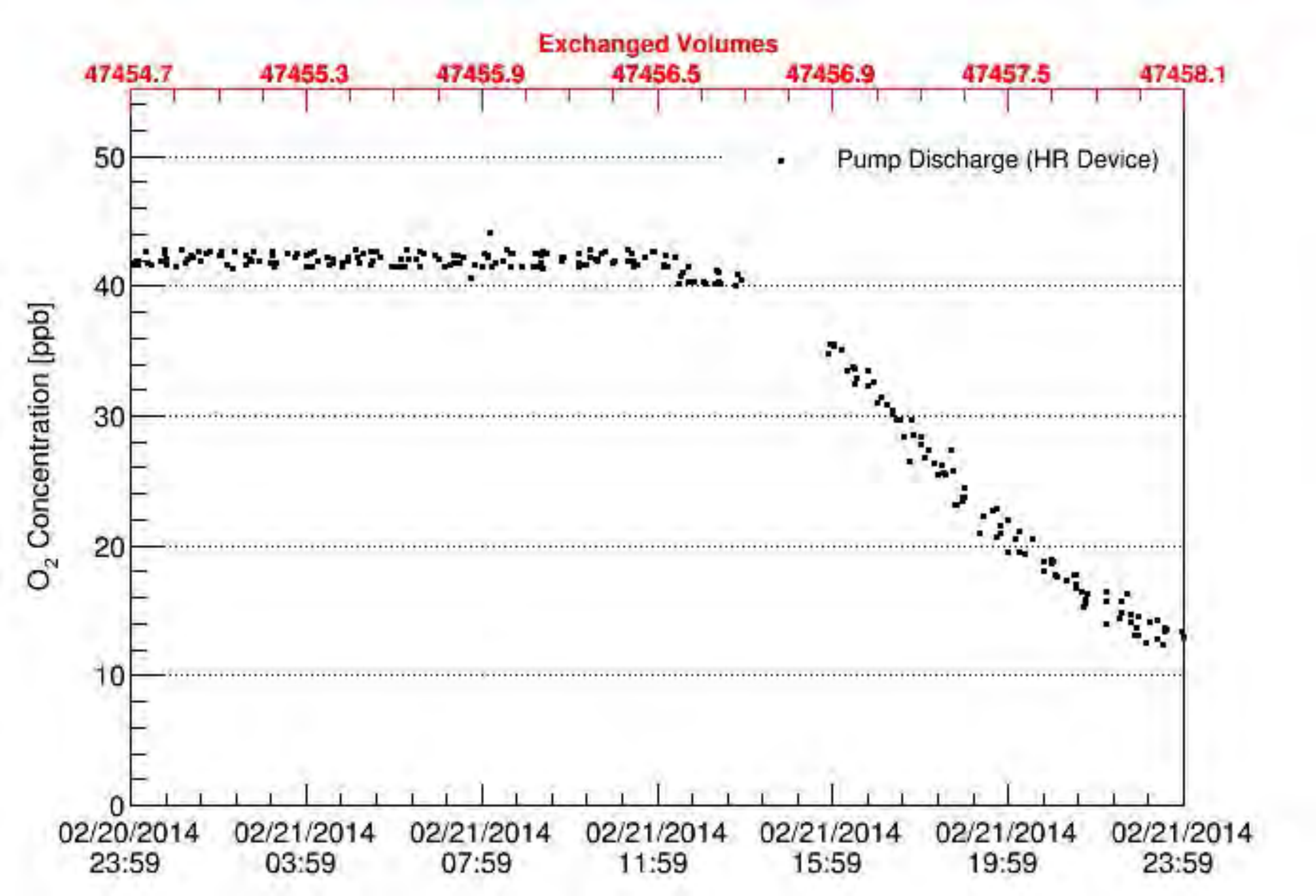}\label{fig:oxygen}
  \caption{Plots of water (left) and oxygen (right) contaminations in MicroBooNE cryogenic system as measured by the gas analyzers as clean-up began for the partially outfitted test. The measurement of the water contamination came from the Tiger Optics Halo+. The measurement of the oxygen contamination came from the Servomex DF-560E.} 
\label{fig:cleanup_gas_analyzers}
\end{figure}

The oxygen contamination level in the argon was monitored using two Servomex gas analyzers, a DF-310E and a DF-560E~\cite{ref:spectris}. Combined, the two oxygen analyzers covered a range of 0.1~ppb to 5000~ppm. The level of nitrogen contamination in the argon was monitored using an LD8000 Trace Impurity Analyzer~\cite{ref:ldetek}. The level of water contamination in the argon was monitored using a Tiger Optics Halo+ gas analyzer~\cite{ref:tiger}. Plots of the water and oxygen contaminations as filtration started up appears in Fig.~\ref{fig:cleanup_gas_analyzers}. 

In addition to the gas analyzers, contamination values ranging between 300 and 50~ppt oxygen equivalent were measured with purity monitors based on the design used in the LAPD and LBNE 35 ton prototype. During this operation, the cryogenic system achieved a measured electron lifetime of better than 3.5~ms. 

\subsection{LAr1-ND Cryogenics} 
A short-baseline neutrino (SBN) program is being developed at Fermilab to study neutrino properties using a combination of detectors: the existing MicroBooNE, the newly proposed LAr1-ND upstream of MicroBooNE, and a refurbished ICARUS T600 downstream of MicroBooNE. The main features of the cryostat and cryogenic systems that are being developed for LAr1-ND are summarized here. It is a truly international effort with US and European participation and two cryogenic groups at Fermilab and CERN to support SBN and long-baseline neutrino (LBN) activities. While the SBN program is independent of the LBN program, it has similar cryogenic requirements and presents similar challenges; the primary challenge being the achievement and maintenance of the liquid argon purity in a membrane cryostat. There is a strong incentive to use existing knowledge and expertise and perform tests and prototyping taking into consideration the needs of both programs. For example the use of external liquid argon recirculation pumps, and a cold roof, kept at temperatures lower than 100K, will be implemented in LAr1-ND.  If successful, these techniques may be part of the design of the LBN far detector. The reason for the external pumps resides in the potentially high maintenance required to operate the submerged liquid argon pumps as mentioned in \S~\ref{sec:lbne35t}. The ullage is known to be the main source of water in the liquid argon and a cold roof could reduce the contamination in the ullage by limiting the potential outgassing. The cryostat and cryogenic systems will be designed and built with the needs of the future LBN program in mind. The cryostat is proposed to be located in a new building located next to the Fermilab SciBooNE enclosure;  that enclosure will house the cryogenic system. The cryogenic piping connecting the cryostat to the rest of the system will run in an underground tunnel linking the two buildings. The cryostat builds on the successful experience of the LBNE 35 ton membrane cryostat prototype described in \S~\ref{sec:lbne35t}. 
\begin{figure}[htbp]
\begin{center}
  \includegraphics[width=4in]{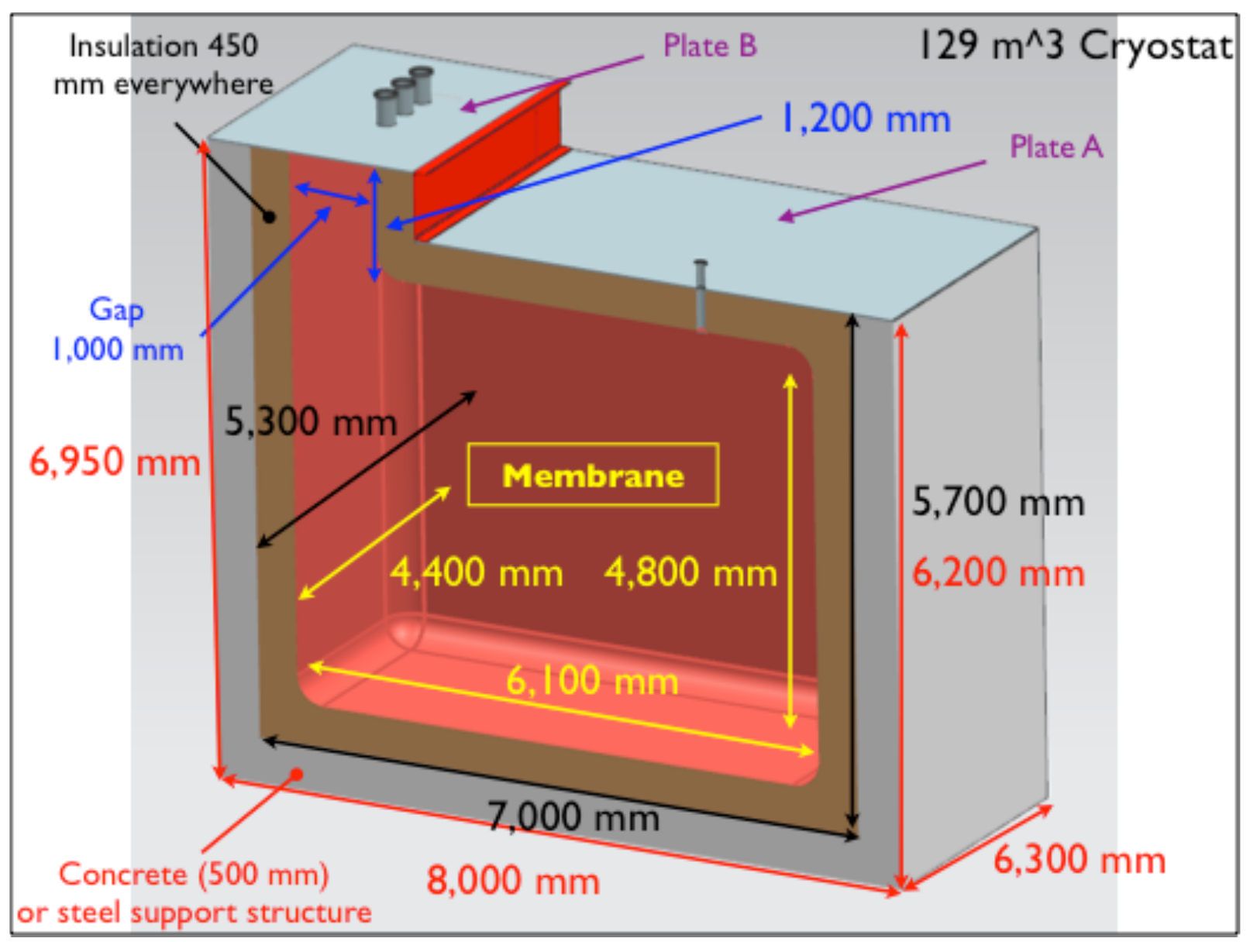}
\caption{3-D model of the LAr1-ND membrane cryostat}
\label{fig:LAr1ND-cryostat}
\end{center}
\end{figure}
Figure~\ref{fig:LAr1ND-cryostat} shows the 3-D model of the LAr1-ND membrane cryostat in its current base configuration with inner dimensions 4.4 m (width) $\times$ 6.1 m (length) $\times$ 4.8 m (height). It will contain about 129 m$^3$ of liquid argon and will have 0.45 m of polyurethane insulation all around.  A concrete or steel support structure will sustain the liquid and gas loads. The current configuration shows two plates at the top: plate A, which is part of the membrane and insulated with polyurethane, and plate B, which is not part of the membrane as it contains all the penetrations and is insulated with radiation shields. The gas argon is foreseen to be contained in the neck region underneath plate B with the liquid argon touching the membrane underneath plate A.  The cryogenic system will use the piston purge method~\cite{ref:LAPD} to remove the impurities from the tank prior to cryogenic operations. It has a goal liquid argon purity of 50 parts per trillion oxygen equivalent contamination, which correspond to a 6 ms electron drift lifetime. The liquid argon will be purified using a combination of molecular sieve and copper beads filters to remove water and oxygen respectively, as was done in the LAPD, the LBNE 35 ton prototype and MicroBooNE. 
\begin{figure}[htbp]
\begin{center}
   \includegraphics[width=5.5in]{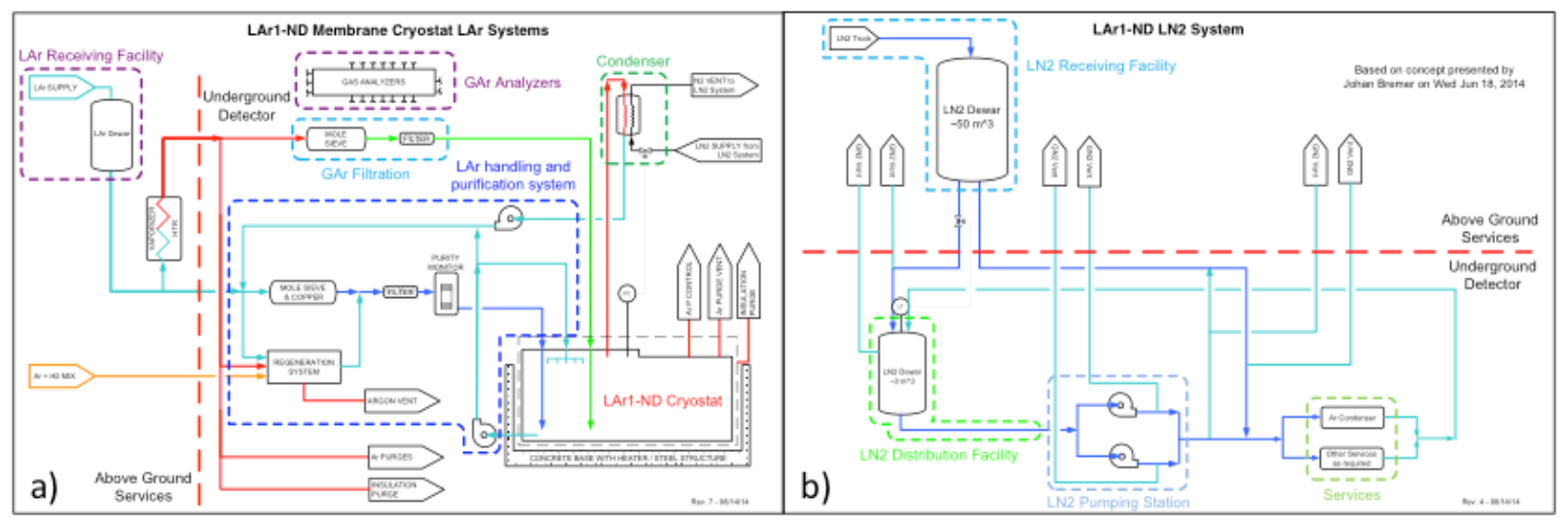}
\caption{Process Flow Diagrams of the LAr1-ND liquid argon (left) and liquid nitrogen (right) systems.}
\label{fig:ProcFlowDia}
\end{center}
\end{figure}

Figure~\ref{fig:ProcFlowDia} shows the Process Flow Diagrams (PFD) for the liquid argon and liquid nitrogen systems. The main features are the receiving facilities, located above ground, with the transfer lines to deliver the cryogens underground; the liquid argon handling and purification system; the gaseous argon filtration to purify the gas after the piston purge; the condenser to re-condense the boil-off argon from the cryostat; the gas analyzers to measure the impurities; and the liquid nitrogen distribution facility and pumping station. All these systems are located underground.

\subsection{Catalytic Purification for  Science and Industry} 

Purification catalysts find major applications in various sectors of the chemical processing industry, like catalytic oxidation, deoxo and oxygen getter catalysts. Removal of oxygen using oxygen getters in particular for the ultra-purification of liquid argon is the primary application for science. This section provides a summary of the current achievements and perspectives from current developments with new materials~\cite{artrip}. Oxygen Removal using Oxygen Getters is a core technology at Research Catalysts.  Several ``standard" getter products are available, including the one most widely used in liquid argon, Engelhard Q-5, as well as several others with higher capacities. The best known of the higher capacity getters is the BASF Catalyst R3-11G.  Additionally, based on testing of these and other materials for oxygen capacity, two new, higher-capacity materials are available to the market, namely GetterMax\textsuperscript{\textregistered} 133 and GetterMax\textsuperscript{\textregistered} 233.  

The oxygen capacity of a given catalyst will vary with operating conditions, commonly indicated by the dynamic capacity of oxygen getters, most notably with temperature but also with contact time, the inlet oxygen concentration, and the allowable oxygen outlet concentration.  Capacities for the most commonly used getters, when operated in gas phase at 25$^\circ$C, range from about 1 liter of oxygen per kilogram of catalyst for Q-5, to around 4 liter per kilogram for R3-11G, to upwards of 10 and even $\ge$ 20 liters per kilogram for the more powerful getters.  These more powerful getters have higher bulk densities, which translates into even greater capacity gains per unit of bed volume.

Copper-based oxygen getters are used in various liquid-phase applications too, including propylene monomer and various solvents including tetrahydrofuran (THF), methanol, and alkanes such as hexane and heptane.  However, industry does not typically make systematic attempts to measure performance of oxygen getters for treating sub-ambient temperatures in either liquid or vapor phase service.  Nevertheless, based on their relative performances in vapor phase at room temperature, one could expect that the GetterMax\textsuperscript{\textregistered} materials, which have shown significantly higher capacity than both R3-11G and Q-5 in benchmarking tests, will provide performance advantages in liquid argon as well.

As indicated above, little is known about the capacities of the oxygen getters in liquid phase service and how those capacities may vary with contact time, temperature, and levels of dissolved oxygen.   Fermilab has reported a capacity factor for Q-5 in liquid argon of about 0.5 gram per kilogram~\cite{ref:LAPD}; in the units more commonly used to benchmark getter capacity, this value is equivalent to about 0.35 liters of oxygen per kilogram of catalyst.  This value is lower than the room temperature value suggesting the capacity in purification of liquid argon may be only about 30\% of that obtained in gas phase at 25$^\circ$ C, at least with Q-5.  

Research Catalysts has expressed interest in expanding its portfolio of oxygen getters and the envelope of knowledge about their performance, particularly in liquids and at sub-ambient temperatures.  It welcomes opportunities to collaborate in this area with the Physics community.

\section{TPC and High Voltage}
\label{sec:TPCHV}
While LArTPC experiments have operated successfully in the past \cite{Amerio:2004ze, Anderson:2012vc}, the technology is being pushed in the design of future experiments by increasing the detector volume and possibly the drift length. Reliable methods are needed to deliver increasingly high voltages (HV) inside the cryostat to the TPC. A better understanding of the liquid argon dielectric properties is necessary, as well as the development of protection systems against possible HV discharges. Prototypes are needed to test in detail the main design features of the envisioned large future detectors.

\subsection{Liquid Argon Dielectric Strength Measurements }

The understanding of liquid argon dielectric properties is crucial for the design of future experiments. Recently, it became apparent that the often-quoted 1 MV/cm liquid argon dielectric strength measurement \cite{Swan1960180, Swan1961448} was not valid at the distance scales important to LArTPC design.  This realization motivated several new studies.

The ETH Zurich group measured the liquid argon dielectric strength in a uniform field over a 1~cm gap distance and a 20~cm$^2$ cathode area \cite{Bay:2014jwa}.  The field was formed using Rogowski profiles that were mechanically polished.  This geometry withstood a field of 100~kV/cm without breakdown over four hours when the liquid argon was below the boiling point at a given pressure.  When the pressure in the vapor space was decreased to bring the liquid argon to the boiling point, dielectric breakdowns were observed at fields as low as 40~kV/cm.  This was interpreted as evidence that bubbles contribute to dielectric breakdown in liquid argon.

The LHEP Bern group also published a result measuring the liquid argon dielectric strength~\cite{Blatter:2014wua}.  They measured the dielectric breakdown voltage in liquid argon, with 1~ppb electronegative impurities, below the boiling point between a 80~mm diameter cathode sphere and a grounded plate for gap spacings up to the centimeter scale.  Liquid argon dielectric breakdowns were observed for fields as low as 40~kV/cm.  For gap spacings greater than 1~cm, they noted an unexpected behavior of breakdown along the feedthrough surface.

A second published study by the group at LHEP Bern described a method to suppress dielectric breakdowns in liquid argon~\cite{Auger:2014eba}.  They used a similar geometry as their previous test, but here, the cathode sphere was coated with a several hundred micrometers layer of natural polyisoprene to inhibit discharges by suppressing field emission.  Breakdown fields were more than an order of magnitude higher with a 450~$\mu$m layer than without while using a 4~cm diameter sphere at a 3~mm gap spacing.

A Fermilab group recently completed measurements of liquid argon dielectric strength with a goal of studying geometry and purity effects~\cite{ref:HVC}.  The test cryostat was plumbed into the MicroBooNE cryogenic system with access to pure argon, gas analyzers to determine the oxygen, water and nitrogen contamination in the argon, and a downstream purity monitor.  A MicroBooNE production high-voltage feedthrough was used to supply high voltage to a sphere-plate geometry for dielectric breakdown studies.  Cathode sphere diameters of 1.3, 5.0 and 76~mm were evaluated in purities ranging from a few parts-per-million (ppm) to tens of parts-per-trillion (ppt) oxygen-equivalent contamination and at anode-cathode spacings up to 25~mm.  Distances above 15~mm for the 5~mm diameter sphere and 10~mm for the 76~mm diameter sphere were not considered in the results because breakdown along the feedthrough was observed for some of the data beyond these distances.

\begin{figure}[]
  \centering
 \includegraphics[width=0.8\textwidth]{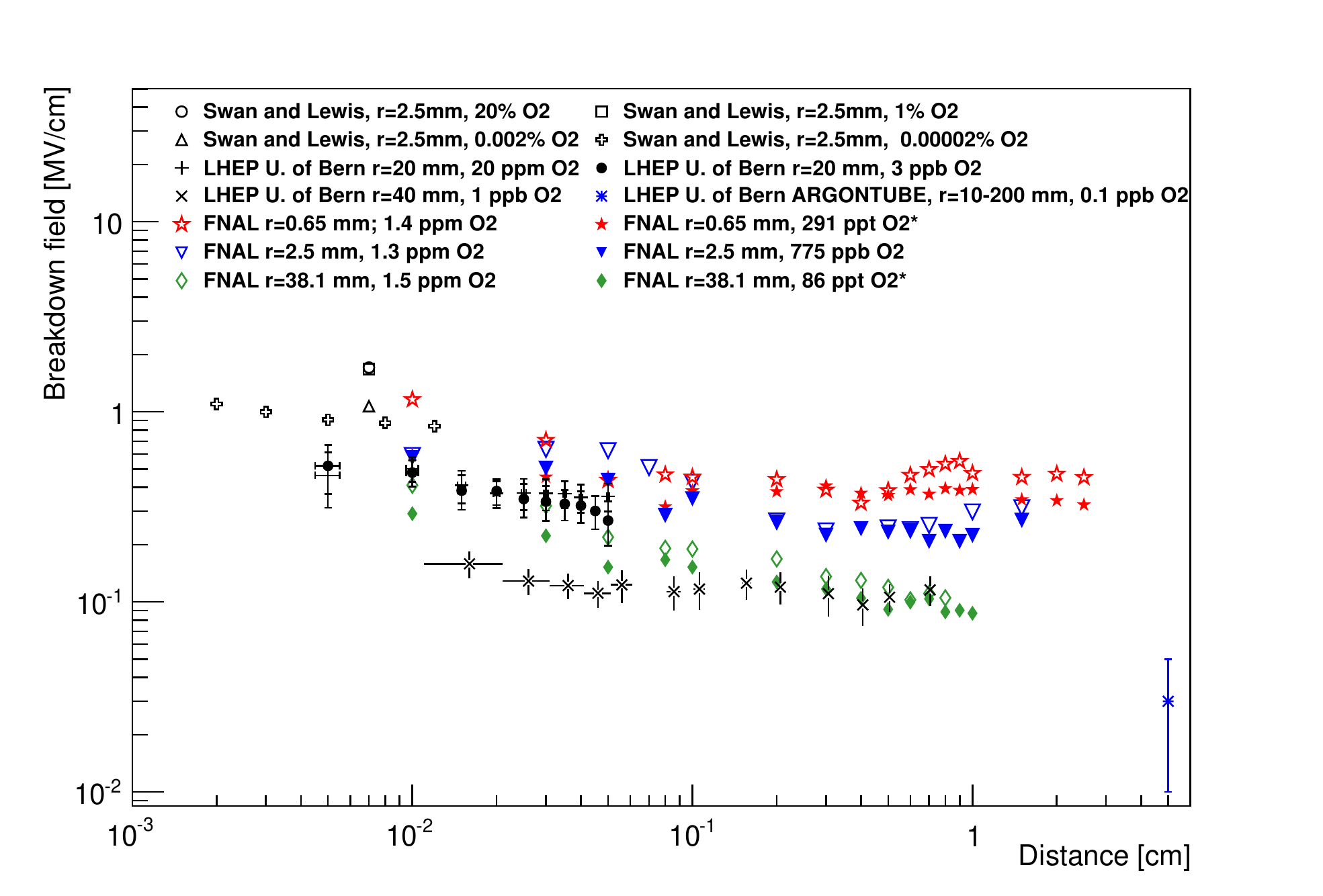}
\caption{A comparison of the several measurements of the magnitude of the electric field at breakdown in liquid argon~\cite{Blatter:2014wua,ref:HVC}. The data points from Fermilab result from the highest and lowest contamination level measurement taken with each probe.
The oxygen values with an asterisk are extrapolated from purity monitor measurements.}
\label{fig:bern}
\end{figure}

A comparison of the Fermilab group's results to earlier data, including those of the Bern group, can be seen in Fig.~\ref{fig:bern}.  The two sets of points for each probe of the Fermilab data show the results from the highest and lowest argon purity data collected.  While the data of the different groups are consistent with each other, a noticeable geometry effect can be seen.

A purity effect was seen by the Fermilab group for the 1.3~mm probe at gap distances greater than 10~mm.  For purities between 0.2 and 1.4~ppm, the breakdown voltage was about 1.5 times greater than for purities between 0.29 and 1.8 parts-per-billion (ppb).  No noticeable purity effect on dielectric strength was observed with the other cathode tips.

\begin{figure}[]
  \centering
 {\includegraphics[width=0.45\textwidth]{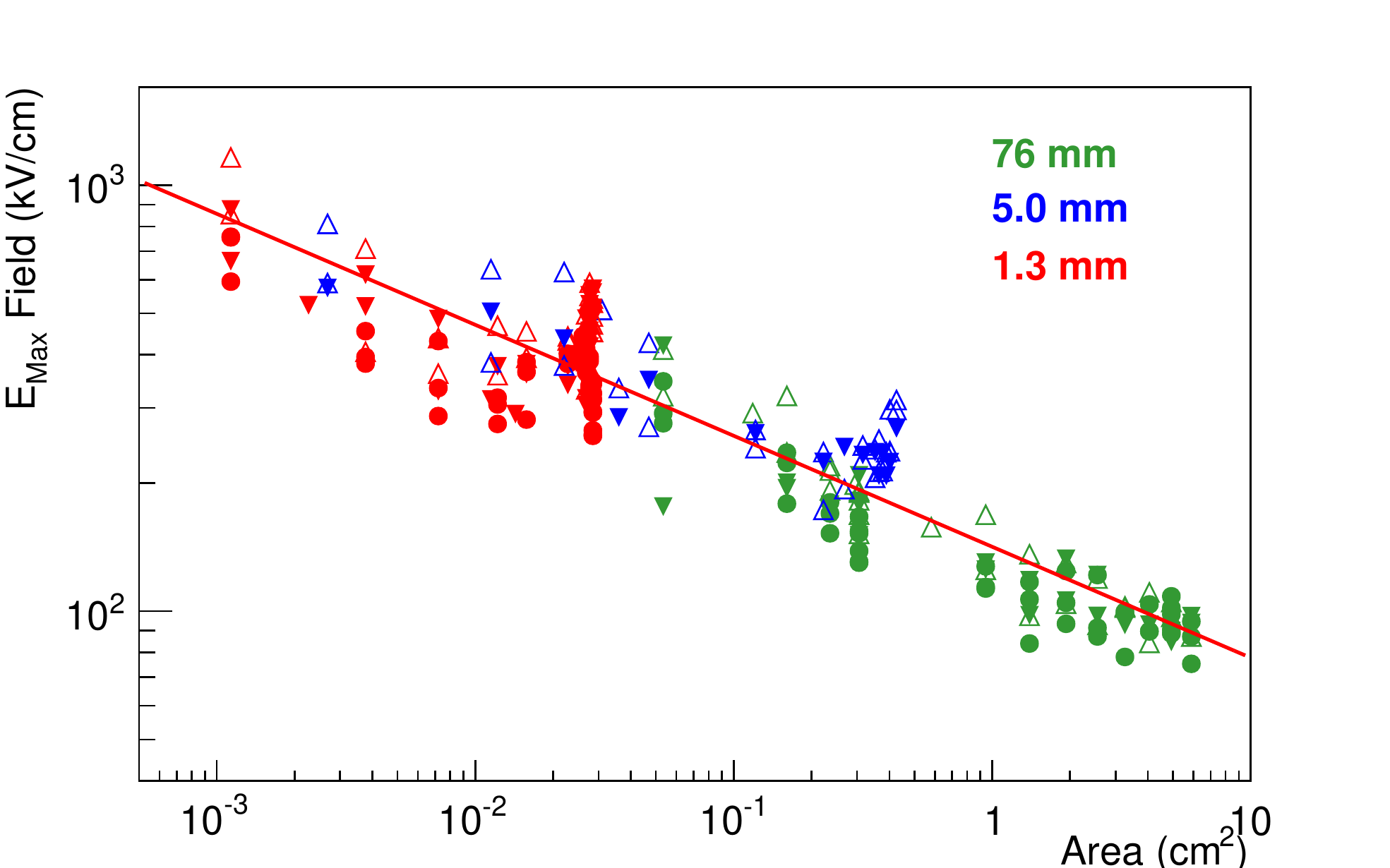}\label{fig:area80}}
 {\includegraphics[width=0.45\textwidth]{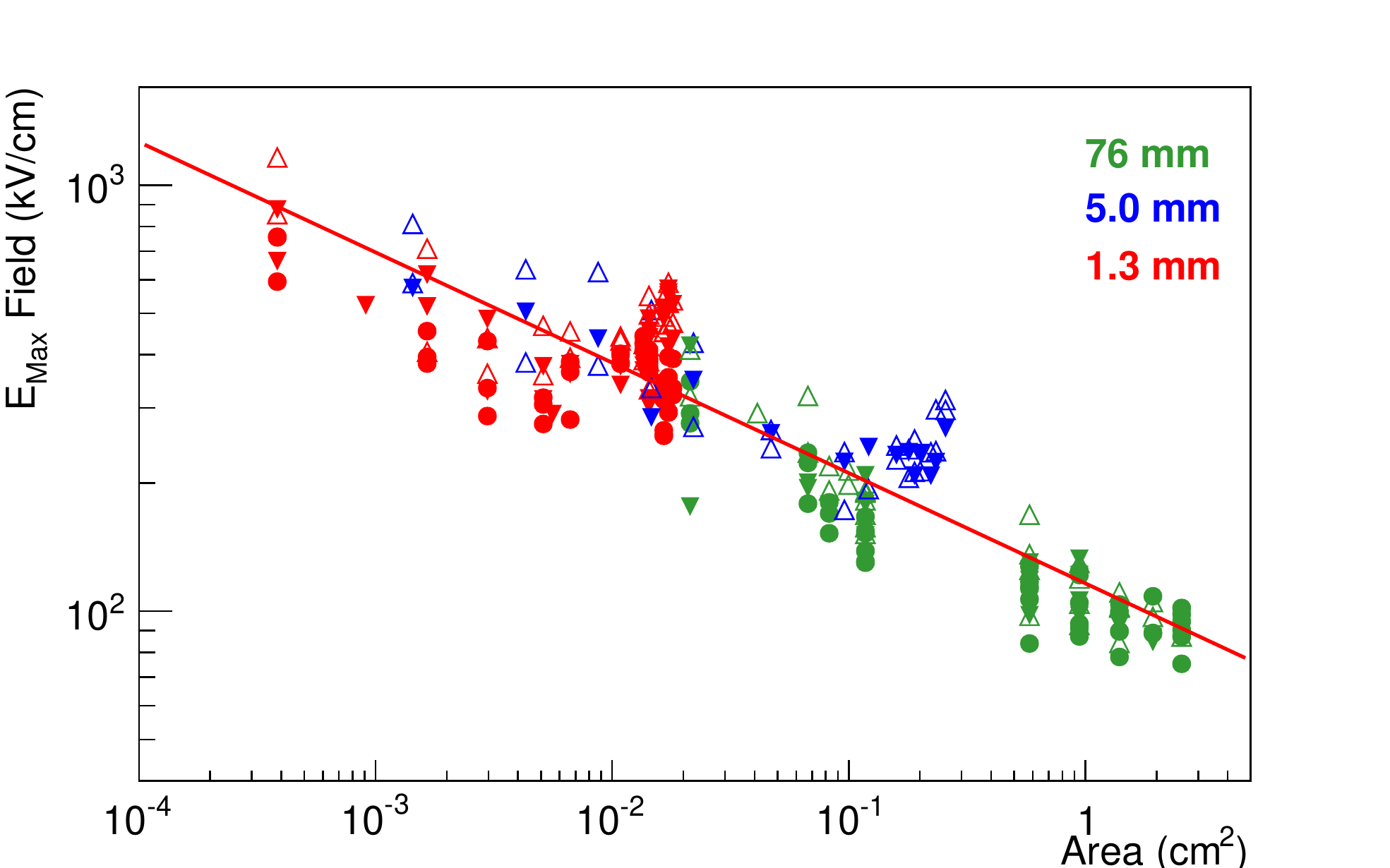}\label{fig:area90}}
\caption{
The average maximum breakdown field versus stressed area of the cathode. (left) The stressed area is defined as the area with an electric field greater than 80\% of the maximum electric field. (right) The stressed area is defined as the area with an electric field greater than 90\% of the maximum electric field.
}
\label{fig:areaa}
\end{figure}

A dependence on peak electric field at breakdown and stressed area was also observed and showed similar behavior across the geometries evaluated in the Fermilab test.  A stressed area or volume dependence was suggested by Gerhold {\it et al.}~\cite{Gerhold1994579} in describing liquid helium data.  Here, a stressed area is defined as the area on the cathode surface that is equal to or greater than some percentage, $\epsilon$, of the peak electric field.  Plots of the peak electric field at breakdown versus the stressed area are shown in Fig.~\ref{fig:areaa} for $\epsilon$ values of 80\% and 90\%.

{\bf Lessons Learned} It appears there are several conditions that can cause liquid argon to have a dielectric strength of less than 1MV/cm.  Those conditions include $i)$ boiling of the argon, creating bubbles in the region of the high electric field, $ii)$ field emission, and $iii)$ large stressed area of the cathode surface.

\subsection{High Voltage Surge Protection Systems for LArTPC Detectors}

Dependencies of the breakdown mechanism on electrode geometry, surface finish, argon purity, and space charge accumulation are only partially understood as the present time \cite{Blatter:2014wua,Swan1960180,Auger:2014eba,ref:HVC}.  These dependencies make it difficult to predict where and how frequently breakdowns will occur.  As such, it is important to take whatever steps are necessary to design TPC field cages to be robust against failure in the event of HV discharges.

As described in~\cite{Asaadi:2014iva}, the effects of a HV discharge from a point on the TPC field cage can lead to a large transient overvoltage, whose relaxation time and distribution over the field cage elements depend upon the distribution of capacitances and resistances in field cage-cathode-cryostat system.  Electrical models of the MicroBooNE~\cite{MicroBooNETDR} field cage showed that such transients could exceed the failure voltage of the originally installed field cage resistors~\cite{Bagby:2014wva}.  

To address this problem, a two-fold solution was implemented.  First, some of the most vulnerable resistors were replaced with more robust components capable of withstanding larger transient voltages.  Second, a high voltage surge protection system was implemented on the field cage, which prevents large overvoltages from evolving between field cage rings in a HV discharge situation.  This use is the first application of a surge protection system in a cryogenic LArTPC detector.

The MicroBooNE surge protection system uses surge arrestors connected between the TPC field cage rings in parallel with the field cage resistors.  A surge arrestor is a device which has a nonlinear voltage-current (V-I) characteristic, and commonly used commercial surge arrestors include gas discharge tubes (GDTs)~\cite{epcos,GDTBourns}  and varistors~\cite{var_material_science}.  Surge arrestors suitable for MicroBooNE are highly insulating at the nominal ring-to-ring voltage, but become conducting when a large overvoltage is applied.  This means that in normal running conditions the surge arrestor does not interfere with the voltage distribution on the field cage or draw a large leakage current.  In the event of a HV discharge, however, charges on the field cage are allowed to redistribute quickly to prevent damaging voltages from being held between field cage rings and damaging field cage resistors.

\begin{figure}[h]
\centering 
\includegraphics[width=0.98\textwidth]{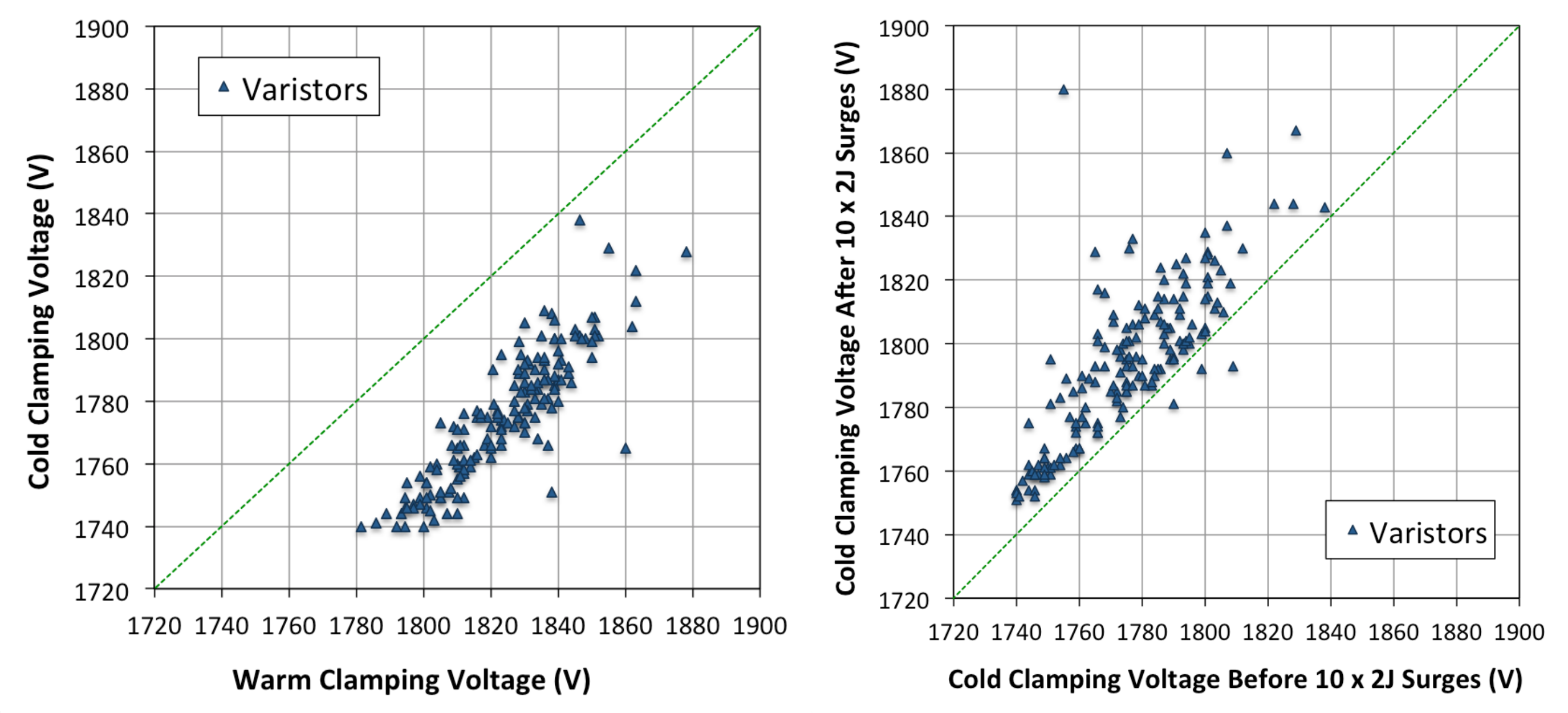}
\caption{Two figures reproduced from \cite{Asaadi:2014iva}. (left) the clamping voltage of a large sample of varistors warm (room temperature) and cold (90 K). (right) The clamping voltage of varistors before and after the application of 10 $\times$ 2 Joule surges\label{fig:FromPaper}  }
\end{figure}

High voltage GDTs~\cite{ep_datasheet} and varistors~\cite{pano} were both tested for the MicroBooNE application, and a detailed report of this testing can be found in~\cite{Asaadi:2014iva}; two plots from that reference are reproduced in Fig.~\ref{fig:FromPaper}.  The devices were tested in cryogenic conditions for their insulating resistance, clamping behavior and robustness under high voltage and high energy surges.  Effects on liquid argon purity and ambient photon background were also studied.  Both classes of device were found to be suitable for surge protection on LArTPC field cages.  Varistors were chosen for MicroBooNE due to their continuous V-I characteristic, which provides a smooth clamping behavior rather than a crowbar action.

\begin{figure}[h]
\centering 
\includegraphics[width=0.98\textwidth]{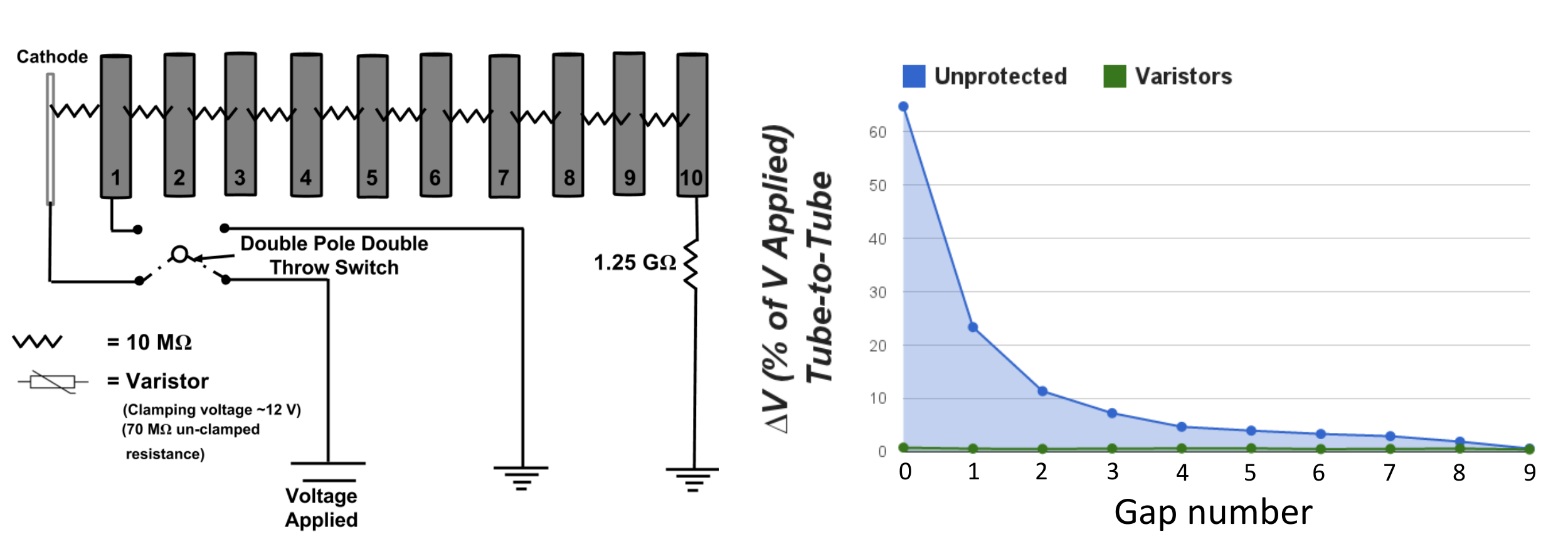}
\caption{Demonstration of clamping properties of varistors on the MicroBooNE TPC. (left) The circuit diagram of "scale model" used to test the surge protection principal.  (right) The maximum evolved ring-to-ring voltages in the protected and unprotected systems. \label{fig:OnTPCTest} }
\end{figure}

As well as ensuring suitability of the devices for operation in a cryogenic environment, the clamping behavior of a varistor based system was investigated {\it in situ} on the MicroBooNE field cage.  Since a 2 kV ring-to-ring voltage could not be applied in air, a lower applied voltage was used, with correspondingly lower voltage varistors and smaller resistors installed across the first ten field cage tubes.  The tenth field cage tube was connected to ground via a 1.25~G$\Omega$ resistance to simulate the resistive effect of the rest of the field cage.  This produces a "scale model" of the MicroBooNE field cage, with the correct geometrical capacitances but lower resistances.  A voltage of up to 1~kV was applied to the cathode and then points in the system were discharged to ground as the HV supply was removed from the cathode using a double-pole-double-throw switch.  As shown in Fig.~\ref{fig:OnTPCTest}, without varistors applied, a large transient overvoltage was seen between adjacent field cage tubes  With varistors applied, the tube-to-tube voltage is effectively clamped and large overvoltages to not occur.  

Other {\it in situ} tests were made implementing a finite HV supply trip time after discharge, and with incomplete discharge of the cathode.  In all scenarios investigated, the surge protection system acted as expected, clamping transient overvoltages between any pair of field cage rings.  

{\bf Lesson Learned} High voltage discharge near the cathode can be a disaster for LArTPCs.  As such, it is important to consider implementing some form of surge protection on the detector.

\subsection{High Voltage Feedthroughs}

High voltage feedthroughs (HVFT) are an important component of LArTPC detectors. Electric fields of $\sim500$~kV/cm are typically used as drift field for the ionization electrons. A 3.5m long drift as in the planned LBNE detector will require a 175~kV nominal operating voltage. A reliable method to deliver the needed high voltage through the cryostat to the TPC cathode has been a major challenge in the LArTPC development. 

Ultra-High-Molecular-Weight-Polyethylene (UHMWPE) cryo-fitted HVFTs have been developed and built for many experiments as listed Table~\ref{HVFT}. An inner conductor, carrying HV, is insulated by UHMWPE from an outer conductor tube, kept at ground~\cite{Amerio:2004ze,LArRDworkshop2013}. 

\begin{table}
\caption{Summary of High Voltage feedthroughs developed for running experiments or test setups}
\centering
\begin{tabular} {lcccc}
\hline\hline
 Project & Design HV & Required HV  & Operated HV & Tested HV \\
            & (kV)          & (kV)             & (kV)              & (kV)         \\
\hline\hline
ICARUS & 75 & 75 & 150 & 150 \\
ZEPLIN II & 25 & 25 & 21 & \\
DarkSide10 & 50 & 36 & 36 & 50 \\
DarkSide50 & 50 & 12 & 12 & 50 \\	
XENON1T & 100 & 100 & Being installed & \\	
XENON1T prototype & 100 & 100 & & 130 \\
mini-CAPTAIN &20 & 20 & & 75 \\
CAPTAIN & 50 & 50 & & 75 \\
LBNE prototype & 200 & 175 & & 200 \\
LBNE 35 ton & 200 & 111 & & 170 \\
\hline
\end{tabular}
\label{HVFT}
\end{table}

The basic principle is to cryo-fit a $\sim 1$~inch long section of the feedthrough. An inner smooth solid core, the insulating UHMWPE, is cryo-fitted into an outer conductor tube with a smooth inner surface in order to create a vacuum-tight seal. The location of this section is such that during operation it is always kept at room temperature. Depending on the manufacturer and on the particular batch, UHMWPE shrinks by $\sim 1.8-2\%$ when cooled to liquid nitrogen temperature. The shrinkage of each given UHMWPE sample is carefully measured, and its inner and outer diameters are precisely machined to specific values to match the outer conductor inner diameter and the inner conductor outer diameter for the cryo-fitting procedure. Leak tests are normally performed where the side of the feedthrough exposed to air is sealed by a helium filled balloon. The requirement is to reach a leak rate less than $1 \times 10^{-11}$ atm-cc/sec after a 10-20 minute wait. The thickness of the UHMWPE insulator is chosen to meet the design needs. In most practical cases, the needed thickness for ease of construction and to obtain a good seal is already much larger than electrically required~\cite{LArRDworkshop2013}.

\begin{figure} 
\centering 
\includegraphics[width=0.98\textwidth]{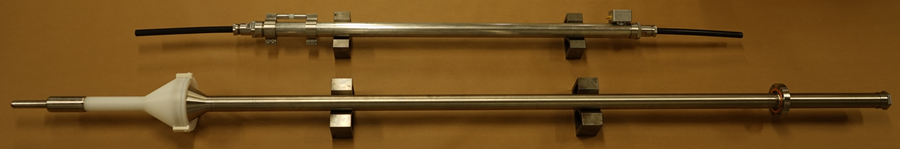}
\caption{The feedthrough for the LBNE 35 ton prototype.}
\label{fig:HVFT35T}
\end{figure}
\begin{figure} 
\centering 
\includegraphics[width=0.98\textwidth]{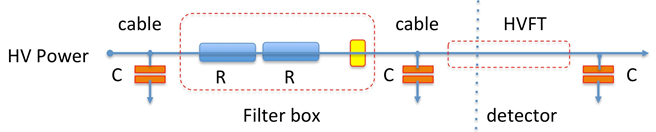}
\caption{Schematics of the filter box circuitry. The value of the resistors need to be customized, the capacitors represent the intrinsic cable and detector capacitances. In addition the filter box contains a corona discharge monitor, represented by the yellow box next to the resistors.}
\label{fig:HVFB}
\end{figure}

The highest voltage feedthrough is the prototype built for LBNE, which has been repeatedly tested up to 200~kV. However an identical feedthrough in design and construction for the LBNE 35 ton prototype detector, seen in Fig.~\ref{fig:HVFT35T}, has never reached 200~kV without discharge above $\sim150$~kV.  A careful examination revealed that a minor mechanical detail, but very important feature, was neglected in the final machining. The three small holes on the outer conductor, designed to avoid argon gas bubble build up at the bottom of the conductor, were missing. This change is the only difference compared to the successful LBNE prototype feedthrough. A distinctive feature of the LBNE and of the LBNE 35 ton feedthroughs, as shown in Fig.~\ref{fig:HVFT35T}, is that the outer conductor flares out at the bottom. The space at the flare is filled by UHMWPE. This feature is thought to be essential in order to reach 200~kV.  In addition the LBNE feedthroughs are equipped with a custom HV filtering box, as shown in Fig.~\ref{fig:HVFT35T} and Fig.~\ref{fig:HVFB}. This filtering box is to minimize the ripple from the HV power supply, which would induce noise on the charge amplifiers reading out the TPC wire planes.

\subsection{LBNE Prototype TPC in the 35 ton Membrane Cryostat}

\begin{figure} 
\centering
\includegraphics[height=0.3\textheight]{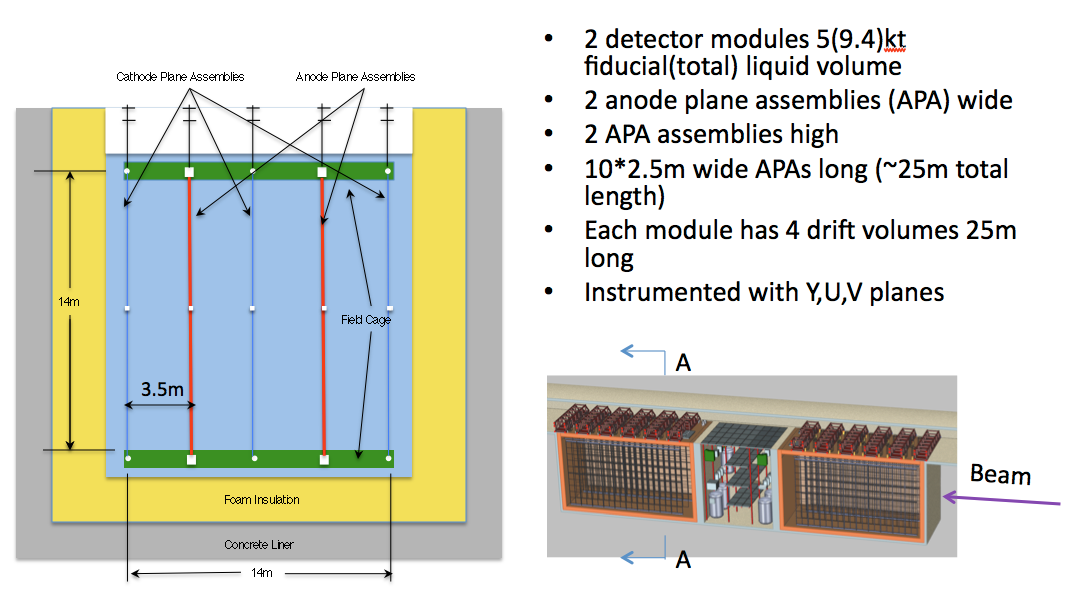}
\caption{LBNE Far Detector design of the anode and cathode configuration (left) and two 5~kiloton (fiducial) volume cryostats (right). }
\label{fig:FD}
\end{figure}

The salient features of the conceptual design of the liquid argon Far Detector (FD) for the LBNE experiment are summarized in Ref.~\cite{Adams:2013qkq}. The FD includes multiple membrane cryostat modules of 5 kiloton fiducial volume. Figure~\ref{fig:FD} shows the layout and the cross-section of the FD. 

%

The original program for the LBNE liquid argon FD had a 1~kiloton prototype cryostat (LAr1) that would have been used to verify the full scale components of the conceptual design. However during a reconfiguration of LBNE in 2012, that prototype was cancelled due to cost considerations. The LBNE 35 ton prototype cryostat was originally built to demonstrate the feasibility of the membrane style cryostat for high purity liquid argon operation. In order to provide a verification of the FD design, the 35 ton cryostat was given a new additional purpose meant to test as many of the FD design choices as possible, within the limited size of the detector. The available volume for a detector in the 35 ton prototype cryostat is limited to roughly $2.7 \times 2.4 \times 2.2$ m$^3$. In addition, entry into the cryostat is through a  75 cm diameter manhole access.

\begin{figure} 
\centering
\includegraphics[height=0.3\textheight]{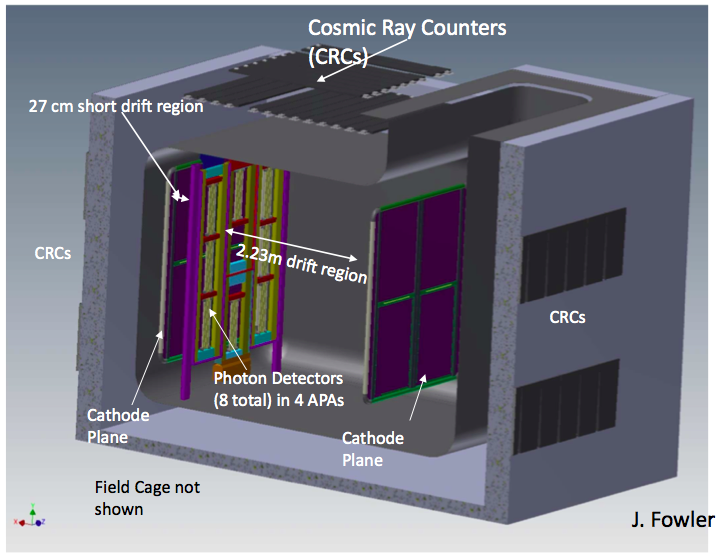}
\caption{The LBNE 35 ton Prototype Cryostat and Detector }
\label{fig:35t}
\end{figure}

The features that were deemed important to the prototype detector design were the double sided nature of the APA wire plane assemblies, and the edges between APAs. This gave rise to the detector shown in Fig.~\ref{fig:35t}. There are two drift regions, one 2.23 m long and the other 27 cm. The APA plane is comprised of 4 APAs, two long on each side with two short APAs in the middle. All are 50.4cm wide, with the long APAs having an active height of 196.3 cm.  The two short APAs are slightly different lengths (top 112.3 cm and bottom 84.3 cm) in order to fit in the same vertical space as the long APAs. The mass of the total active drift volume is 10.2 tons of liquid argon. 

%
%
%
%

\section{Electronics, Data Acquisition, and Triggering}
\label{sec:Electronics}
\subsection{Front-end Electronics}

The BNL group is continuing the development of a complete front-end signal processing chain that operates in liquid argon. Two ASICs, an analog front-end ASIC and an ADC ASIC, have been designed based on new device models and design rules developed for a 180 nm TSMC CMOS process.  This process allows the circuits to operate throughout the temperature range from above room temperature to below 77~K~\cite{sli}, with essentially invariant performance except for a reduction of input referred noise by about a factor of about 2 from 300~K to 77~K. The use of CMOS technology and appropriate circuit design allows channel-to-channel and chip-to-chip variations of operating parameters to be very tightly controlled. For example, charge gain and channel calibration capacitance vary by about 0.1\%, allowing simple charge calibration of all channels of an operating detector.  Approximately 2000 of the 16 channel, low noise, low power analog front-end ASICs have been fabricated, and they are now deployed in multiple LArTPC experiments. These experiments include the MicroBooNE and LArIAT~\cite{lariat} experiments at Fermilab, ARGONTUBE~\cite{Ereditato:2013xaa} at the University of Bern and CAPTAIN~\cite{CAPTAIN} at LANL. These ASICs have been fabricated into the LBNE 35t prototype TPC electronics, and they are planned for use in the LAr1-ND experiment at Fermilab as well. In collaboration with CERN, ICARUS is considering using cold CMOS analog front-end ASIC too. The MicroBooNE experiment has successfully installed 8,256 channels of cold front-end electronics on a TPC that is sealed inside a cryostat. Based on this experience, the BNL group is making minor revisions to the analog front-end ASIC to improve the robustness and to simplify system design, principally by improving protection of inputs against electrostatic discharge, and by adding an programmable precision internal calibration pulse generator and a smart reset circuit. 

A low power, clockless, 16 channel, 12 bit prototype ADC ASIC has also been designed, fabricated, and fully characterized in the lab. The effective resolution is 11.6 bits and the differential non-linearity is less than 4 least significant bits (LSBs) over 99\% of the ADC range at both room temperature and 77 K. The performance of this ADC ASIC meets the requirements of the LBNE far detector. A new ADC ASIC revision has been evaluated and will be used in the LBNE 35 ton prototype TPC.

\begin{figure}
\centering
\includegraphics[width=4.in]{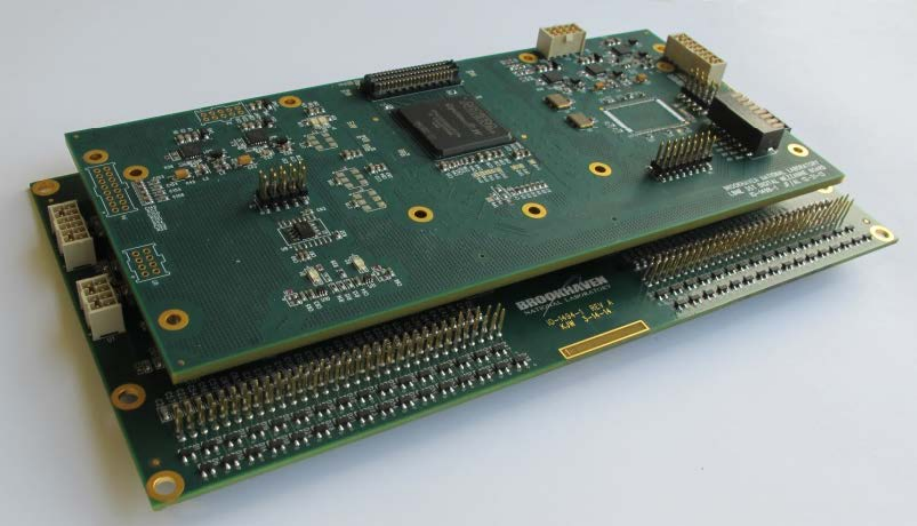}
\caption{\label{fig:prototypefeb}  A prototype prototype front-end motherboard assembly developed for the LBNE 35 ton prototype TPC.}
\end{figure}
The testing and operation of commercial Field Programmable Gate Arrays (FPGAs) at 77 K has been investigated. A candidate FPGA, the ALTERA Cyclone IV GX, has been identified as suitable for use on front-end board at 77 K. This FPGA will aggregate data from 8 ADC ASICs, and multiplex and transmit the data out of the cryostat. It can also perform digital signal processing, such as zero suppression and data compression. Studies of long term operation of commercial low dropout (LDO) voltage regulators at 77K have identified few suitable devices. One candidate LDO regulator has been running at 77 K successfully for more than 12 months. Both FPGA and LDO regulator candidates will be used to equip the LBNE 35 ton prototype TPC. A prototype front-end motherboard assembly, with an analog motherboard and an FPGA mezzanine board, has been designed and used in tests of the LBNE 35 ton prototype TPC Anode Plane Arrays (APAs). A full APA readout chain, including front-end motherboard assembly and a compact DAQ system, has been constructed and is being used to characterize APA assemblies. A picture of the front-end motherboard assembly is shown in Fig.~\ref{fig:prototypefeb}. Since the operating lifetime of commercial FPGAs at 77 K has not been established, and a digital ASIC for 77 K operation is being designed at Fermilab, the front-end motherboard assembly is designed to allow simple swapping of the FPGA and digital ASIC mezzanine boards. This design of the front-end motherboard is also planned for use in the LAr1-ND experiment and in the LBNE far detector, with incremental improvements as the technology develops and with optimization for the particular TPC configuration.

\subsection{Cold Digital Electronics (COLDATA)}

With the front-end ASIC and ADC in liquid argon, it becomes desirable to also place digital electronics near the front-end in the liquid.  After digitization the data can then easily be sparsified, compressed or otherwise processed, and multiplexed for transmission out of the cryostat. These operations all reduce the amount of information that needs to be transmitted, and the number of cables required to transmit it. A reduction of cables in the cryostat results in less contamination of the ultra-pure liquid argon caused by outgassing and cryostat penetrations. Cryostat design and readout design are decoupled.  As an example, a single APA would require 2400 lines to transmit individual wires, but digitized at 2 Msps the same signals would require only 30 each of 2Gbps lines.  This number can be reduced further by sparsification and compression.

Although FPGAs have been demonstrated to operate successfully in liquid argon, the operating lifetime, limited by hot carrier injection, is not known and is difficult to determine for a commercial device designed for near room temperature environments.  These considerations have motivated the need for a digital multiplexor ASIC that is designed for operation at 77K for use in the LBNE far detector. This work has begun by a team consisting of Grzegorz Deptuch and James Hoff at Fermilab, and of Tao Zhang, Tianwei Liu, Guoying Wu and Ping Gui at Southern Methodist University (SMU), the goal of which is a COLDATA  (COLd DAta Transmission ASIC)~\cite{jrhoff,gwu}.    This chip will provide bidirectional data transfer to and from the BNL ADCs: an uplink  at 2Gbps and slow down link for control and setup, all with adequate redundancy.  Several decision points are presently being studied and evaluated: $i)$ should zero suppression and/or data compression be implemented on chip?  $ii)$ which CMOS technology should be chosen, dependent on device models and CMOS electronics lifetime requirements? $iii)$ should the digital chip operate synchronously or asynchronously? $iv)$ do low drop-out voltage (LDO) regulators also need to be designed for long lifetime operation in liquid argon? $v)$ what level of redundancy is desirable, i.e. duplication of uplinks, reserve uplinks, clock links, lost clock / lost sync operation? $vi)$ what testability functions are desirable, i.e. reading back control and status data, generation of test data for uplink testing?

Current work on specifying and identifying COLDATA components has begun. A local clock oscillator device must be identified and evaluated for cryogenic operation.  Candidate devices have been identified and are being evaluated. Transistor simulation models and device parameters for the cryogenic temperature range are not available, and therefore test devices must be fabricated and characterized to obtain them. Statistical data characterizing variation of model parameters and statistical models of process and standard library digital cells are provided by foundries only near room temperature. These essentials for full custom, synthesized logic design and physical design verification need to be developed and verified for reliable use. 

A preliminary investigation of the lifetime at cryogenic temperature of MOSFETs produced in the Global Foundries CMOS digital process for two technology nodes, 130~nm and 65~nm, has been completed.  An automated test setup for device parameter measurements at cryogenic temperatures has been constructed at Fermilab.  It consists of a cryogenic vacuum chamber with a cryocooler and a computer control and measurement system. It implements an accelerated stress test method, which is the standard technique used to investigate the hot carrier degradation of MOSFET devices. This test applies voltages to the device that are higher than the nominal operating condition, and observes the time for a device parameter such as VTH, ID, and gm to degrade, typically by 10\%, from its initial value. The basic device physics quantitatively relates the damage at a shorter time at higher electric field to that in a longer time at lower electric field. This relation allows the measured lifetime under stress to be extrapolated to a lifetime for acceptable device performance under a defined operating condition. The new automated test system allows devices to be tested for longer times than it was possible before (in excess of 100 ks) and at lower stress voltages to achieve more accurate lifetime predictions. This system has been used to test 37 different MOSFET transistors of several widths and lengths in 130nm and 65nm technologies. Similar results were found for both technologies.

{\bf Lessons Learned}
\begin{itemize}
\item The smallest length transistors at stress condition of VGS$=$VDS have the shortest lifetime.
\item The width dependence of degradation at 77 K is not evident.
\item The degradation at stress condition of VGS$=$VDS is worse than that of VGS$=(1/2)$VDS for small length transistors.
\end{itemize}

\begin{figure}
\centering
\includegraphics[width=4.5in]{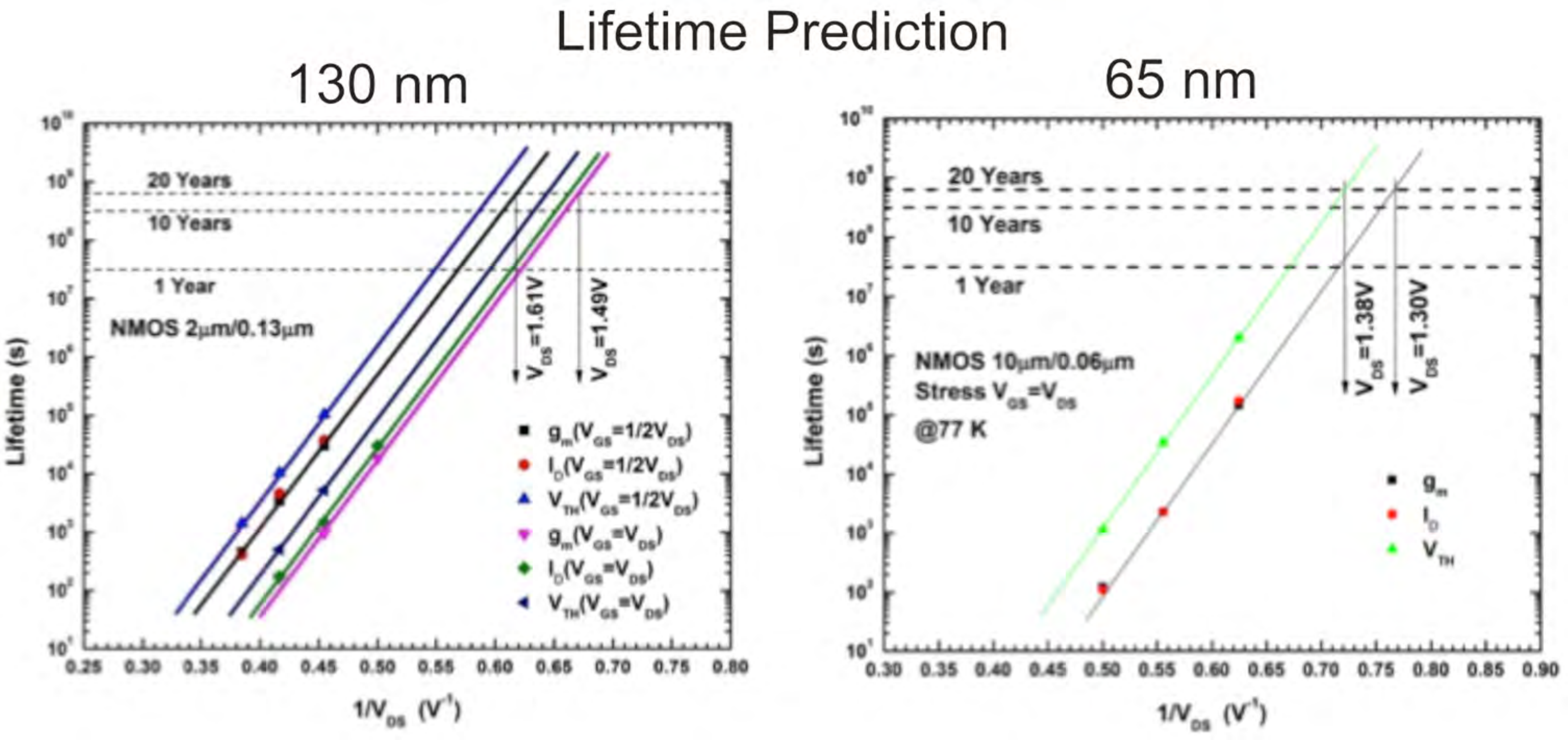}
\caption{\label{fig:elpred}  The lifetime as a function of the inverse of the drain-source voltage of the transistor at 77 K for devices fabricated in the 130~nm (left) and 65~nm (right) process.  Lifetime is defined as the time to a 10\% reduction of transconductance and drain current and by a 50~mV increase of threshold voltage.}
\end{figure}

Devices fabricated in either technology can be operated with long lifetimes at 77 K.  For the 130 nm devices, with a nominal supply voltage of 1.5 V, to achieve a 20 year lifetime, the power supply voltage should be less than 1.49 V at the worst stress condition. For the 65 nm devices, with a nominal supply voltage of 1.2 V, to achieve a 20 year lifetime, the power supply voltage only needs to be less than 1.30 V at the worst stress condition.  Figure~\ref{fig:elpred} shows the results of these measurements, with lifetime increasing to acceptable values as VDS in decreased.
 
With these positive results, a R\&D phase will now begin to extract basic transistor parameters at cryogenic temperature, including statistical variations, and add them to models for design of basic blocks, to develop a minimal digital library for synthesizable architecture of the COLDATA chip, and to investigate system architecture, and logic architecture required by data zero suppression and compression. The first functional blocks to be developed in collaboration with SMU will be a low-power PLL, a serializer and clock dividers for a 2 Gbps serializer.  Following the final decision to replace the present FPGA with the custom digital chip, work will begin on the design, prototyping and evaluation of major functional blocks, such as a line driver, a multi-channel input multiplexor with phase de-skewing, blocks for data formatting, and depending on results from the various LArTPC experiments in progress, blocks for implementing zero suppression and data compression.  This path will ensure a digital processing ASIC operating in liquid argon with a lifetime greater than 20 years.

\subsection{Data Acquisition}

A DAQ system that is quite similar to a standard collider system but using only commercially available computers and networking equipment can meet the needs of a large LArTPC.  Such a system will provide all available information including below threshold data for all triggered events.  Since all the data in a region of interest is read out, zero suppression thresholds can be set quite high to suppress noise and signals from radioactive decay. The current LBNE design is the result of the work of several people in the collaboration and the Fermilab Scientific Computing Department, including G. Barr, K. Biery, M. Graham and R. Rechenmacher.

Data rates for underground LArTPC detectors are dominated by radioactive decay of Ar$^{39}$ and Kr$^{85}$.  However, the mean track lengths in liquid argon for these decays are 0.28~mm and 0.39~mm respectively and the mean ionizations are 30\% and 40\% respectively of a minimum ionizing particle.  Most TPC designs have wire spacings more than 1~mm so only a small amount of charge will be collected on at most two wires.  Thus, a fairly simple zero suppression algorithm at the front end should eliminate most of these events.  The following discussion assumes that noise and decay event rates are very small.

From a DAQ perspective, physics events of interest have three defining characteristics: $i)$ they are quite rare, $ii)$ each event will be extensively studied and $iii)$ events occur in a localized region of the detector so not all of the detector needs to be read out.   The low event rate coupled with modern computer networking means that a software trigger is feasible.  Such a trigger allows easy upgrades over the course of the experiment and also allows the introduction of new triggers for possible new physics.  

Since each individual event may be of keen interest, the DAQ should read out all the event data including channels that are below a nominal zero suppression threshold.  To first order this goal can be achieved by recording data from wires that are near those that are above the threshold.  Developing an algorithm to select neighboring wires that works well for all possible events appears to be a challenging task.   A better approach is to read out an entire region around a triggered event.  Since events are localized to a subset of the detector and large LArTPCs are segmented into drift volumes, it is quite feasible to read out all the data for 1 or 2 drift times for 2 or 3 drift volumes near the triggering event.  This design is similar to a collider experiment where all detector data is stored for a short period of time while a trigger is formed and then the data is read out for triggered events.  The main difference is that in the LArTPC system the work is all done in software. 

Interesting events may come from processes such as supernova events or proton decays that are not synchronized to any clock system.  Thus, the data stream from the front-end electronics is divided into event blocks that are made up of data from the current drift time and one or more previous drift times.  This process allows finding events that cross a drift time boundary. 

\begin{figure}
\centering
\includegraphics[width=5.in]{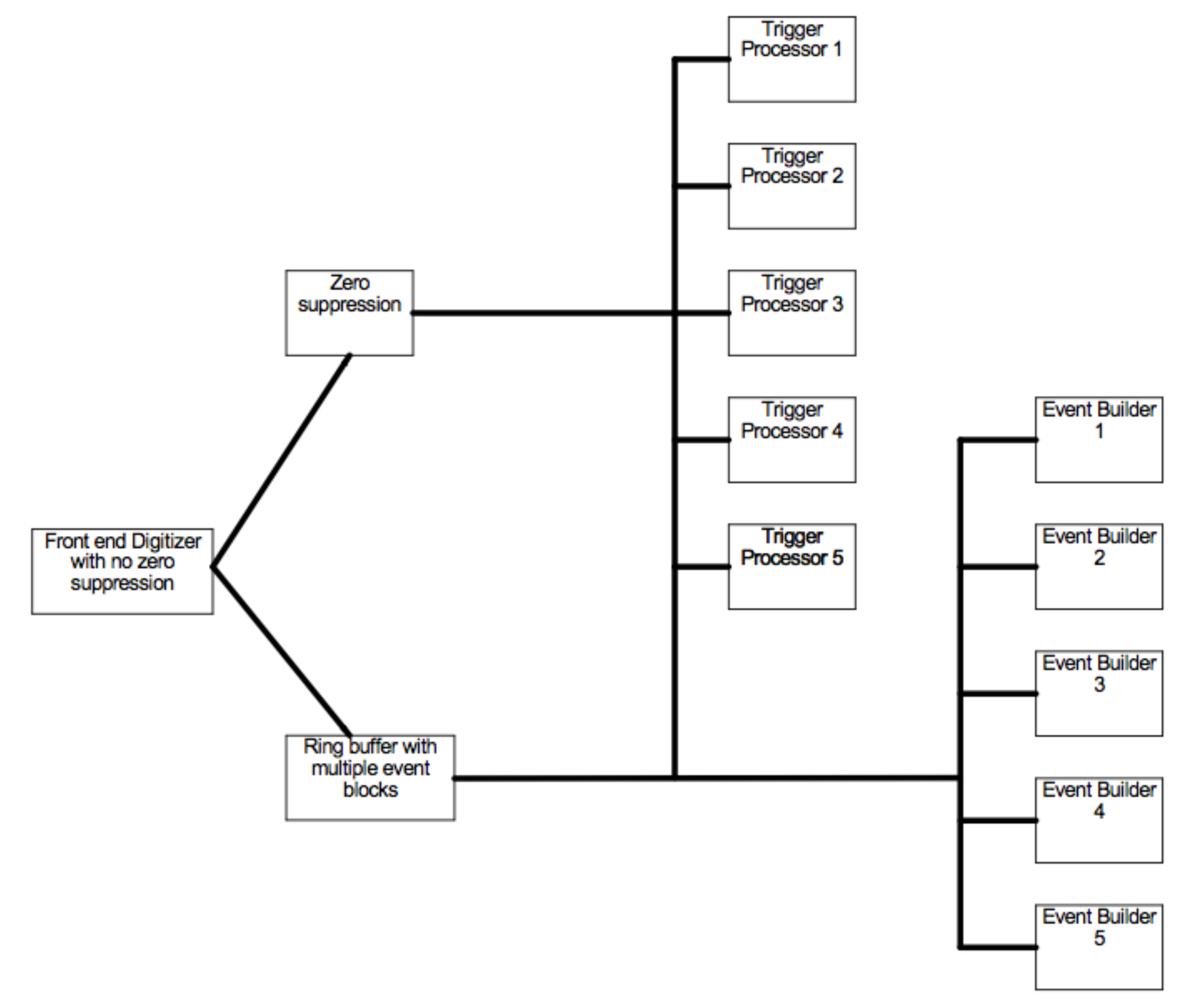}
\caption{\label{fig:block}  Simplified block diagram of the LBNE DAQ system.  Digitized data from the front-end system is sent simultaneously to the zero suppression system and the ring buffer system.  Data that is above the zero suppression threshold is sent to the trigger processors.  If a trigger pattern is found, the trigger processor looks up the readout pattern for this type of event and its location and broadcasts the data to all the ring buffers.  The selected ring buffers then extract the associated event block and send it to the event builders.  All data is synchronized by the time stamp system, which is not shown.}
\end{figure}

A block diagram of such a system is shown in Fig.~\ref{fig:block}.  All front-end data is simultaneously processed by a zero suppression algorithm and stored in a large ring buffer.  Zero suppressed data blocks are sent to trigger farm using a round robin scheduler.  When an event is found, the trigger processor determines the detector region to read out from a look up table and broadcasts this information along with the event block number to all the front-end processors over the network.  The selected front ends then extract the block number from the ring buffer and send the data to an event processor farm over a second network using a second round robin scheduler.

Round robin schedulers are used for both the trigger and event schedulers.  Since front-end buffer size is limited, there must be enough processors in the two networks so that a processor is free when its turn comes.  Since processors and network equipment is not very expensive, this appears to be feasible.  If not, one could easily implement a "routing master" computer that would keep a list of free processors and send the address of the next free processor to the front ends.  This adds considerable complexity to the system so it is likely to be a higher cost option.

The use of two different networks for triggers and events may not be needed.  Network bandwidth continues to increase and most network switches have internal buffers to locally buffer large blocks of data being sent to the same processor.  

\subsection{Time Stamping and Detector Synchronization}

The global positioning system (GPS) based clock provides synchronization with the accelerator and other detectors in the experiment and also provides a unique event number~\cite{novatime}.  The addition of encoding signals on the clock line checks the experiment wide synchronization at the end of every drift time.  Including a similar signal for the test pulse synchronizes the test pulse across the entire detector.

Not all physics events are associated with an accelerator cycle and some events such as those associated with a supernova need to be correlated with other detectors, an event numbering system based on the time of day is needed.  Events also need to be correlated with accelerator operation.  A system based on the satellite GPS meets all of these needs.  A 32 MHz system clock locked to the GPS system needs a 64 bit time stamp to allow a unique time number for all events over a 20 year operation of the experiment.  This time stamp is applied at the start of each trigger and event block and it is the only identifier used for these entities.  It is also applied to the accelerator cycle so that detector events can be correlated with the accelerator beam.

This system would operate in the same manner as the current NOvA system~\cite{novatime}.  A time stamp number would be sent to all the front end systems some time before the actual time and then a synch pulse would be sent at the appropriate GPS time to latch the time stamp number into the local time stamp register.

Multi-kiloton scale LArTPCs will have a large number of channels spread out over a large area.  All of these channels need to be kept synchronized, which can be done by encoding a synch pulse on the clock line.  That is, the start of every drift time block or sub set would be encoded on the clock line so that every event block would be guaranteed to have the same starting time.  Test pulses also need to be sent synchronously to all of the front-end digitizers.  Encoding a second signal on the clock line can accomplish this goal as well.  This pulse would be ANDed with a local enable signal so any pattern could be sent to the entire detector.

\section{Scintillation Light Detection}
\label{sec:Photons}


When charged particles pass through liquid argon, excited Ar$_2^*$ dimers are produced either through self-trapped exciton luminescence or through recombination luminescence.  In both cases different populations of singlet (${}^1\Sigma_{\mathrm{u}}$) and triplet (${}^3\Sigma_{\mathrm{u}}$) state Ar$_2^*$ dimers are produced, which radiatively decay with mean lifetimes of $\sim$6 ns and $\sim$1.5 $\mu$s, respectively. The photon that is emitted has a wavelength narrowly peaked at 128~nm. While detection of such vacuum ultraviolet (VUV) photons is typically a challenge, the facts that liquid argon is transparent to this scintillation light and that copious numbers of photons, $\mathcal{O}(10000)$ per MeV, are produced mean these photons are a valuable signal.

In liquid argon, de-excitation of the singlet state Ar$_2^*$ dimer provides a bright, prompt signal for fast timing. Additionally, the ratio of the fast singlet signal to the slow triplet signal depends on the ionization density from the incident charged particle and can provide a method of particle identification through pulse-shape discrimination. Detection of VUV photons is generally facilitated through use of photoluminescent compounds to shift the wavelength of the photons to a value more easily detected.  Wavelength shifters (WLS) like 1,1,4,4-tetraphenyl-1,3-butadiene (TPB) absorb short-wavelength photons and re-emit in the visible range with a peak around 430 nm. Typical approaches involve coating a surface in front of a cryogenic photodetector with a layer of the WLS. These wavelength shifters must be handled carefully, though, as they will degrade when exposed to UV light.

\subsection{Test Stands Available for R\&D}\label{sec:TestStands}

Much progress has been made in understanding the scintillation properties of liquid argon and optimizing various photon detection system designs at a variety of test stands across the US, ranging from small dewars to large multi-user facilities. 

Fermilab hosts a valuable facility for research and development in liquid argon. It provides cryogenic services and technician support for liquid argon operations. The facility provides ultra-pure argon to the cryostats, and a condenser is mounted on the cryostats to enable closed system operation without loss of argon through boil off.  The facility boasts a large general purpose liquid argon dewar,  ``TallBo.''  TallBo is a 460-liter dewar of similar design in which, for example, the larger-scale LBNE detector design studies described below in \S\ref{paddles} were conducted. 

Beyond Fermilab, the Cryogenic Detector Development Facility (CDDF) at Colorado State University will be operational soon and provide another platform for testing of large-scale liquid argon-based detector technologies. The dewar was designed to accommodate photon detector paddles which span the full two-meter-wide LBNE far detector anode plane assemblies, and can support testing of a variety of detector technologies.

\subsection{Recent Advances in Photon Detection}\label{sec:Advances}



A number of novel approaches to photon detection are currently being tested.  The detection of VUV photons in liquid argon is currently being advanced through four primary approaches: $i)$ WLS coupled with photomultiplier tubes, $ii)$ WLS-coated and reflective detector walls, $iii)$ large-area picosecond photodetectors, and $iv)$ light guides coupled to silicon photomultipiers.

\subsubsection{PMT-Based Photon Detection} \label{pmt}

A straightforward approach for collecting scintillation photons is to place cryogenic PMTs behind the the anode wire planes in a LArTPC.  The ICARUS T600 detector was the first to use this approach.  It used TPB evaporatively deposited on 8 inch diameter PMTs having a quantum efficiency of 18\% for blue light and a global quantum efficiency, defined as the number of photoelectrons (PEs) detected for each incident 128 nm photon, of 4--5\%~\cite{antonello}.  A total of 74 PMTs were installed in the ICARUS detector providing 1--2\% coverage of the anode plane surface area, which itself only subtends a fraction of the total solid angle at any point in the TPC.

\begin{figure}[ht]
  \begin{center}
    \includegraphics[width=0.45\textwidth]{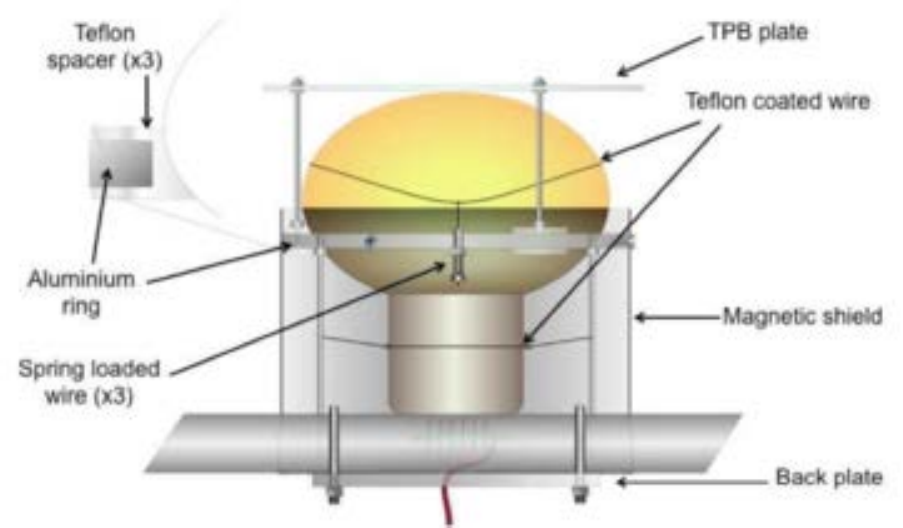}
    \includegraphics[width=0.45\textwidth]{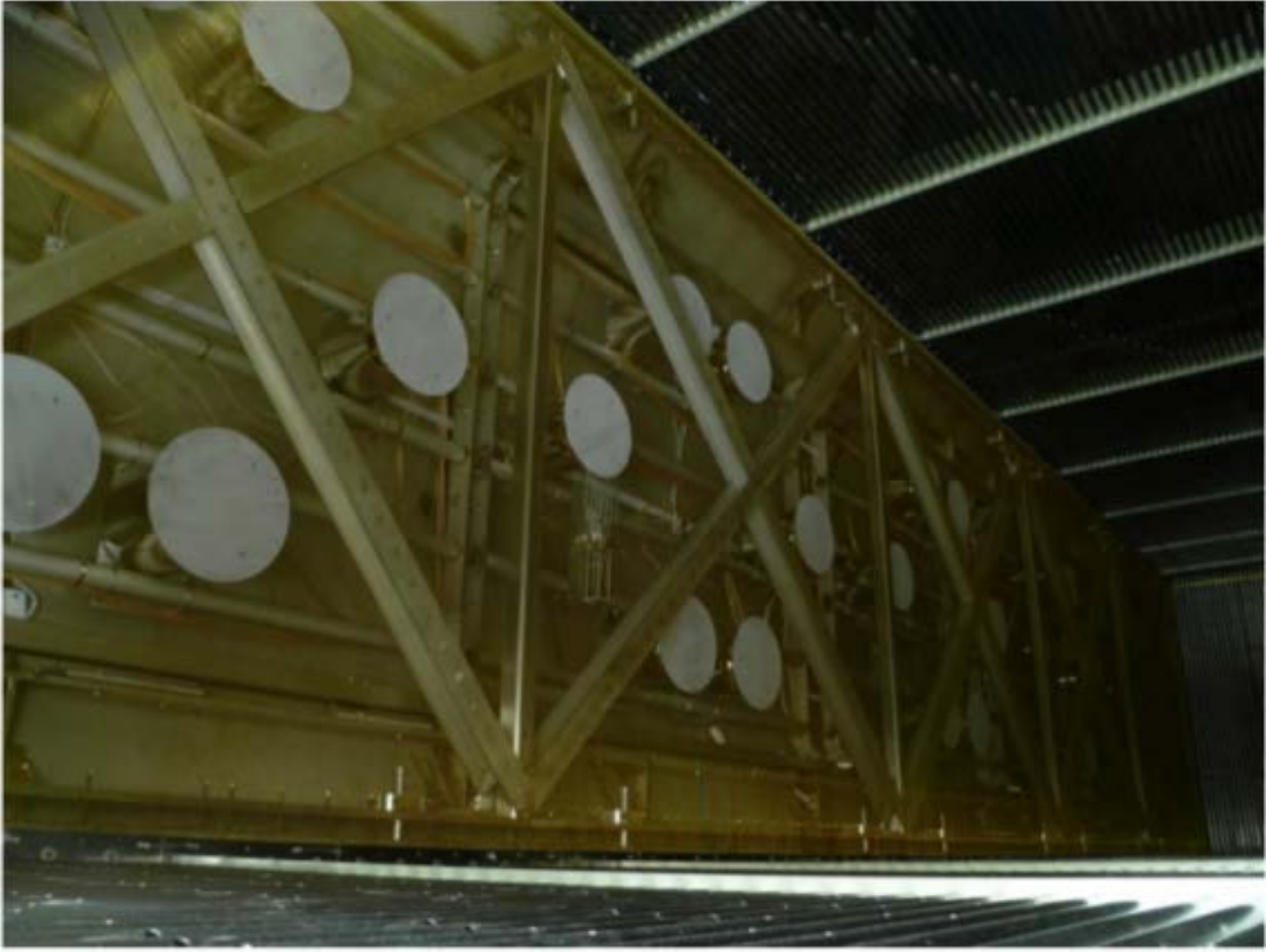}
    \caption{(left) Schematic illustrating the design of the PMTs for the \uboone\ photon detection system and (right) a photograph of the installed photon detection system in the \uboone\ TPC.}
    \label{fig:uBooNE}
  \end{center}
\end{figure}

The \uboone\ experiment has adopted a similar approach to liquid argon scintillation light detection and installed 8" PMTs 15 cm behind the TPC anode wires. In contrast to the ICARUS design, \uboone\ did not apply WLS to the PMT face but instead mounted thin TPB-coated acrylic disks with a 12 inch diameter directly in front of their PMTs, yielding assemblies with global quantum efficiencies of $\mathcal{O}$(1\%).  This simplifies the coating procedure, allows for a straightforward way implementation of quality assurance, and eases the challenges associated with storing, handling, and installing materials with TPB-based coatings.  MicroBooNE has 32 PMTs covering 9\% of the anode plane surface area.  Figure~\ref{fig:uBooNE} illustrates the basic PMT design (left) and their installed configuration (right).

The CAPTAIN detector, described in \S~\ref{sec:TestBeams}, envisions placing between 16 and 24 1" cryogenic PMTs behind the anode and cathode wire planes, covering 1\% of the anode plane surface area.  CAPTAIN also plans to install between 2 and 4 3" cryogenic PMTs.    As in \uboone, a TPB coating will be applied to an acrylic sheet placed in front of the PMTs to ease the preparation, storage, and installation of TPB-coated materials.

\subsubsection{Enhanced VUV Photon Collection from Detector Walls \label{LArIATphoton}}

\begin{figure}[ht]
  \begin{center}
    \includegraphics[width=0.45\textwidth]{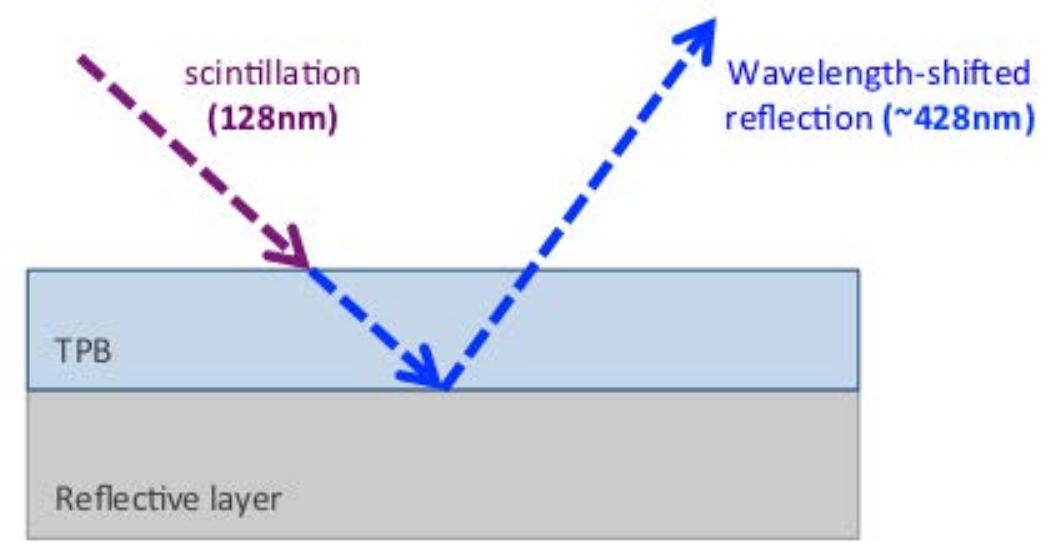}
    \includegraphics[width=0.45\textwidth]{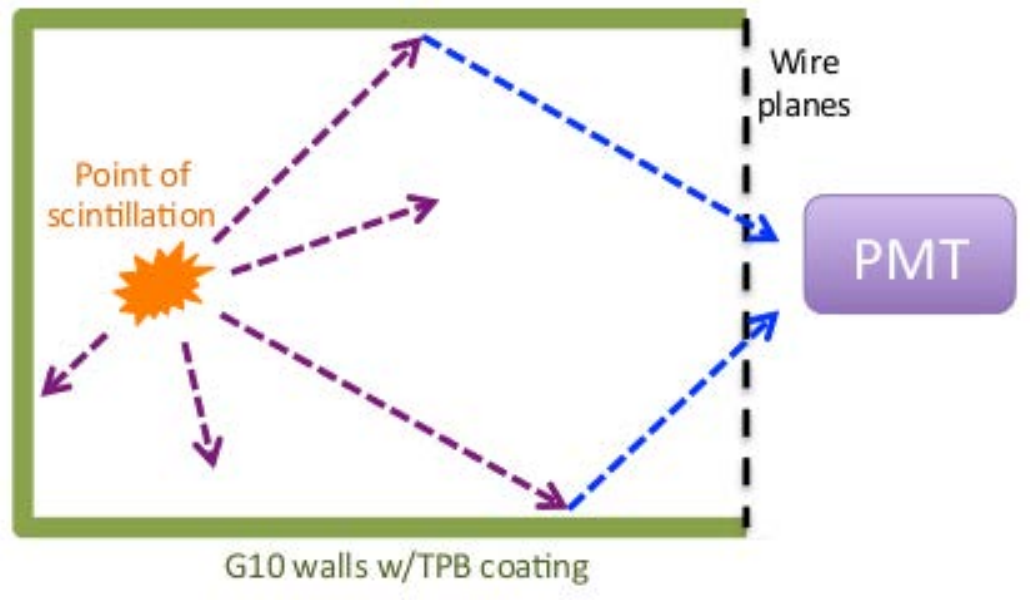}
    \caption{(left) The inner walls of the LArIAT TPC are coated with a layer of TPB on a thin reflecting foil. (right) The wavelength-shifted scintillation light is reflected back into the volume and collected by a PMT.}
    \label{fig:LArIATLight}
  \end{center}
\end{figure}

Photon detection systems based on PMTs have proven effective but are generally limited in geometric coverage. The LArIAT program, described in \S~\ref{sec:TestBeams}, is exploring a reflector-based approach to enhance of the total amount of light collected by a PMT-based photon detection system. Instead of placing the photodetector behind a TPB-coated plate, the walls of the LArIAT TPC are covered with a foil onto which a thin layer of TPB has been deposited by an evaporative coating process. The foils are efficient at reflecting this shifted light, which increases the total amount of light collected by a photocdetector viewing the TPC interior.  A schematic of how this technique works is shown in Fig.~\ref{fig:LArIATLight}.

The WArP~\cite{warp}, ArDM~\cite{ArDM}, and DarkSide~\cite{darkside} experiments have demonstrated the ability of TPB-coated reflectors to enhance light yield efficiency, achieving efficiencies of $\mathcal{O}$(10\%). The ratio of fast to prompt signals measured by such a system was shown to provide sufficient time resolution to separate nuclear recoil events from minimum ionizing events. LArIAT will be the first deployment of a reflector-based photon detection system in a single-phase LArTPC detector, providing results on the performance of this approach in the presence of a TPC field cage.

The experiment expects to achieve a light collection efficiency of $\sim$0.1\%, about an order of magnitude larger than the light collection efficiency of conventional PMT-based photon detectors in the $\mu$BooNE and ICARUS detectors. Additionally, LArIAT will collect the shifted/reflected scintillation signal with two cryogenic PMTs and two silicon photomultipliers in order to compare the relative performance of the two photodetector technologies.

\subsubsection{Progress Toward a Cryogenic LAPPD}

\begin{figure}[ht]
  \begin{center}
    \includegraphics[width=0.45\textwidth]{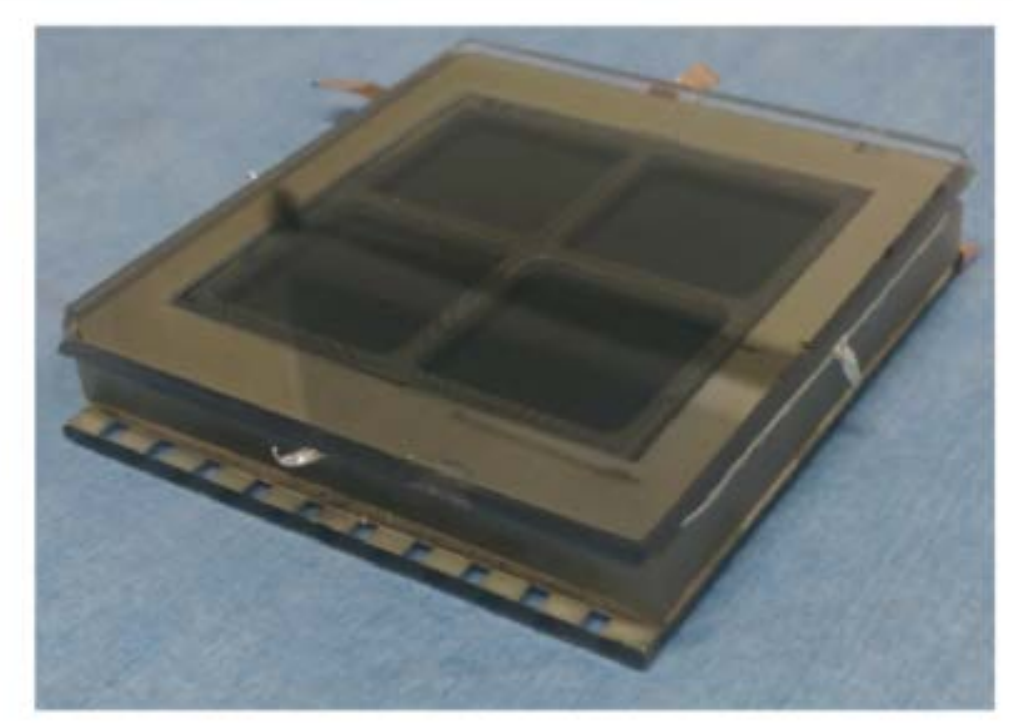}
    \includegraphics[width=0.45\textwidth]{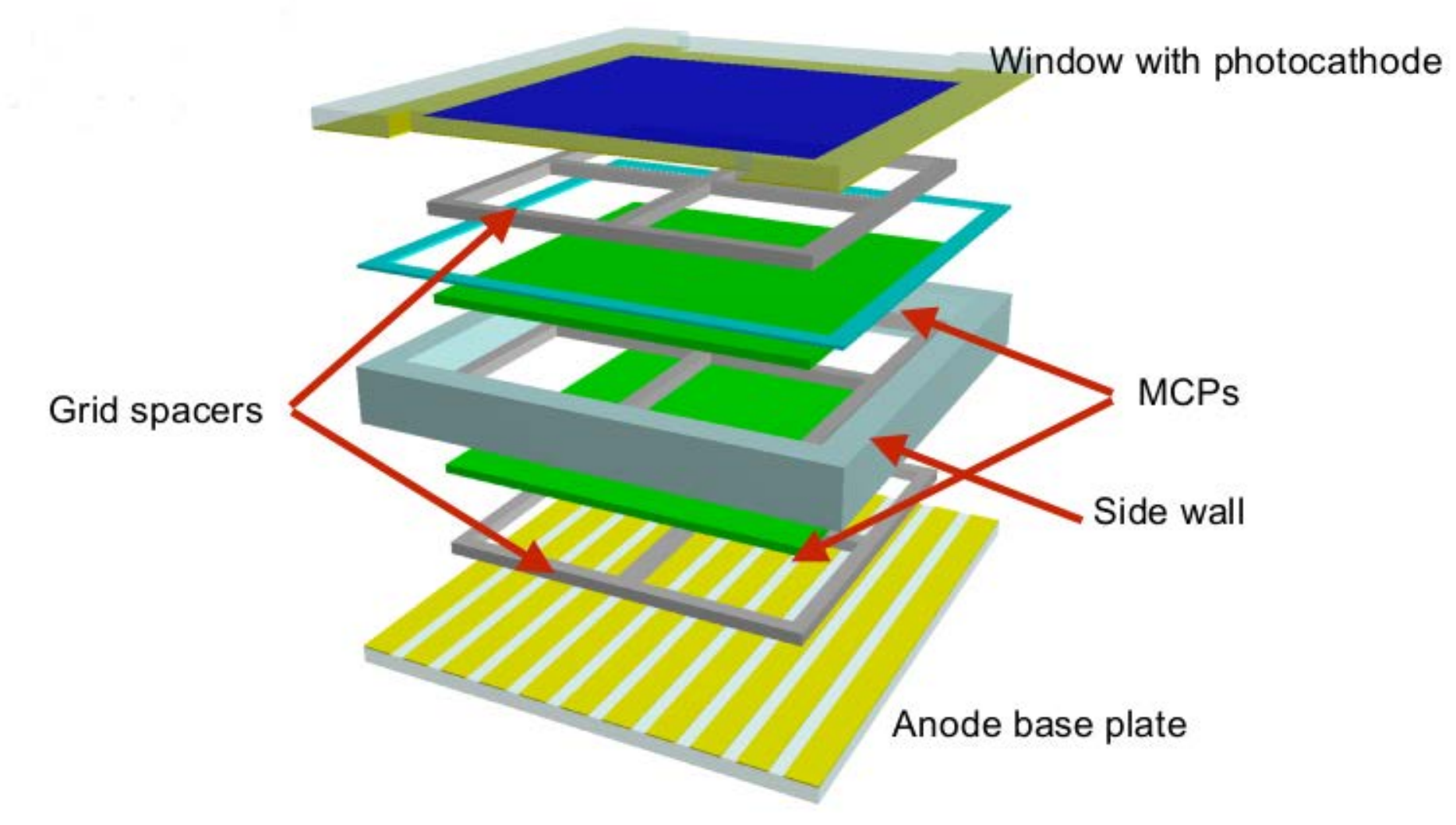}
    \caption{(left) A 6$\times$6 cm$^2$ ``small tile'' LAPPD and (right) exploded view of its components.}
    \label{fig:LAPPD}
  \end{center}
\end{figure}

A large-area picosecond photodetector (LAPPD) is a fast photodetector based on the microchannel plate (MCP) design~\cite{lappd}. LAPPDs have been deployed in a variety of high-energy physics experiments. They offer high spatial and temporal resolution, large geometric coverage, high quantum efficiency, and a low cost. These features make LAPPDs attractive for use in large area liquid argon photon detection applications. The LAPPD development group at Argonne National Laboratory (ANL) is instituting a systematic program of photocathode research and analysis with the goal of operating LAPPDs in a cryogenic environment. They are currently studying a 6$\times$6 cm$^2$ design based on borosilicate glass capillary arrays made by Incom, Inc.\ and functionalized using an industrial batch method of atomic layer deposition (ALD) is shown in Fig.~\ref{fig:LAPPD}. This combination offers a novel and inexpensive manufacturing method that allows for large formats, greater than 20 cm$^2$, and a tunable response.

Through fast and slow dunk tests in liquid nitrogen followed by vacuum tests, the group has investigated the robustness of these LAPPDs at 87 K and found that both the indium seal on the cathode and glass frit bond on the anode side remain intact after cryogenic cycling. The ANL group is also investigating the resistance of the MCP versus temperature in order to find the optimal ALD recipe for cryogenic operation. Going forward, they plan to deploy a 6$\times$6 cm$^2$ LAPPD that has been functionalized for cryogenic operation, along with appropriate readout electronics, into a dewar with TPB-coated walls to study the response of the LAPPD to wavelength-shifted light induced by the liquid argon scintillation signal.

\subsubsection{Light Guides for Large-Area Photon Detection \label{paddles}}

Large-area photon detector design based on acrylic light guides coupled to silicon photomultipliers (SiPM) are an attractive design for multi-kiloton scale detectors like LBNE.  These unprecedented large volumes present new challenges for the design of an affordable and effective photon detection system. This system must provide event timing of better than 1~$\mu$s  in order to determine event positions and correct track energy measured by the TPC. It must also provide an ability to trigger on non-beam events for investigations of proton decay and non-accelerator neutrinos. Furthermore, to be sensitive to supernova burst neutrino physics the photon detection system must also be able to detect neutrinos with energies on the order of 10 MeV.

\begin{figure}[ht]
  \begin{center}
    \includegraphics[width=0.75\textwidth]{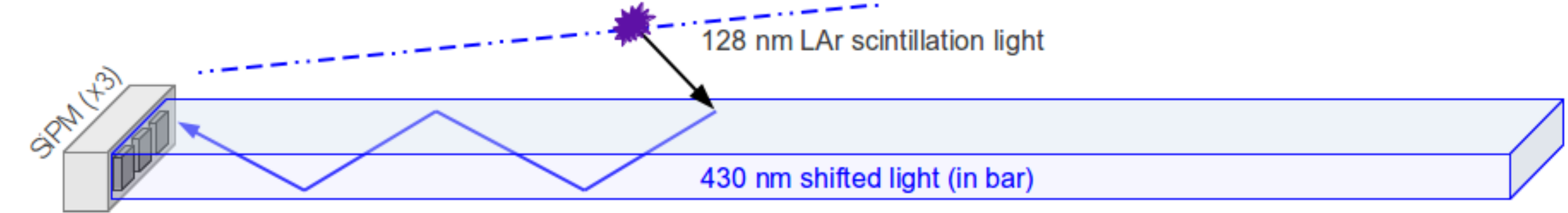}
    \caption{Illustration of a single light guide. A WLS-coated acrylic bar converts VUV scintillation photons to visible and propagates them via total internal reflection to an array of SiPMs at the end.}
    \label{fig:lightguide}
  \end{center}
\end{figure}

The LBNE collaboration is pursuing this option for photon detection. In the current baseline design, each photon detector module consists of four acrylic bars coated with a WLS which collects VUV scintillation photons and converts them to $\sim$430 nm.  The photons propagate via total internal reflection to the ends of the bars. An array of three 6$\times$6 mm SiPMs on each acrylic light guide, with a peak response at 420 nm, detects the waveshifted light. This segmented design provides some position-dependent response, helping to localize the trigger signal while maintaining a large sensitive surface area with a low photocathode area.  The basic design of one light guide is illustrated in Fig.~\ref{fig:lightguide}.

A number of waveguide designs are currently being investigated by the LBNE collaboration. The base design is an acrylic bar of dimensions 50.8 cm $\times$ 2.54 cm $\times$ 0.6 cm with a surface coat of WLS, either TPB or 1,4-bis-(o-methyl-styryl)-benzene (bis-MSB). Three methods of applying the surface coat have been investigated: $i)$ a design developed at Indiana University where a solution of TPB or bis-MSB in methylene chloride is spray-painted onto an acrylic bar from a pressurized vessel, after which the bar is flash-heated rapidly to melt a thin outer layer of acrylic; $ii)$ a design developed at the Massachusetts Institute of Technology where a solution of TPB and acrylic dissolved in toluene is applied to the surface of an acrylic bar by hand-painting; $iii)$ another design by the MIT group where an acrylic bar is dipped into this solution of TPB, acrylic, and toluene to leave a thin layer of acrylic with embedded TPB on the surface of the bar.  In addition to these surface-coating methods, two commercial vendors have been contracted to produce acrylic or polystyrene bars  with TPB or bis-MSB doped throughout the bulk of the plastic at a concentration of 1\% during the polymer casting process.

\begin{figure}[ht]
  \begin{center}
    \includegraphics[width=0.75\textwidth]{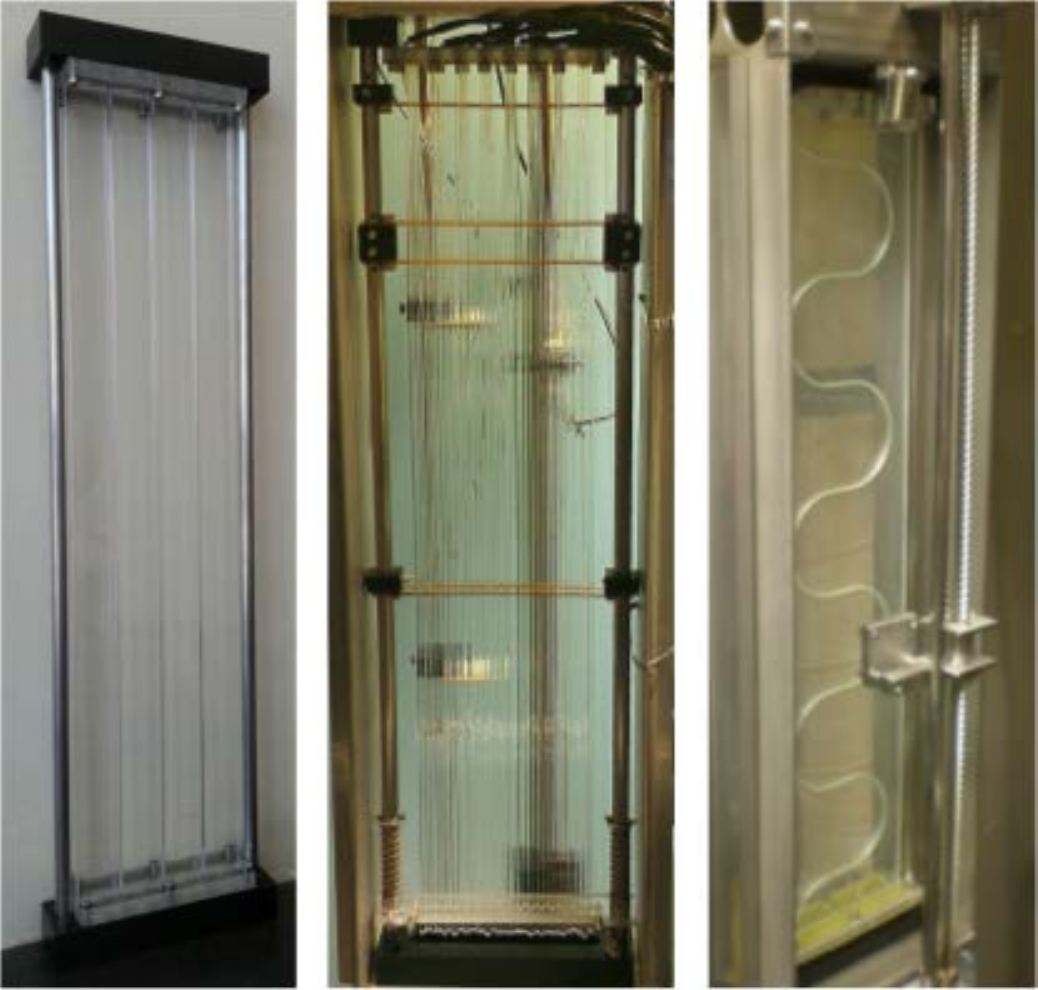}
    \caption{Examples of the three light guide-based designs under investigation for LBNE. From left to right are acrylic bars coated with a WLS compound, TPB-doped polystyrene fibers, and a TPB-coated acrylic plate with imbedded WLS fiber.}
    \label{fig:LBNE}
  \end{center}
\end{figure}

Two additional module designs are also under development. Colorado State University uses thirty-two 3$\times$3 cm$^2$ polystyrene fibers doped with TPB to cover approximately the same active area as four bar-based light guides. This design has the advantage of mapping four fibers at a time completely onto the surface of the 6$\times$6 cm$^2$ SiPM package, reducing the number of SiPMs required for the same geometric coverage. The group at Louisiana State University is developing a 20.3 cm-wide acrylic plate into which a Y11 WLS fiber has been embedded. The surface of this plate is then coated in TPB so that the scintillation light is converted to 430 nm, then captured and converted by the fiber and propagated to its ends. One SiPM is positioned at each end of the fiber, with the potential to further reduce the number of photodetectors required and provide a position measurement based on the relative signal strength at each SiPM. Figure~\ref{fig:LBNE} shows the three options under consideration for LBNE.

To facilitate acquisition of data from this SiPM-based photon detection system, a fast custom readout system tailored to the output characteristics of SiPMs is being developed by the High-Energy Physics electronics group at ANL. The SiPM Signal Processor (SSP) features single photoelectron-capable resolution, a 14-bit dynamic range, timing resolution better than 5 ns, a 13~$\mu$s data buffer to collect both the prompt and late signals, and highly configurable digital signal processing implemented in firmware. Four SSPs have been completed and deployed at institutes for local testing of photon detector designs.

Design comparisons were carried out in tests at the TallBo facility at Fermilab. The cryostat is large enough to accommodate up to sixteen different lightguide designs in an ultra high purity liquid argon environment. Tests of various coat thickness options of TPB and bis-MSB on flash-heated acrylic bars have been performed, allowing for the identification of the optimum amount of each waveshifter for that design. Another set of tests compared several bar-based lightguide modules and the polystyrene fiber design side by side. The comparison identified acrylic dip-coated with TPB, polystyrene doped with TPB, and one flash-heated bar coated with bis-MSB as the most promising designs. This run also deployed the SSP to read out one module, allowing a detailed analysis of the time-resolved structure of the scintillation signal induced by cosmic-ray muons.

\subsection{Effects of Contaminants on Scintillation Light \label{contaminants}}

While ultra-pure liquid argon features a long attenuation length at its scintillation wavelength, the presence of impurities can both reduce this attenuation length and quench the production of scintillation light. Some of these impurities also impact electron lifetime in the electric field of the TPC. Contaminants are particularly difficult to remove when they are present in the source gas, similar in boiling point to liquid argon, or not removable by regenerable filtering techniques.

The particular composition of impurities depends on the source gas. Industrially-produced argon is collected from air, and typically includes some amount of nitrogen, oxygen, and water. Commercial ultra-high purity (UHP) argon is manufactured by industrial distillation followed by purification with molecular sieves and filters. This method typically produces UHP liquid argon with small residual nitrogen at the ppm level and oxygen at below ppb level concentrations. Argon sourced from $\mathrm{CO}_2$ contained in underground wells may lack the radioactive ${}^{39}\mathrm{Ar}$ isotope, making it ideal for liquid argon-based dark matter applications. However, this underground-sourced argon typically contains additional residual helium, methane, and $\mathrm{CO}_2$ impurities.

Of these contaminants, nitrogen is a particularly difficult impurity to remove. While it is not a concern for TPC performance, nitrogen both quenches the production of VUV scintillation photons~\cite{acciarri} and shortens their attenuation length~\cite{jones1, jones2}.  The absorption effect for meter-scale detectors is significant at concentrations above 2~ppm. Furthermore, the presence of nitrogen above this level also shortens the mean lifetime of the triplet state, degrading a photon detector system's pulse-shape discrimination capability. Fortunately, typical industrially-sourced UHP argon contains sub-ppm concentrations of nitrogen. Careful procedures during filling and operation of the cryostat should be sufficient to maintain a pure liquid argon environment.

The quenching and absorption effects of methane on liquid argon scintillation light has recently been studied along with a search for evidence of reemission in the visible spectrum~\cite{methane}.  Both absorption and quenching of the VUV signal from a polonium source in liquid argon was found and was unaccompanied by any observed re-emission in visible wavelengths.  For distances of $\mathcal{O}$(10 cm)  absorption lengths of $\mathcal{O}(10^{-3})$/ppb were found and absorption was shown to dominate over quenching.  This result indicates the maximum residual methane concentration tolerable for liquid argon experiments dependent on photon detection which rely on underground-sourced argon.

Instead of being lost to a contaminant, the $\mathrm{Ar}_2^*$ excitation energy can also be deliberately transferred to a dopant, which then decays by photon emission. A promising dopant for signal enhancement through this transfer mechanism is xenon. The photon emitted by an excited xenon atom has a wavelength of 175~nm. This transfer could prove advantageous in certain photon detection applications because of the higher sensitivity of some TPB-based coatings to VUV photons of that wavelength. At xenon concentrations between 100~ppm and 1000~ppm, a large fraction of the triplet state $\mathrm{Ar}_2^*$ emission seems to be replaced by a signal at timescales of $\mathcal{O}(10~\mathrm{ns})$.

These two effects make xenon a candidate for moderate enhancement of photon detection in liquid argon. However, xenon is an expensive element, making its deployment as a dopant at concentrations as low as 100~ppm in a large-volume LArTPC an expensive proposition. The current estimated budget for the LBNE far detector's photon detection system is \$10~million. For comparison, the cost of adding 10~ppm of xenon into the far detector cryostat is \$5~million. An additional 50\% photon detector coverage would provide a more substantial enhancement in light collection efficiency than the equivalent cost of xenon dopant.  While further studies need to be conducted, current results seem to indicate that xenon doping does not produce a cost-effective enhancement of the scintillation yield~\cite{xenon}.

\subsection{Characterization of Silicon Photomultipliers for Photon Detection Systems}

While operated at cryogenic temperatures, SiPMs offer several appealing features suitable for use in a photon detection system. The single-microcell response features a long exponential tail but a sharp rising edge with a few nanosecond rise time. The amplitude of the prompt rising edge and the integrated charge collected by the SiPM both exhibit a quantized and discrete integer photoelectron response, making measurement of the number of photons incident on the SiPM straightforward. Furthermore, the noise rate in the SiPM just above the breakdown voltage drops from several kHz to a few Hz at 87~K, meaning they can be operated with a bias voltage of a few tens of volts in a regime with simultaneously high gain and low noise.

The LBNE photon detector prototype test at the TallBo facility provided the first in-situ test of the SSP prototype photon detector readout electronics. The system operated with negligible electronics noise, similar gain to warm temperature operation, and reliable data quality. The single-photoelectron response of each SiPM was easily derived from in-situ data, and the time-resolved structure of the scintillation signal from cosmic-ray muons was successfully measured with this system despite the 500 ns long exponential tail of the SiPM response. The discrete and quantized response of SiPMs to incident photons at cryogenic temperatures makes signal calibration simple.

{\bf Lesson Learned} SiPMs have been demonstrated to perform well in liquid argon and offer several appealing features including low noise, high gain, and fast timing. 

\subsection{Design Considerations from Physics and Engineering}\label{sec:Engineering}

\subsubsection{Physics-Driven Design Requirements}

The scintillation light in neutrino LArTPC detectors can be used to trigger detector readout, determine the event start time ($t_0$) for non-accelerator events, reject cosmic backgrounds, and supplement charge-based event reconstruction such as particle identification.  The specific requirements of a photon detection system vary between applications.  For example, light collection systems play a pivotal role in rejecting cosmic backgrounds in surface LArTPCs.  To do so requires the matching of TPC tracks to light flashes reconstructed by the photon detection system.  Therefore, it is not only the amount of light collected but also the granularity of the photon detection system that matters in order to localize the light signal.  On the other hand, underground LArTPCs are sensitive to more types of non-accelerator events, such as supernova neutrinos, which must be triggered.  Here, some position resolution can be sacrificed in order to collect more light at a reduced cost. In both cases it is important to collect enough light to determine the interaction time, $t_0$.

\subsubsection{Timing Resolution}

The fundamental limit on the timing resolution of a liquid argon photon detection system is dependent on the arrival time of the first detected photon, typically less than the 6~ns mean lifetime of the singlet state scintillation component when many photons are detected. This fact presents the potential for more sensitive measurements of event timing.  The MiniBooNE experiment highlights an opportunity for improved timing resolution.  It has demonstrated the capability of resolving the radio frequency structure of the Fermilab Booster neutrino beam. Such a measurement may provide an important handle on background rejection and could be a key handle on light dark matter searches. A nanosecond timing precision could also provide for other extended capabilities of a photon detection system, like enhanced cosmic background rejection or even separation of Cherenkov light from the liquid argon scintillation signal.

\subsubsection{New Approaches to Geometric Coverage}

The growing size of liquid argon-based particle detectors is driving photon detection systems toward larger geometric coverage. A variety of designs are being explored with the goal of mitigating the corresponding rise in cost. Interference with the TPC must be also minimized, requiring photon detection system designs that avoid both mechanical and electrical interference with TPC operations.

The enhanced collection of scintillation photons using coated detector walls, described above, is a promising option for enhancing the collection of VUV photons using conventional detectors like PMTs. The total number of VUV photons converted to visible is larger, potentially providing more light to be collected by a traditional array of PMTs without a wavelength-shifting coating. This method does present a possible loss of position resolution since photons arriving at the photocathodes directly from the source may be swamped by reflected light. Materials for the reflective walls must also be carefully chosen to avoid buildup of charge near the TPC. 

The LBNE collaboration is testing another scalable design for a large-area photon detection system, based on light guide paddles (\S\ref{paddles}). The entire paddle surface is sensitive to the VUV photons while internally-conducted light is detected at the end of the paddle. Currently 50-cm paddles are being tested, with the goal of exploring designs up to 2 meters in length. Readout is accomplished by small-area SiPMs, helping to reduce the photocathode area and cost. This modular design affords some degree of position resolution while providing large and scalable surface area coverage. The photon detector modules will be located outside the TPC drift field in the anode plane assemblies, aided by the small physical profile of the SiPM detectors.

\section{Calibration and Test Beams}
\label{sec:TestBeams}

Liquid argon calibration and test beam efforts are now better coordinated than in previous years.  The common aim is to develop techniques and make measurements that will be useful in future liquid argon experiments, and to make this global knowledge base available to the wider community. The previous LArTPC R\&D workshop summary~\cite{LArRDworkshop2013} gave an overview of the main calibration and test beam efforts ongoing in the liquid argon community. This section summarizes in more detail the types of calibrations that will be done in MicroBooNE and LBNE, and gives an updated status and plans for test beam experiments such as CAPTAIN, LArIAT, and CERN WA105 (LBNO-Demo).

\subsection{Calibration}

An overview of the MicroBooNE, CAPTAIN, LArIAT, and LBNE TPCs, readout electronics, and cryogenic systems were presented in the previous workshop summary~\cite{LArRDworkshop2013}. Here we do not repeat those details, but instead discuss the specifics of planned calibrations using these detectors.

Charged particles traveling through the argon lose energy by both ionization and by excitation, resulting in scintillation light. The ratio of the two is well understood, and depends on the strength of the drift electric field. However, other factors in the detector environment, such as purity and temperature, can also affect the amount of charge and light collected by the detector readout. Calibration of this response is a crucial component to proper reconstruction and identification of particle tracks in LArTPCs.

\subsubsection{Charge}

The initial cloud of charge will disperse due to diffusion and some fraction of the charge will be captured on impurities in the argon.  Controlling the the purity, the temperature, and the drift field is an important initial step in maintaining accurate calibration of LArTPCs. 

In MicroBooNE, there are three purity monitors and a gas analyzer to measure the levels of oxygen, water, and nitrogen, the most relevant contaminants affecting the charge and scintillation light detection. In addition, temperature probes will monitor the TPC and bulk liquid temperatures, and heaters have been installed on the outside of the cryostat for temperature control, to ensure uniformity of the argon temperature throughout the full volume of the TPC. Any distortion in the drift electric field can lead to bent tracks; distortions may be due to space charge buildup of the slow-moving ions drifting opposite to the direction of the ionization electrons, or due to mismatch in the components of the field cage resistive divider chain.

\begin{figure}
\centering
\includegraphics[width=5.in]{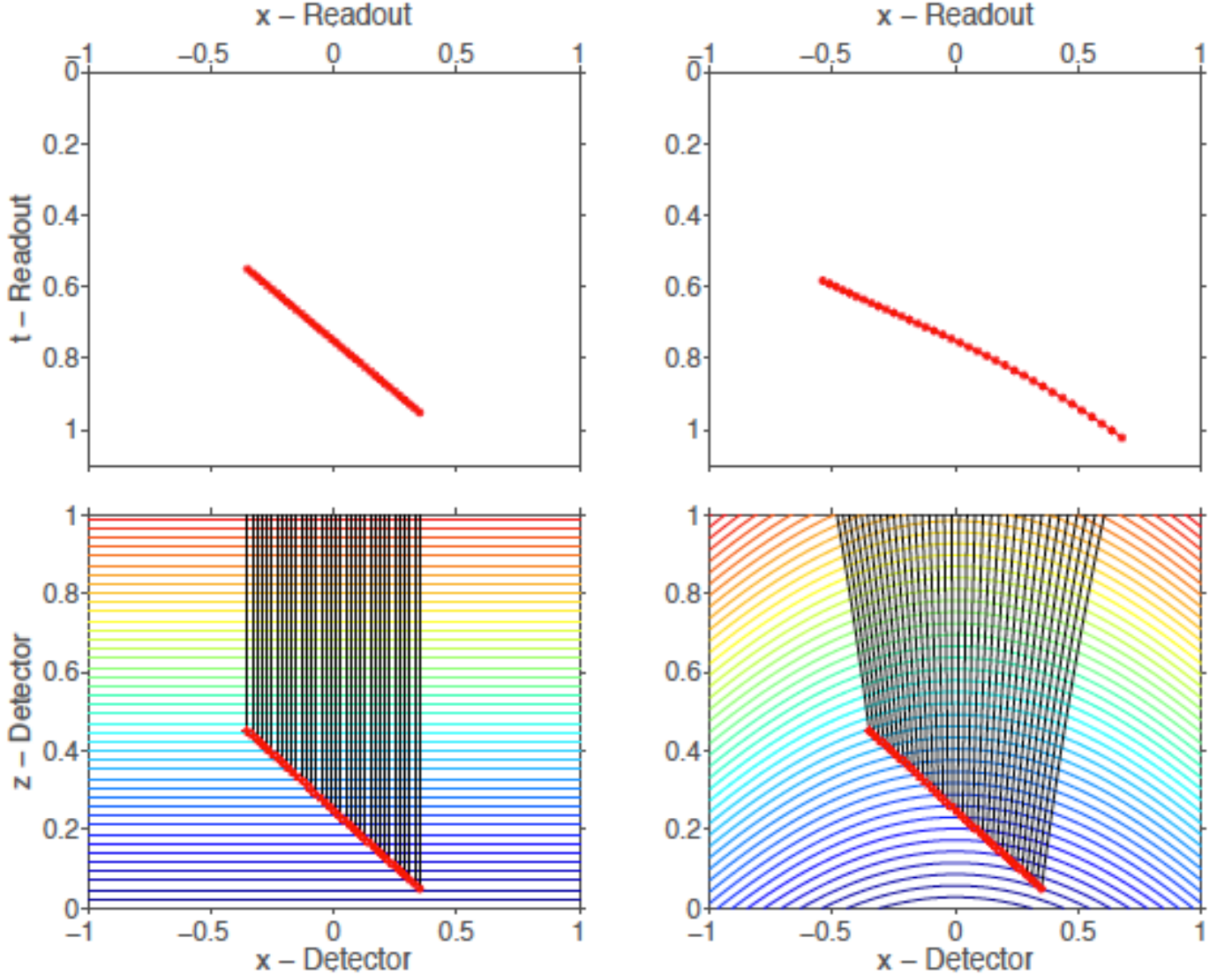}
\caption{\label{fig:bent-track} 
For a uniform electric field (bottom left panel), ionization charges from a straight track are drifted to the readout wire plane along the shortest path, resulting in a straight track read out in wire number vs. time (top left panel). For a distorted electric field (bottom right panel), ionization charges from a straight track may be drifted along a less direct path to the readout wire plane, resulting in a track that appears curved in the wire number vs. time view (top right panel).
}
\end{figure}

Figure~\ref{fig:bent-track} demonstrates the effect of a distorted electric field. For a uniform electric field (bottom left panel) in detector coordinates of wire number (x-Detector) {\it{vs.}} drift distance (z-Detector), ionization charges from a straight track are drifted to the readout wire plane along the shortest path, resulting in a straight track read out in wire number {\it{vs.}} time (top left panel). For a distorted electric field (bottom right panel), ionization charges from a straight track may be drifted along a less direct path to the readout wire plane, resulting in a track that appears curved in the wire number {\it{vs.}} time view (top right panel).

\begin{figure}
\centering
\includegraphics[width=3.in]{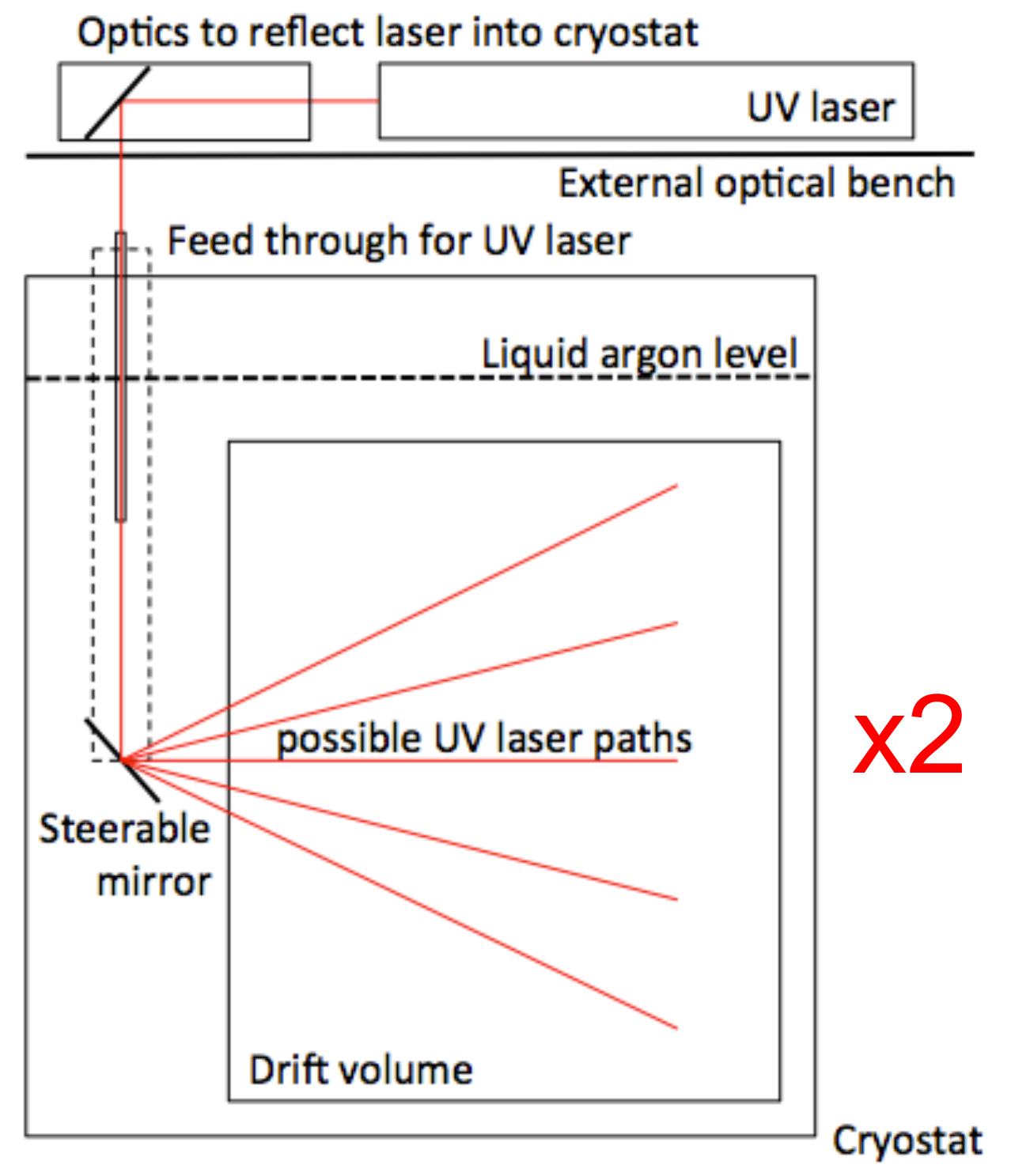}
\caption{\label{fig:ub-laser} 
The MicroBooNE UV laser calibration system with steerable mirror, situated at each end of the TPC (only one end shown here).  }
\end{figure}

In MicroBooNE, the uniformity of the electric field will be mapped using 266~nm UV lasers situated at each end of the TPC, shown in Fig.~\ref{fig:ub-laser}. The lasers ionize the argon along their paths, leaving trails of ionization charge that {\em{should}} be straight. Any distortion in the laser tracks can be corrected back to straight lines, thus creating a calibration map to compensate for any inhomogeneities in the electric field. The proof of principle was successfully demonstrated in ARGONTUBE at LHEP Bern, presented in the previous workshop of this series. The design and implementation of the fully-featured production system installed in MicroBooNE benefited from the ARGONTUBE demonstration.

\begin{figure}
\centering
\includegraphics[width=3.in]{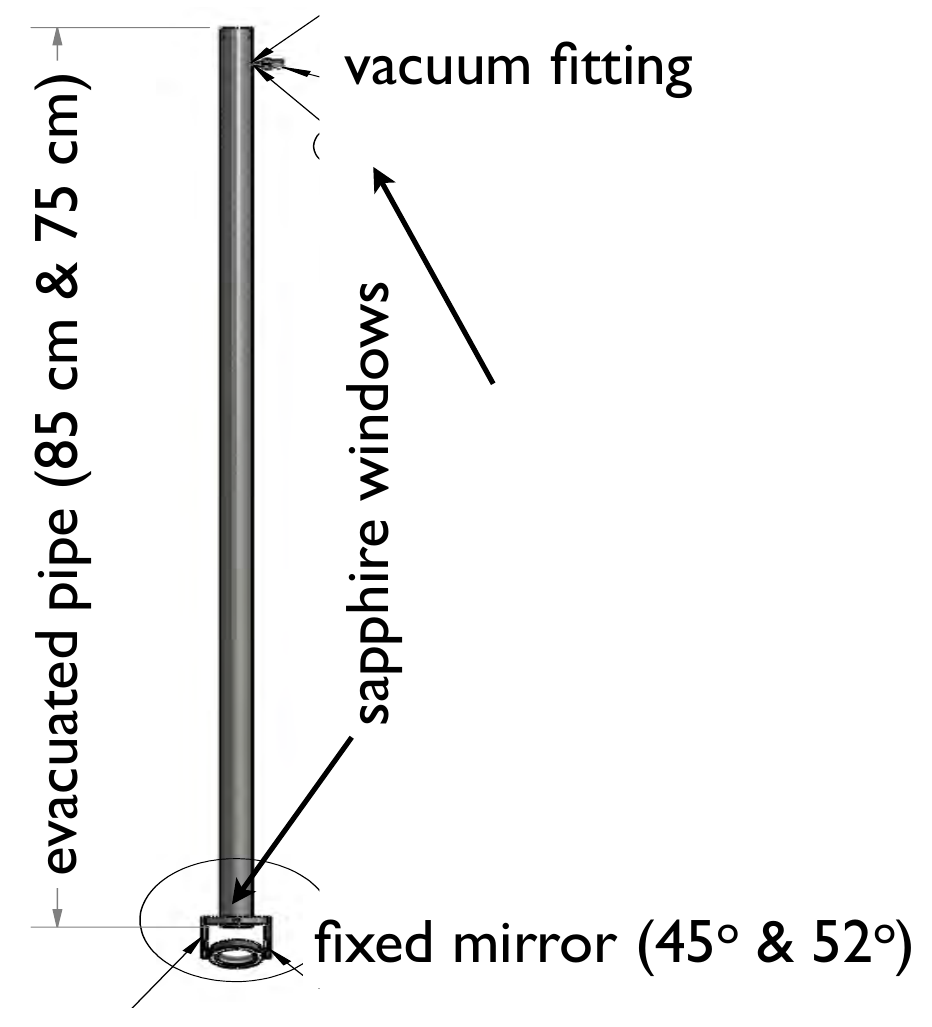}
\caption{\label{fig:CAPTAIN-laser} 
The prototype LBNE UV laser system installed in CAPTAIN. 
}
\end{figure}

LBNE also plans to have a UV laser calibration system. A small-scale prototype of this system is being implemented in CAPTAIN (and Mini-CAPTAIN). It is similar to the MicroBooNE laser system, but with a simplified fixed-incidence-angle mirror design that is rotatable through a small range of azimuthal angles, as seen in Fig.~\ref{fig:CAPTAIN-laser}. The CAPTAIN system will be used to validate the LBNE optical feedthrough design and the rotary control and encoder systems. It will also provide the opportunity to test proposed calibration concepts and to understand potential damage to the TPC. In view of future use in much larger detectors such as LBNE, this system will also verify whether the required pointing accuracy can be achieved of better than the wire spacing for distances up to about 20~m, and whether the beam divergence can be kept small enough to be able to ionize liquid at 30~m from the source.

Charge calibration is also necessary in the readout electronics for the induction and collection plane wires. For each readout channel, MicroBooNE will measure the baseline, noise, crosstalk, gain, linearity, and signal shape/rising time. A calibration capacitor is incorporated into the cold electronics ASIC front-end; injected calibration pulses will be used to characterize the cold ASIC behavior and the behavior of all downstream components, including cables, feedthroughs, warm electronics, and digitization. This calibration scheme will also be used in CAPTAIN, which has the identical cold electronics, readout electronics, and DAQ chain as MicroBooNE. 

Finally, cosmic rays can also be used to perform a number of calibration studies. The average reconstructed charge of the cosmic rays, which should be a Landau distribution, will be used to characterize the calorimetric reconstruction performance of LArTPCs. Additionally, Michel electrons from the decay of stopping cosmic muons have a very well-understood energy spectrum with a sharp cutoff just above 50~MeV; these can be used as a calibration source for both the TPC and the light collection system in MicroBooNE and future LArTPC experiments.

\subsubsection{Light}
\label{sec:testbeamlight}

Since a large fraction of the energy deposited in the argon by charged particles will be in the form of scintillation light, it is desirable to collect as much of it as possible in order to perform accurate calorimetric energy reconstruction. The prompt fraction of the light will also provide a $t_{0}$ to mark the beginning of an event, which will be especially critical for non-beam analyses in LArTPCs.

The MicroBooNE optical system consists of 32 8-inch photomultiplier tubes (PMTs) and 4 light-guide paddles. Before installation, a few of the PMTs were subjected to detailed characterization of stability, linearity, and absolute collection efficiency, described further in~ \cite{uBPMTtests}. The light-guide paddles are a candidate light collection detector for LBNE, installed as an R\&D test in MicroBooNE.

\begin{figure}
\centering
\includegraphics[width=4.in]{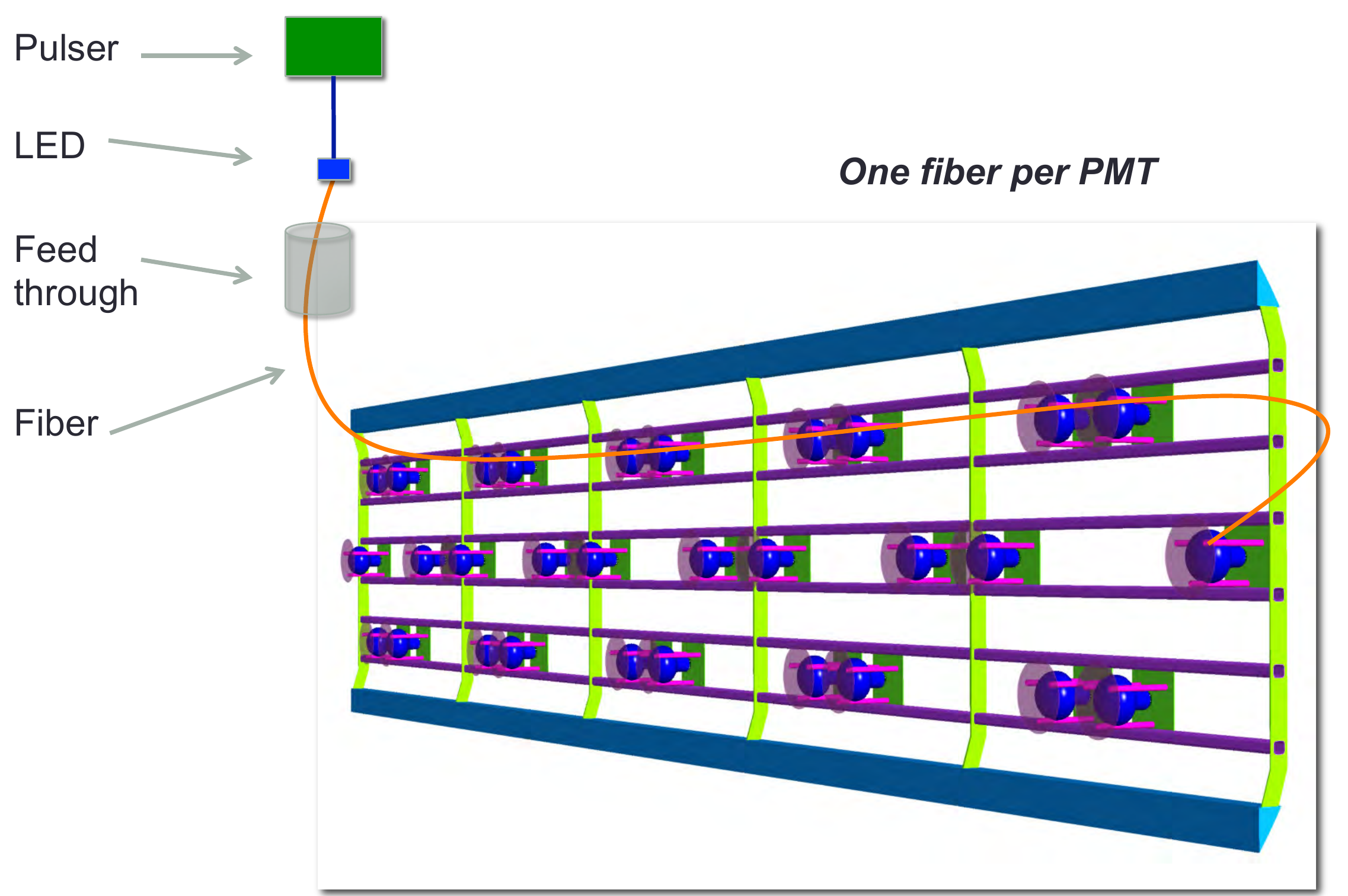}
\caption{\label{fig:uB-flasher} 
Representation of the MicroBooNE optical fiber calibration system.
}
\end{figure}

Calibration of the light collection system in MicroBooNE will be achieved by pulsing a series of quartz optical fibers coupled to a 450~nm LED situated outside the cryostat, shown in Fig.~\ref{fig:uB-flasher}. An external pulse delivered simultaneously to all PMTs and light-guides via these optical fibers will enable routine gain and timing calibrations of the entire light collection system. Trigger efficiencies and readout deadtimes will also be characterized using this system.

\begin{figure}
\centering
\includegraphics[width=4.in]{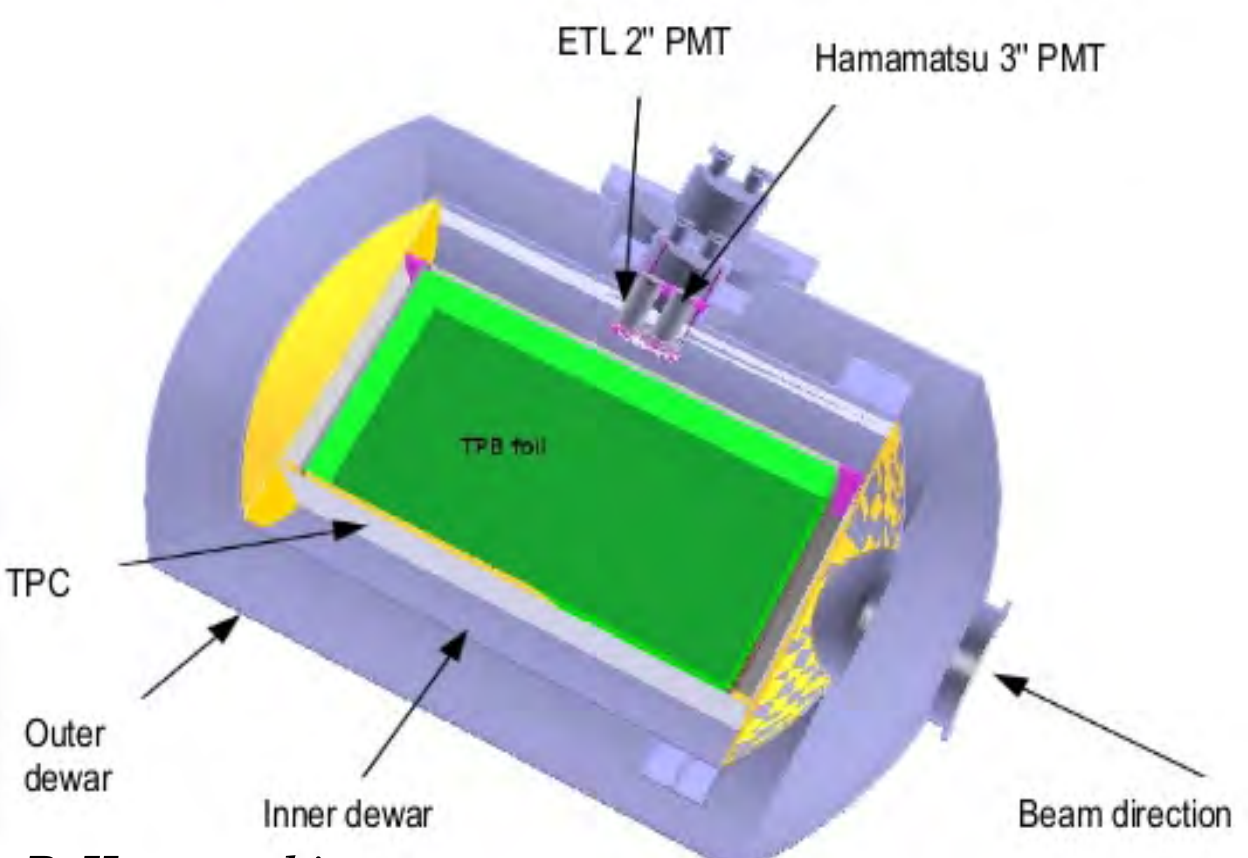}
\caption{\label{fig:lariat-lightsystem} 
3D CAD model cutaway of LArIAT system with wavelength-shifter-lined TPC walls and PMTs (and SiPMs, not shown) viewing the interior volume.  }
\end{figure}

In LArIAT, a reflector-based light collection system is being implemented.  The system is similar to that used in some dark matter experiments, but is a first for neutrino LArTPCs. To maximize the amount of light collected by the two PMTs and three SiPMs in the system, the interior walls of the TPC are covered with a reflective film coated with wavelength-shifting TPB. The PMTs and SiPMs will observe the TPC volume through the wire planes to collect both direct and reflected light, enhancing the overall collection efficiency. The setup is shown in Fig.~\ref{fig:lariat-lightsystem}.  Particle identification and calorimetry will benefit from the higher light yield, and new studies on muon sign determination using scintillation light will also be possible. 

\subsection{Test Beams}

Calibrations and detector characterizations that are made possible by operating LArTPCs in charged and neutral particle test beams are another important facet of building the global knowledge base for the wider LArTPC community. In this area, there are three main efforts: CAPTAIN, LArIAT, and LBNO-Demo (WA105). 

CAPTAIN's primary aim is to characterize LArTPC response to low energy neutrons. As a first step, the CAPTAIN collaboration is building mini-CAPTAIN, a 400~kg TPC instrumented with 1000 channels. This prototype device is currently under construction, and will run in a neutron beamline at the Los Alamos Neutron Science WNR facility in 2015. Using time-of-flight upstream and downstream scintillator paddles, a visible interaction in the LArTPC can be correlated with a neutron of specific kinetic energy. The collaboration plans to study neutron interactions on liquid argon in terms of topology, multiplicity and identity of visible particles in the final state, and kinematic properties. The focus will be on primary neutron interactions of known incident neutron energy to create a catalog and a probability distribution of possible event topologies as a function of incoming neutron energies. Ultimately, this catalog will be useful for identifying outgoing neutrons in neutrino interactions, where the neutron energy can then be obtained by PDF, resulting in better reconstruction of incoming neutrino energy.

\begin{figure}
\centering
\includegraphics[width=6.in]{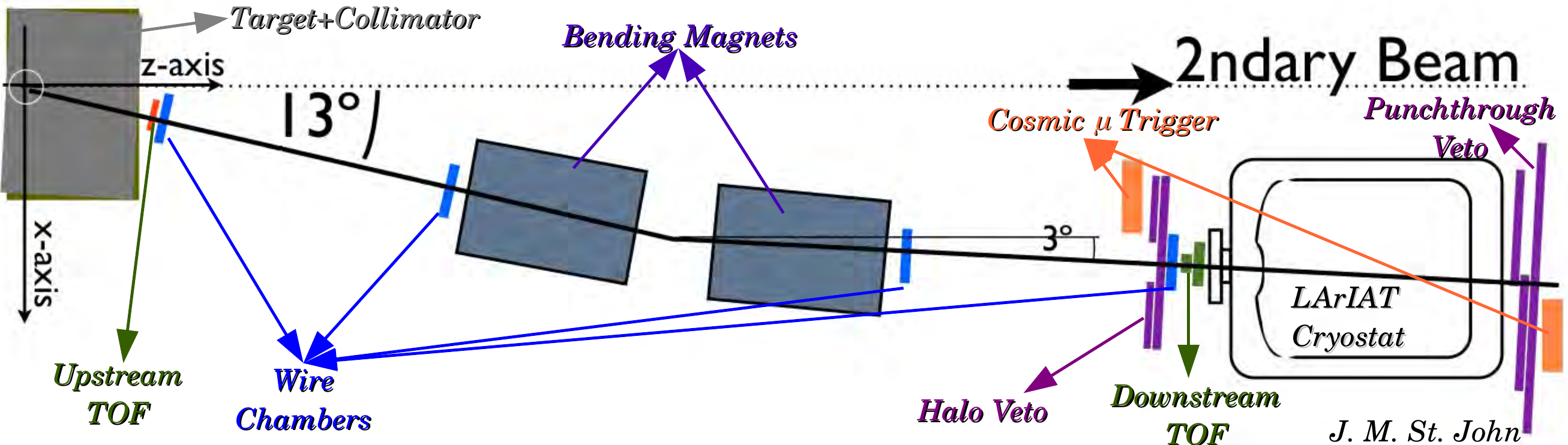}
\caption{\label{fig:lariat-tertiarybeam} 
LArIAT tertiary beam configuration. Tunable secondary pion beam impinges on tertiary target at left side of image. Exiting tertiary beam is collimated in direction of bending magnets. TOF and wire chambers tag particles entering the LArIAT cryostat, and veto counters enable tagging of particles outside the cryostat beam window (halo) or exiting particles (punchthrough).}
\end{figure}

LArIAT's focus is on charged particles in the energy range that will support MicroBooNE and LBNE, as described in detail in the previous workshop summary. New cold electronics have been produced using the BNL ASIC developed for MicroBooNE, and the TPC was assembled with new wire planes. Notably, after almost a decade of not operating, the MCenter beamline at the Fermilab Test Beam Facility has been refurbished and the secondary beam is fully commissioned. Efforts are now on tuning the LArIAT tertiary beam, seen in Fig.~\ref{fig:lariat-tertiarybeam}, while commissioning the DAQ using beamline wire chambers and TOF paddles. An engineering run using all beam detectors began in September 2014 and lasted through the Fermilab shutdown.  The detector was installed in the beam enclosure during the Winter 2015 and the experiment will take data during Spring 2015.

CERN's WA105 test beam experiment, also known as LBNO-Demo, is a dual-phase LArTPC. The drift electrons produced in the liquid phase are extracted from the liquid into the gas phase with the help of an electric field. Two grids, situated just below and above the liquid level, define the appropriate extraction field. The extracted charge is then amplified in the gas phase with a Large Electron Multiplier (LEM). The concepts of dual-phase liquid argon LEM-TPC detectors have been demonstrated previously in smaller prototypes~\cite{ArDM,Badertscher:2012dq,Badertscher:2010zg,Badertscher:2013wm}, and a full-scale dark matter experiment using this technology, called ArDM, is currently running in the underground laboratory in Canfranc, Spain~\cite{ArDM}. 

\begin{figure}
\centering
\includegraphics[width=6.in]{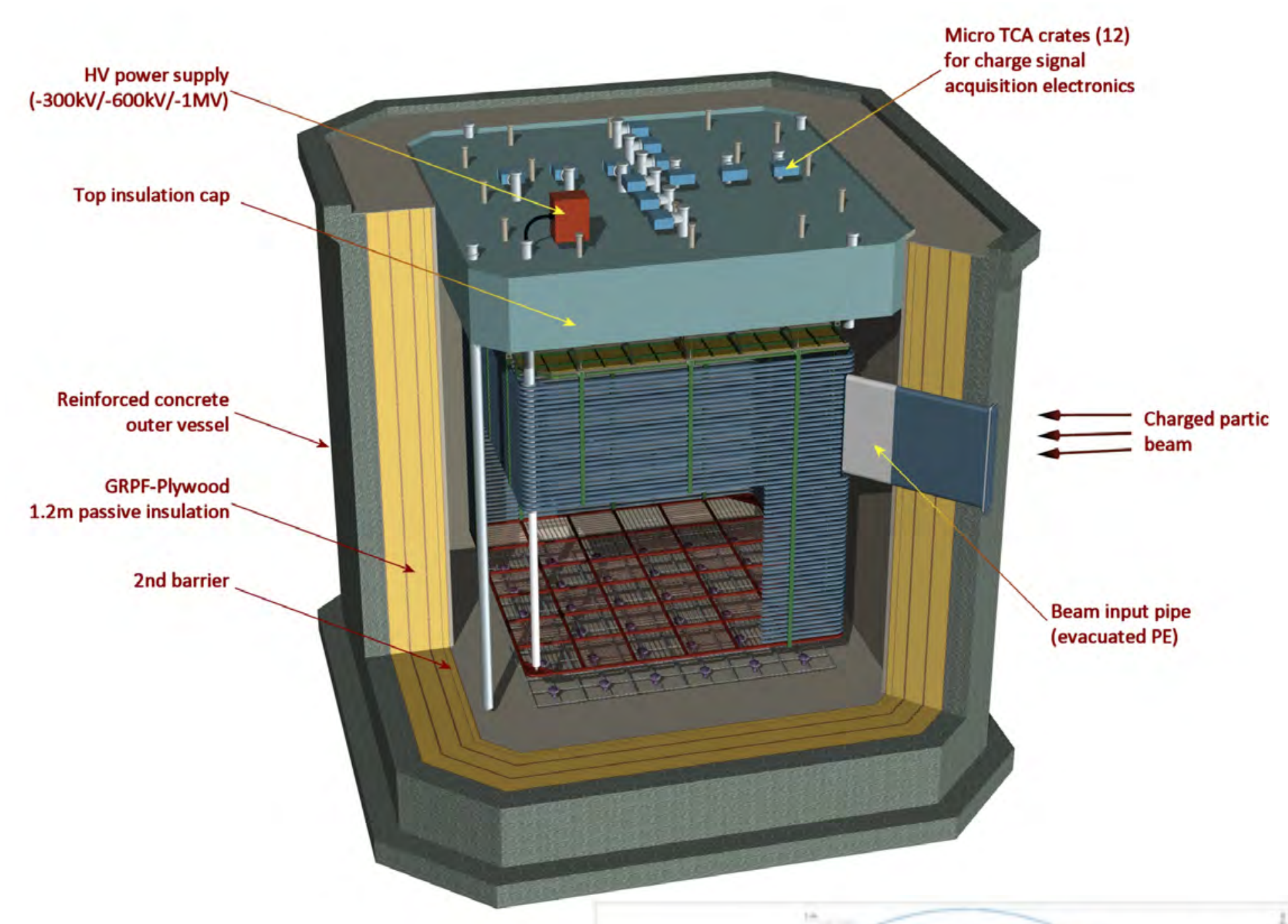}
\caption{\label{fig:WA105-detector} 
LBNO-Demo (WA105) LAr LEM-TPC detector in membrane cryostat.
}
\end{figure}

The goal of LBNO-Demo is to demonstrate the scalability of these detectors. The design consists of a $6\times6\times6~{\mathrm{m}}^3$ ($\sim 300$-ton) dual-phase LAr LEM-TPC with 2D-collection readout anode. Its size will allow full containment of showers up to $10$~GeV. The detector will be placed inside a non-evacuated membrane cryostat, and the experiment will be installed in a $1-20~{\mathrm{GeV/c}}$ charged hadron beam at CERN. Figure~\ref{fig:WA105-detector} shows a cutaway of the cryostat and LAr LEM-TPC. Test beam running will enable development of automatic event reconstruction, testing of neutral current background rejection algorithms on $\nu_{e}$-free events, measurement of charged pion and proton cross sections on argon, and characterization of calorimetric performance. As an intermediate step, a $3\times1\times1~{\mathrm{m}}^3$ prototype is currently under construction, with a timescale of 2014-2015.

The test beam activity across the world has evolved over the past year into a multi-pronged program with complementary branches to address pertinent questions regarding LArTPC performance and calibration. CAPTAIN will characterize LArTPC performance in a beam of low- and medium-energy neutrons ($<1$~GeV) at the Los Alamos Neutron Science Center. Additional possibilities for later running include studying low energy neutrino interactions ($<50$~MeV) from a stopped pion source, and higher energy neutrino interactions ($3-9$~GeV) in the NuMI neutrino beam. LArIAT will operate in a charged particle beam at Fermilab, collecting data in the $\sim200$~MeV to $2$~GeV range relevant to MicroBooNE and LBNE. Finally, LBNO-Demo will extend these studies to higher energies in a charged particle beam at CERN, collecting data in the $1-20$~GeV range. The collection of these complementary datasets will provide invaluable input to future experiments studying long- and short-baseline neutrino oscillations, atmospheric neutrinos, and burst supernova neutrinos.

\section{Software}
\label{sec:Software}
\subsection{LArSoft:  Simulation and Reconstruction Software}

LArSoft~\cite{larsoft} is a software suite for simulating, reconstructing, and analyzing data from LArTPCs.  Its purpose is to collect tools and interfaces to external software packages that perform important functions that are needed by all experimental groups. Currently ArgoNeuT, LArIAT, MicroBooNE, LBNE, and LAr1-ND are using LArSoft. LArSoft allows the sharing and re-use of liquid argon TPC code, and also reduces the setup and learning required to use the external software in a way consistent with the requirements of the experimental collaborations. An important example of this latter use is the interface to  Geant4~\cite{geant4}, which in general requires a significant investment in the time of physicists to define the detector geometry, to set up the appropriate physics processes to simulate, and to interface it to particle generators and to produce useful output.  LArSoft has a uniform interface to geometry specification via reading of a GDML file, and it manages the instantiation of the geometry in Geant4.  LArSoft contains interfaces to the CRY~\cite{cry} cosmic-ray generator, the GENIE~\cite{genie} neutrino-nucleus generator, and also manages the detector response to each Geant4 step.  It also provides flexibility in terms of which event generators can be used through the support of a text based event generator that uses the StdHEP syntax for describing particles.

LArSoft is built on the {\it art} framework~\cite{art}, which is developed and supported by Fermilab's Scientific Computing Division (SCD).  The {\it art} framework drives the event loop, and steers the execution of the various processing steps via configuration files.  Event data are immutable and are stored in ROOT format.  The framework provides I/O utilities for reading and storing data products and tracking their history.  There is also a feature-rich event display facility. LArSoft is also supported by Fermilab's Scientific Computing Division with regular input from experimental stakeholders.


The LArSoft management team at Fermilab has been refactoring the code base, rationalizing the configuration files, and optimizing the resource usage.  Some data structures and algorithmic choices that perform adequately for a small liquid argon TPC perform poorly when scaled up to a very large detector with hundreds of thousands of readout channels.  Optimization of the memory needed to represent the geometry of the detector, as well as that required for the data and the CPU needed to process it has been a focus of the LArSoft management team.  Not only must the software be optimized for execution on available computing resources, but it must also meet the requirements of the experiments and be easy to understand and use. 

{\bf Lessons Learned:} 
\begin{itemize}
\item Transitioning to using new tools takes time and effort, and requires effort both on the part of the code management and the experimental stakeholders.  
\item Interactions between stakeholders must also be carefully managed, with appropriate tools in order to minimize the impact of disruptive changes from one area of developing software to another while maximizing the transfer of desired changes. 
\item Liquid argon software is still in a formative stage, with large improvements in accuracy, speed, and usefulness possible; new developers can make significant contributions.  
\item A mixture of personnel is needed on a large software project -- managers, librarians, stakeholders, and developers, and these categories must be staffed by students, postdocs, laboratory staff, and faculty.
\end{itemize}

\subsection{Simulation of Liquid Argon TPC's}

LArSoft provides a general interface to specifying a LArTPC detector geometry,  though use-case specific needs, such as multiplexing readout DAQ channels, or  experiment-specific photon readout, must still be met with experiment-specific code.

Because ionizing particles in liquid argon generate tens of thousands of drifting electrons and scintillation photons per MeV of energy deposited, LArSoft is configured to handle the charge drifting and photon propagation parametrically. All detector geometries are broken into volume elements, ``voxels'', that force the Geant4 steps on the boundaries, in order to enforce a fine-grained simulation. The energy deposited in a voxel is then used to compute a number of electrons and photons produced in that voxel, using either  Birks' Law~\cite{birks} or the Modified Box Model~\cite{modifiedbox}, or  {\tt NEST}~\cite{nest}.  The {\tt NEST} model has been tuned to available noble-liquid detector data, and simulates the statsitical anticorrelation between the electron and photon production.  The drift velocity is parameterized in terms of the argon temperature and the electric field, which is a constant in LArSoft for the time being but models including space charge buildup are being developed.  Timing offsets between readout planes are configurable.

The attenuation due to electron lifetime is adjustable by a configuration parameter, as is the diffusion.  The spreading of a cloud of drifting electrons impinging on  the wires of the detector is simulated numerically by sampling from a Gaussian distribution.  The response of the wires to drifting charge is simulated from a pre-parameterized function of time describing the impulse response of induction or collection-plane wires.  Parameterizable noise is then added, and the data product is ADC digits as a function of clock tick.  LBNE also applies a zero-suppression algorithm which is lossy in that it suppresses ADC readings below a threshold, while MicroBooNE applies the lossless Huffman encoding.

Photons are produced in three possible modes: fast scintillation with a time constant of $\approx 6$~ns, slow scintillation with a time constant of 1.6~$\mu$s, and  Cherenkov emission.  Photons undergo Rayleigh scattering on their path through the liquid argon, and they may be absorbed or reflected on uninstrumented surfaces. Some wavelength shifter is typically necessary to detect these photons.  LArSoft simulates nearly all of these processes but lacks a simulation of the wavelength shifter, and also lacks  purity and electric-field-dependent quenching. Photons may be propagated using Geant4, which provides our best model of the physical processes, but is very slow computationally given the tens of thousands of quanta per MeV deposited.  Instead, a library is built which lists the probabilities that a photon emitted at a point in space in the detector will be detected by a specified photon detector, for example a PMT in the case of MicroBooNE, or a SiPM in the case of LBNE.


%

Simulations require significant amounts of careful work to be useful.  Currently, as data from operating experiments are lacking to tune the simulations, authors must rely on test stand, prototype, and older experimental data as a starting point.  LArSoft makes it easy to share and incorporate simulation ideas, though the detector-specific details and challenges will require a certain division of the code between shared and specific elements.  Sometimes it is good to have separate versions of routines that do the same thing for different detectors -- e.g. the digitizaiton routine for the LBNE 35 ton prototype need not be identical to that of the far detector.  MicroBooNE needs its own version as well which is protected from changes by other experimental collaborations' software efforts to match simulations to data.

\subsection{Signal Processing and Reconstruction}

Ionization liberated by charged particles traveling through the TPC will drift along electric-field lines towards the finely segmented anode planes.  Proper biasing of the voltages on the anode planes allows the ionization to drift through the first Induction layers before ultimately stopping on one of the wires in the collection layer.  The electrical current that is created in the wires of the TPC due to this drifting charge is the base signal that is used to study the interactions that occurred inside of the TPC.  Simulating and analyzing the current in the TPC wires requires models of recombination, diffusion, the electric field in the anode region, and knowledge of the response of the electronics connected to the wires.  Each of these facets of signal development, with varying degrees of sophistication, are implemented in the LArSoft simulation package and can be tailored to the specifications of a given LArTPC detector.  Treatment of signal development will continue to be refined as new insights are gleaned from the current generation of LArTPC experiments.

\begin{figure}[htbp] 
   \centering
   \includegraphics[width=3.5in]{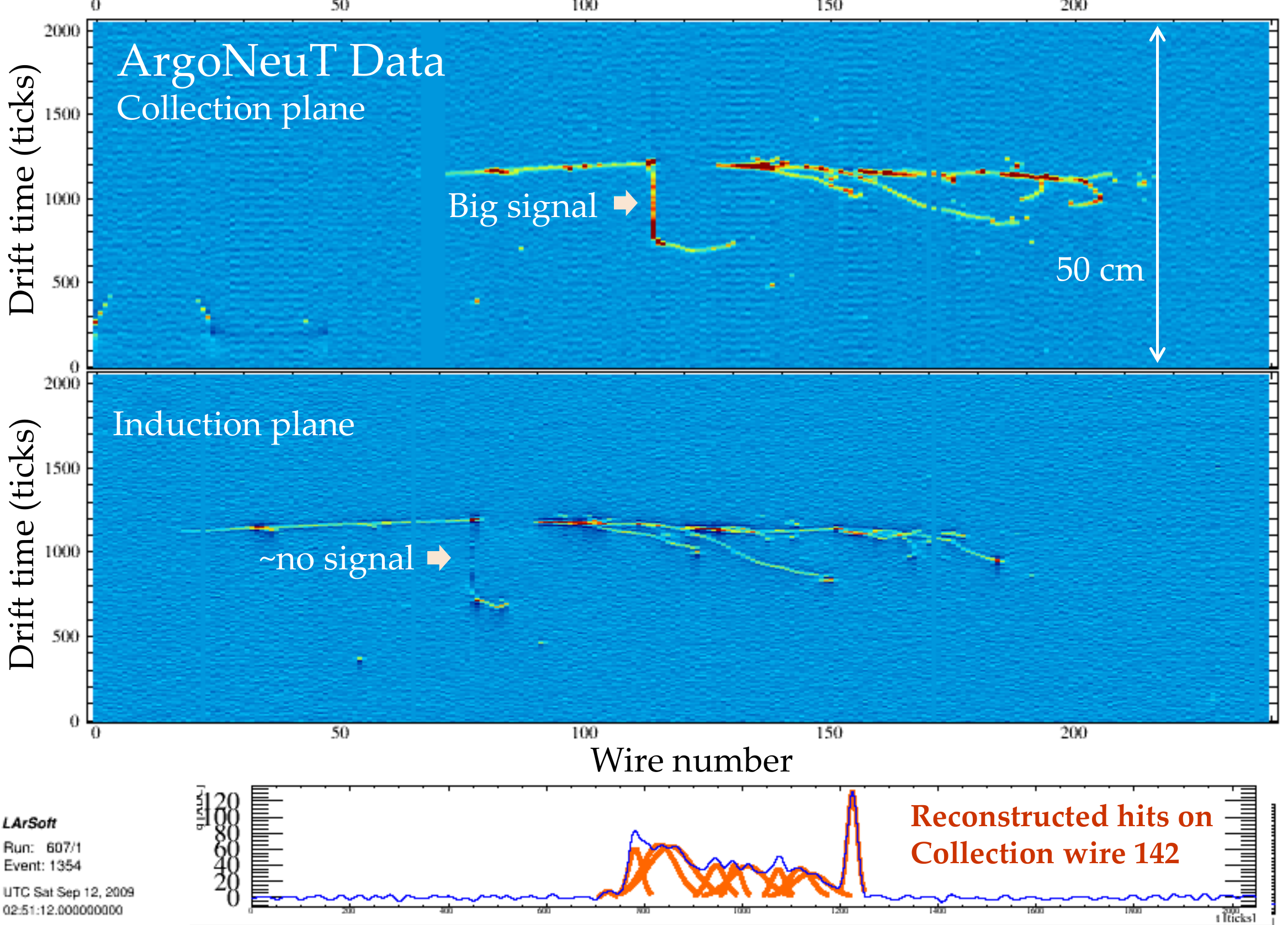} 
   \caption{ArgoNeuT data event demonstrating complicated pulse shape when a track is approximately traveling towards a single wire.}
   \label{fig:event}
\end{figure}

Signal processing algorithms can be used to manipulate the data collected by the TPC, for example allowing removal of potentially noisy regions of frequency space, or converting the induction plane signals from their inherent bipolar shape to a unipolar shape similar to those on the collection plane.  This shape conversion enables common hit-finding and reconstruction code to run on all LArTPC planes.  The performance of reconstruction algorithms will depend on how the signal processing scheme is optimized for these parameters.  Signal processing algorithms should be robust enough to handle even pathological signals, such as the example shown in Fig.~\ref{fig:event}.  

Reconstruction algorithms in LArSoft are utilized to take the signals collected by the TPC and light-collection systems and use them to infer the properties of any particles present inside the detector during operations.  The reconstructed particles can be further analyzed to determine the flavor and kinematics of any neutrinos that interacted in the detector.

\begin{figure}[htbp] 
   \centering
   \includegraphics[width=4in]{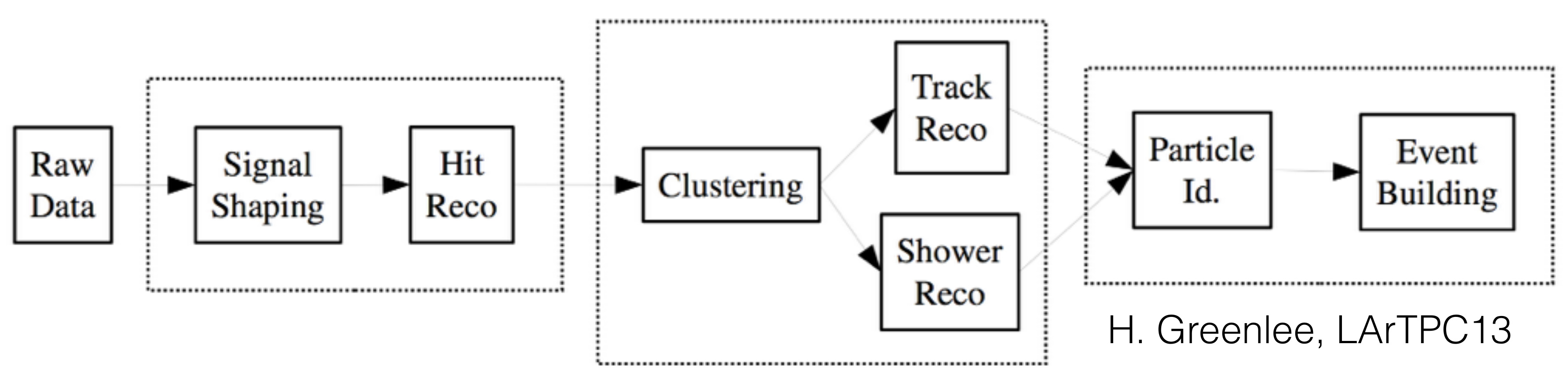} 
   \caption{Example flowchart of reconstruction modules in LArSoft.}
   \label{fig:reco}
\end{figure}

The prevailing philosophy of reconstruction in LArSoft is to have many different algorithms that each perform one specific task, also allowing for the option to have different algorithms that each attempt to do the same task in a different manner.  Objects produced by one algorithm can be saved to the event record and used by downstream algorithms.  The sequence of reconstruction modules that run during a job are configurable by the user, as are parameter settings within those modules.  Figure \ref{fig:reco} is a simplified flow-chart of a typical LArSoft reconstruction job.  

The typical reconstruction job follows a ``bottoms-up" approach that begins with isolating hits within individual wires, followed by combining proximal hits into clusters, and then combining clusters from different views into three-dimensional track objects.  Many variations on this approach have been investigated, and the optimum approach will very likely depend on the analysis being pursued.  Tracks formed using TPC data can be further analyzed for their calorimetric properties, which is crucial information that is used to identify the track's particle type and also to reconstruct the energy and flavor of neutrinos that interacted in the detector. Track information can also be combined with information from light-collection systems to help remove out-of-beam activity.  

In the past year many new reconstruction algorithms have been introduced into LArSoft, and emphasis has been placed on developing metrics to gauge the performance of individual reconstruction methods.  Though it is clear that LArTPC detectors offer exceptional capabilities for studying neutrino physics, it is also clear that developing reconstruction software to take full advantage of the LArTPC technology is a major challenge.  The challenge is intrinsic to the projections of images provided by the detector, and converting this information into a fully formed three-dimensional picture.  As experiments get larger in size, and the electronics channel count grows from thousands to hundreds of thousands, software needs to be able to cope with the increased amounts of information to handle.  Many algorithms in LArSoft were found to scale poorly with channel count, and with increased drift times, and so have been reformulated to work more efficiently.  Code optimization needs to be constantly monitored as new software is developed.


\section{World Wide R\&D Efforts}
\label{sec:World}

As indicated in the preceding sections, the development of LArTPC technology is being pursued actively throughout the world. The development is being carried on both at the level of individual institutions as well as through major research laboratories.  In this section we highlight some of the efforts that have not already been mentioned in this document.

\subsection{Development in Continental Europe}

CERN is focusing on developing a Neutrino Platform to support development of neutrino detectors.  Several aspects of this platform focus on LArTPC development.  One aspect is the refurbishment of the ICARUS T600, now known as WA104, in order to expose it to the Booster neutrino beam at Fermilab.  Another aspect is the development of a dual phase LArTPC to be exposed in a test beam, WA105, which was described in \S~\ref{sec:TestBeams}. The WA105 effort may also include the exposure of a prototype of the LBNE detector to the test beam.  As part of these efforts, CERN has been partnering with Fermilab to develop common solutions for the cryogenic systems needed to support LArTPC experiments.  The goal of developing these common solutions is to create standardized components of the cryogenic systems that can be replicated for a variety of experiments, including LAr1-ND, WA104, and WA105.  CERN has committed significant resources to these projects in order to provide the necessary infrastructure for them by 2016, including the construction of a large neutrino test area capable of delivering charged particle beams to these detectors.

The University of Bern in Switzerland had proposed a novel and modular approach for the design of large mass LArTPCs called ArgonCube, which also includes a significant set of R\&D studies to create a next generation of detectors. This detector design is the result of the evolution of the research program conducted over several years starting with the ArgonTube project.  The program also benefits from a series of notable contributions from other institutes, such as Fermilab and Brookhaven.  The ArgonCube program aims at the construction and operation of a series of detectors of increasing dimensions and complexity. The ultimate goal of the program is a feasibility study of modular LArTPC detectors with masses up to several tens of kilotons. The central idea is to construct and position identical and separate modules in a common bath of liquid argon. Each module features a relatively short drift length and a fully independent TPC with its own readout, light detection system, cryogenics, and services. Module walls are made thin to provide transparency to electromagnetic and hadronic showers as well as to neutrino produced primary particles. This detector configuration should also allow the optimal use of the liquid argon with a relatively large fraction of active volume, as compared to other implementations of the technology. The short drift length will permit less stringent requirement on the liquid argon purity, hence allowing the concentration of development efforts on other topics, such as the development of novel pixel readout schemes instead of the traditional wire plane readout.  In a pixel readout, a single readout electrode would be segmented into pixels of a size comparable to the wire spacing in the standard approach. The number of pixels for equal spatial resolution will be two or three orders of magnitude higher than the number of sense wires, with a corresponding increase of the number of signal channels, data rates and power dissipation. This scaling would make such a solution untenable except for small detectors and a readout scheme that reduces these issues to a reasonably low level must be found. Multi-prong development efforts will be an iterative process with multiple phases of prototypes including cosmic rays and charged particle test beams, as well as data analysis to reach the optimum configurations. This group is implementing these plans in a three-phased approach that builds on their previous work and the lessons learned by the construction of MicroBooNE and ArgoNeuT. Ultimately the goal is to produce a prototype that will be exposed to a charged particle beam at CERN and that can demonstrate the success of such a modular approach.

\subsection{Development in the United Kingdom}

Activity in development of liquid argon technology for neutrinos has ramped up significantly in recent years in the UK, with at least 10 institutions now active. All these institutions are now members of the LBNE and LAr1-ND projects at FNAL and some are participating in MicroBooNE. The UK funding agency, STFC, recently awarded construction resources for participation in LAr1-ND.  The agency also supports engineering design and DAQ resource support for LBNE, including for participation in the 35 ton prototype at FNAL.   Broadly a 'northern hub' of institutes comprising Lancaster, Liverpool, Manchester and Sheffield plus UCL in London, have particular interest in hardware development.  The Sheffield group pioneered early UK work on liquid argon technology in the area of single phase readout concepts using light readout of thick GEMs.  Liverpool has recently made major progress here, working with CCD readout and using both single and dual phase systems.  They demonstrated CCD images of alpha tracks using secondary scintillation in liquid argon.   Manchester has developed expertise in TPC wire technology through involvement in the SuperNEMO double beta decay experiment.  This expertise is now being turned towards participation in the LAr1-ND project; they are developing new wire winding robotics for LAr1-ND.  For this work the three groups are together funded to build part or all of the anode plane and cathode plane assemblies (APA, CPA).  The Lancaster group have developed a cryogenic testing facility for these assemblies and UCL has taken its experience in the dark matter field to develop HV feed throughs for LAr1-ND.  Significant engineering R\&D has also developed around all of this work, for instance at Sheffield new designs of APA frames have been established, built and tested.  This work is feeding  into new designs for the much larger LBNE far detector APAs and the CERN platform program.  The earlier work on CCD operation in liquid argon has also allowed development of a multi-camera array that is to be deployed inside the LBNE 35 ton prototype at Fermilab for monitoring of potential HV corona.   An important aspect of the UK efforts outside detector hardware has been development of DAQ systems by Oxford and RAL, with input also from Sussex on event building.  Oxford are taking a lead on DAQ for the LBNE 35 ton prototype.  The remaining UK groups, particularly Cambridge and Warwick, are focussing mainly on event reconstruction software.

\section{Summary}
\label{sec:Summary}

The workshop successfully brought together experts from around the world to discuss the latest developments toward the goal of producing multi-kiloton scale LArTPCs.  The final session of the workshop was a discussion of $i)$ the development of stable high voltage delivery systems for longer drift distances than currently achieved, $ii)$ new directions in scintillation photon collection, and $iii)$ the development of magnetized LArTPCs.

The high voltage delivery system development continues to challenge many experiments.  During the workshop, it was agreed that a more complete picture of the physics leading up to discharges of high voltage systems needs to be much better understood.  The geometry required for creating a discharge, as well as the precursors to a discharge need to be documented so that future experiments can avoid designs that are susceptible to discharge.  These studies are especially important when pushing the limits of drift distance as the amount of high voltage delivered is directly proportional to the drift distance.  One may also ask whether it is necessary to push drift distances beyond the current scale, a question that assumes a different approach to instrumenting the liquid argon volume than having a single, monolithic volume.  Even at smaller scales though, an experiment needs to have extremely high confidence that the high voltage delivery system will not discharge.

The discussion of scintillation photon collection focussed on the potential gains of adding dopants to the liquid argon to enhance light production at longer wavelengths and the possibility of improving timing resolution of the light detectors.  Adding a dopant such as xenon could improve the production of photons in wavelength regions better suited to the response of some wavelength shifting materials.  Such dopants could also shorten the timescale associated with the late light considerably.  Both effects are described in \S~\ref{contaminants}.  Improving the timing resolution of the photon detectors may give one access to the prompt Cherenkov light produced by charged particles traversing the argon.  That information could be incorporated into reconstruction algorithms to further enhance signal to background rejection for many analyses ranging from accelerator based neutrino oscillation studies to searches for more exotic phenomena such as light dark matter particles.

The desire of many researchers is to magnetize LArTPCs to provide both charge sign discrimination of charged particles as well as kinematic information about those particles.  These features would certainly enhance searches for CP violation in the neutrino sector, as well as improve the precision of the other neutrino oscillation parameters.  Two potential methods for magnetizing these detectors are the use of a solenoid magnetic design or a super-conducting double-racetrack configuration with super-conducting wire looped at the top and bottom of the cryostat.  Both options require significant development of the systems and then studies of their performance. Coupling a magnetic field with the position resolution of LArTPCs would truly make LArTPCs the digital analog to the bubble chamber.

\section{Acknowledgments}
The speakers at this workshop represented many colleagues from both collaborations and local university groups.  The work they presented was supported by a variety of funding agencies.  The following statements of acknowledgement were supplied by the speakers and have not been edited.

We thank the ICARUS collaboration for sharing their experiences and lessons learned in building and operating the first large LArTPC.  The High Energy Astrophysics Group at Indiana University is supported by the U.S. Department of Energy Office of Science with grant DE-FG02-91ER40661 to Indiana University and LBNE project funding from Brookhaven National Laboratory with grant BNL 240296 to Indiana University.  The LArIAT collaboration is supported by the U.S. Department of Energy Office of Science and the National Science Foundation.  The CAPTAIN detector has been designed and is being built by the Physics and the Theory divisions of Los Alamos National Laboratory under the auspices of the LDRD program.  T. Strauss spoke on behalf of the Albert Einstein Center, Laboratory of High Energy Physics of the University of Bern.  The MicroBooNE and LBNE collaborations have participated in the development of cold electronics as supported by the U.S. Department of Energy Office of Science.

\bibliographystyle{JHEP}
\bibliography{sumbib}

\providecommand{\href}[2]{#2}\begingroup\raggedright\begin{thebibliography}{10}

\bibitem{workshop}
{\tt https://indico.fnal.gov/conferenceDisplay.py?ovw=True\&confId=8381}.

\bibitem{Anderson:2011ce}
{\bf ArgoNeuT} Collaboration, C.~Anderson et~al., {\it {First Measurements of
  Inclusive Muon Neutrino Charged Current Differential Cross Sections on
  Argon}},  {\em Phys.Rev.Lett.} {\bf 108} (2012) 161802.

\bibitem{ref:LAPD}
M.~Adamowski et~al., {\it {The Liquid Argon Purity Demonstrator}},  {\em JINST}
  {\bf 9} (2014) P07005.

\bibitem{lbnedune}
At the time of this workshop, the long baseline neutrino experiment planned in
  the US was called LBNE. It has since been renamed to DUNE, however we kept
  the name in this document consistent with what was used at the time of the
  workshop.

\bibitem{MicroBooNETDR}
MicroBooNE Technical Design Report {\tt
  http://www-microboone.fnal.gov/publications/TDRCD3.pdf}.

\bibitem{lar1ndsbnd}
As of the time of submission for this document, the short baseline neutrino
  experiment planned in the US called LAr1-ND changed its name to SBND. We kept
  the name in this document consistent with what was used at the time of the
  workshop.

\bibitem{ref:purityMonitor}
G.~Carugno et~al., {\it {Electron lifetime detector for liquid argon}},  {\em
  Nucl. Instrum. Meth. A} {\bf 292} (1990) 580.

\bibitem{ref:HVC}
{\bf MicroBooNE} Collaboration, R.~Acciarri et~al., {\it {Liquid Argon
  Dielectric Breakdown Studies with the MicroBooNE Purification System}},  {\em
  JINST} {\bf 9} (2014), no.~11 P11001.

\bibitem{ref:sigmaaldrich}
Sigma-Aldrich, P.O. Box 14508, St. Louis, MO 63178 USA.

\bibitem{ref:basf}
BASF Corp., 100 Park Avenue, Florham Park, NJ 07932 USA.

\bibitem{ref:spectris}
Spectris plc, Heritage House, Church Road, Egham, TW20 9QD, United Kingdom.

\bibitem{ref:ldetek}
LDetek Inc., 271 Saint-Alphonse Sud, Thetford Mines, Quebec, Canada G6G 3V7.

\bibitem{ref:tiger}
Tiger Optics LLC, 250 Titus Ave, Warrington, PA, 18976-2426.

\bibitem{artrip}
Information on catalysts provided by D. Artrip of Research Catalysts, {\tt
  http://www.catalyst-central.com}.

\bibitem{Amerio:2004ze}
{\bf ICARUS} Collaboration, S.~Amerio et~al., {\it {Design, construction and
  tests of the ICARUS T600 detector}},  {\em Nucl.\ Instrum.\ Meth.} {\bf A527}
  (2004) 329.

\bibitem{Anderson:2012vc}
{\bf ArgoNeuT} Collaboration, C.~Anderson et~al., {\it {The ArgoNeuT Detector
  in the NuMI Low-Energy beam line at Fermilab}},  {\em JINST} {\bf 7} (2012)
  P10019.

\bibitem{Swan1960180}
D.~W. Swan and T.~J. Lewis, {\it Influence of electrode surface conditions on
  the electrical strength of liquified gases},  {\em J. Electrochem. Soc.} {\bf
  107} (1960) 180.

\bibitem{Swan1961448}
D.~W. Swan and T.~J. Lewis, {\it The influence of cathode and anode surfaces on
  the electric strength of liquid argon},  {\em Proc. Phys. Soc.} {\bf 78}
  (1961), no.~3 448.

\bibitem{Bay:2014jwa}
F.~Bay et~al., {\it {Evidence of electric breakdown induced by bubbles in
  liquid argon}},  \href{http://xxx.lanl.gov/abs/1401.2777}{{\tt
  arXiv:1401.2777}}.

\bibitem{Blatter:2014wua}
A.~Blatter et~al., {\it {Experimental study of electric breakdowns in liquid
  argon at centimeter scale}},  {\em JINST} {\bf 9} (2014) P04006.

\bibitem{Auger:2014eba}
M.~Auger et~al., {\it {A method to suppress dielectric breakdowns in liquid
  argon ionization detectors for cathode to ground distances of several
  millimeters}},  {\em JINST} {\bf 9} (2014) P07023.

\bibitem{Gerhold1994579}
J.~Gerhold, M.~Hubmann, and E.~Telser, {\it Gap size effect on liquid helium
  breakdown},  {\em Cryogenics} {\bf 34} (1994), no.~7 579 -- 586.

\bibitem{Asaadi:2014iva}
J.~Asaadi et~al., {\it {Testing of High Voltage Surge Protection Devices for
  Use in Liquid Argon TPC Detectors}},  {\em JINST} {\bf 9} (2014) P09002.

\bibitem{Bagby:2014wva}
L.~Bagby et~al., {\it {Breakdown voltage of metal-oxide resistors in liquid
  argon}},  {\em JINST} {\bf 9} (2014), no.~11 T11004.

\bibitem{epcos}
{\tt
  http://www.epcos.com/blob/174146/download/\\5/ueberspannungsableiter-und-schaltfunkenstrecken.pdf}.

\bibitem{GDTBourns}
{\tt http://www.bourns.com/pdfs/bourns\_gdt\_whitepaper.pdf }.

\bibitem{var_material_science}
T.~K. Gupta, {\it Application of zinc oxide varistors},  {\em Journal of the
  American Ceramic Society} {\bf 73} (1990), no.~7 1817--1840.

\bibitem{ep_datasheet}
{\tt http://www.epcos.com/inf/100/ds/a71h45xx2590.pdf }.

\bibitem{pano}
{\tt http://industrial.panasonic.com/www-data/pdf/AWA0000/AWA0000CE2.pdf }.

\bibitem{LArRDworkshop2013}
B.~Baller et~al., {\it {Liquid Argon Time Projection Chamber Research and
  Development in the United States}},  {\em JINST} {\bf 9} (2014) T05005.

\bibitem{Adams:2013qkq}
{\bf LBNE} Collaboration, C.~Adams et~al., {\it {The Long-Baseline Neutrino
  Experiment: Exploring Fundamental Symmetries of the Universe}},
  \href{http://xxx.lanl.gov/abs/1307.7335}{{\tt arXiv:1307.7335}}.

\bibitem{sli}
S.~Li et~al., {\it {LAr TPC Electronics CMOS Lifetime at 300K and 77K and
  Reliability under Thermal Cycling}},  {\em IEEE Trans. on Nucl. Sci.} {\bf
  60} (2013) 4737.

\bibitem{lariat}
{\bf LArIAT} Collaboration, F.~Cavanna et~al., {\it {LArIAT: Liquid Argon In A
  Testbeam}},  \href{http://xxx.lanl.gov/abs/1406.5560}{{\tt arXiv:1406.5560}}.

\bibitem{Ereditato:2013xaa}
A.~Ereditato et~al., {\it {Design and operation of ARGONTUBE: a 5 m long drift
  liquid argon TPC}},  {\em JINST} {\bf 8} (2013) P07002.

\bibitem{CAPTAIN}
{\bf CAPTAIN} Collaboration, H.~Berns et~al., {\it {The CAPTAIN Detector and
  Physics Program}},  \href{http://xxx.lanl.gov/abs/1309.1740}{{\tt
  arXiv:1309.1740}}.

\bibitem{jrhoff}
J.~R. Hoff et~al., {\it {Lifetime Studies of 130 nm nMOS Transistors Intended
  for Long-Duration, Cryogenic High-Energy Physics Experiments}},  {\em IEEE
  Trans. on Nucl. Sci.} {\bf 59} (2012) 1757.

\bibitem{gwu}
G.~Wu et~al., {\it {Degradations of Threshold Voltage, Mobility, and Drain
  Current and the Dependence on Transistor Geometry For Stressing at 77 K and
  300 K}},  {\em IEEE Trans. Devices and Materials Reliability} {\bf 14} (2014)
  477.

\bibitem{novatime}
A.~Norman et~al., {\it {The NOvA Timing System: A system for synchronizing a
  long baseline neutrino experimen}t},  {\em Journal of Physics Conf. Ser.}
  {\bf 396} (2012) 012034.

\bibitem{antonello}
{\bf ICARUS} Collaboration, M.~Antonello et~al., {\it {The trigger system of
  the ICARUS experiment for the CNGS beam}},  {\em JINST} {\bf 9} (2014)
  P08003.

\bibitem{warp}
{\bf WArP} Collaboration, R.~Brunetti et~al., {\it {WArP liquid argon detector
  for dark matter survey}},  {\em New Astron.Rev.} {\bf 49} (2005) 265--269.

\bibitem{ArDM}
A.~Badertscher et~al., {\it {ArDM: first results from underground
  commissioning}},  {\em JINST} {\bf 8} (2013) C09005.

\bibitem{darkside}
{\bf DarkSide} Collaboration, T.~Alexander et~al., {\it {DarkSide search for
  dark matter}},  {\em JINST} {\bf 8} (2013) C11021.

\bibitem{lappd}
{\bf LAPPD} Collaboration, J.-F. Genat, {\it {Development of Large Area,
  Pico-second resolution Photo-Detectors and associated readout electronics}},
  .

\bibitem{acciarri}
{\bf WArP} Collaboration, R.~Acciarri et~al., {\it {Effects of Nitrogen
  contamination in liquid Argon}},  {\em JINST} {\bf 5} (2010) P06003.

\bibitem{jones1}
B.~Jones et~al., {\it {A Measurement of the Absorption of Liquid Argon
  Scintillation Light by Dissolved Nitrogen at the Part-Per-Million Level}},
  {\em JINST} {\bf 8} (2013) P07011.

\bibitem{jones2}
B.~Jones et~al., {\it {Erratum: A Measurement of the Absorption of Liquid Argon
  Scintillation Light by Dissolved Nitrogen at the Part-Per-Million Level}},
  {\em JINST} {\bf 8} (2013) P09001.

\bibitem{methane}
B.~Jones et~al., {\it {The Effects of Dissolved Methane upon Liquid Argon
  Scintillation Light}},  {\em JINST} {\bf 8} (2013) P12015.

\bibitem{xenon}
C.~G. Wahl et~al., {\it Pulse-shape discrimination and energy resolution of a
  liquid-argon scintillator with xenon doping},  {\em JINST} {\bf 9} (2014)
  P06013.

\bibitem{uBPMTtests}
T.~Briese et~al., {\it {Testing of Cryogenic Photomultiplier Tubes for the
  MicroBooNE Experiment}},  {\em JINST} {\bf 8} (2013) T07005.

\bibitem{Badertscher:2012dq}
A.~Badertscher et~al., {\it {First operation and drift field performance of a
  large area double phase LAr Electron Multiplier Time Projection Chamber with
  an immersed Greinacher high-voltage multiplier}},  {\em JINST} {\bf 7} (2012)
  P08026.

\bibitem{Badertscher:2010zg}
A.~Badertscher et~al., {\it {First operation of a double phase LAr Large
  Electron Multiplier Time Projection Chamber with a 2D projective readout
  anode}},  {\em Nucl.\ Instrum.\ Meth.} {\bf A641} (2011) 48.

\bibitem{Badertscher:2013wm}
A.~Badertscher et~al., {\it {First operation and performance of a 200 lt double
  phase LAr LEM-TPC with a 40x76 cm$^2$ readout}},  {\em JINST} {\bf 8} (2013)
  P04012.

\bibitem{larsoft}
{\tt https://cdcvs.fnal.gov/redmine/projects/larsoftsvn/}.

\bibitem{geant4}
S.~Agostinelli et~al., {\it Geant4 -- a simulation toolkit},  {\em
  Nucl.Instrum.Meth.} {\bf A506} (2003) 250--303.

\bibitem{cry}
C. Hagmann, D. Lange, J. Verbeke, D. Wright, ``Cosmic-ray Shower Library (CRY),
  UCRL-TM-229453 (2012); {\tt http://cnr07.llnl.gov/simulation/main.html}.

\bibitem{genie}
C.~Andreopoulos et~al., {\it {The GENIE Neutrino Monte Carlo Generator}},  {\em
  Nucl. Instrum. Meth.} {\bf A614} (2010) 87--104.

\bibitem{art}
{\tt https://web.fnal.gov/project/ArtDoc/Pages/home.aspx}.

\bibitem{birks}
J.~B.~Birks, Proc. Phys. Soc. A {\bf 64}, 874 (1951); J.~B.~Birks, The Theory
  and Practice of Scintillation Counting, Pergamon, London (1964).; T.~Doke et
  al., Nucl. Instrum. Methods A {\bf 269}, 291 (1988).

\bibitem{modifiedbox}
J.~Thomas and D.~A. Imel, {\it Recombination of electron-ion pairs in liquid
  argon and liquid xenon},  {\em Phys. Rev. A} {\bf 36} (1987) 614.

\bibitem{nest}
M.~Szydagis et~al., {\it {NEST: A Comprehensive Model for Scintillation Yield
  in Liquid Xenon}},  {\em JINST} {\bf 6} (2011) P10002.

\end{thebibliography}\endgroup

\end{document}